\documentstyle[eqsecnum,prd,aps,amssymb]{revtex}

\def\be{\begin{equation}}
\def\ee{\end{equation}}
\def\baray{\begin{eqnarray}}
\def\earay{\end{eqnarray}}
\def\dlr{\partial_ r}
\def\dlth{\partial_\theta}
\def\dlph{\partial_\varphi}

\def\ofthph{(\theta,\varphi)}
\def\ofr{(r)}
\def\ofrthph{(r,\theta,\varphi)}
\def\ofphthz{(\varphi,\theta,0)}
\def\unit{{\mbox{\boldmath $e$}}}
\def\xibf{\mbox{\boldmath $\xi$}}
\def\xbf{{\bf x}}

\def\bfpi{\mbox{\boldmath $\pi$}}
\def\bfnabla{\mbox{\boldmath $\nabla$}}

\def\bfzeta{{\bf\zeta}}
\def\bfchi{{\bf\chi}}
\def\bfsigma{{\bf\sigma}}
\def\bfxi{{\mbox{\boldmath $\xi$}}}
\def\bfchi{{\mbox{\boldmath $\chi$}}}

\def\bfzeta{{\mbox{\boldmath $\zeta$}}}
\def\bfgamma{{\mbox{\boldmath $\gamma$}}}
\def\bfbeta{{\mbox{\boldmath $\beta$}}}
\def\bfsigma{{\mbox{\boldmath $\sigma$}}}
\def\bftau{{\mbox{\boldmath $\tau$}}}
\def\bfkappa{{\mbox{\boldmath $\kappa$}}}
\def\beq{\begin{equation}}
\def\endeq{\end{equation}}
\def\opL{{\bf L}}
\def\bfM{{\bf M}}
\def\half{{1\over 2}}

\def\sixth{{1\over 6}}
\def\third{{1\over 3}}

\def\rttwo{\sqrt{2}}
\def\sumlam{\sum_\Lambda}

\def\sumjinf{\sum_{j=0}^{\infty}}

\def\sumlamotth{\sum_{\Lambda_1\Lambda_2\Lambda_3}}

\def\Lo{L^{(1)}_{\Lambda_1\Lambda_2\Lambda_3}(r)}
\def\Li{L^{(i)}_{\Lambda_1\Lambda_2\Lambda_3}(r)}
\def\Lt{L^{(2)}_{\Lambda_1\Lambda_2\Lambda_3}(r)}
\def\Lth{L^{(3)}_{\Lambda_1\Lambda_2\Lambda_3}(r)}
\def\Lf{L^{(4)}_{\Lambda_1\Lambda_2\Lambda_3}(r)}

\def\rtti{{1\over\sqrt 2}}

\def\covdel{{\bf\nabla}}

\def\xiuilcdj{\xi^i_{\enskip ;j}}
\def\xiuilcdk{\xi^i_{\enskip ;k}}
\def\xiuilcdl{\xi^i_{\enskip ;l}}

\def\xiuklcdj{\xi^k_{\enskip ;j}}
\def\xiuklcdk{\xi^k_{\enskip ;k}}

\def\xiullcdk{\xi^l_{\enskip ;k}}
\def\xiullcdl{\xi^l_{\enskip ;l}}

\def\xiullcdl{\xi^l_{\enskip ;l}}

\def\delxi{\left( \grad \cdot \xivec \right)}

\def\gummbar{g^{m\bar m}}

\def\gull{g^{ll}}
\def\glmmbar{g_{m\bar m}}

\def\glll{g_{ll}}
\def\Gullmmbar{\Gamma^{l}_{m\bar m}}

\def\Gumllm{\Gamma^{m}_{lm}}

\def\Gumlmm{\Gamma^{m}_{m m}}
\def\Gumlmmbar{\Gamma^{m}_{m\bar m}}
\def\Gumbarlmbarmbar{\Gamma^{\bar m}_{\bar m\bar m}}

\def\eveclrhat{{\bf e}_{\hat r}}
\def\eveclthhat{{\bf e}_{\hat \theta}}
\def\eveclphhat{{\bf e}_{\hat \varphi}}
\def\Ylm{Y_{ lm}}

\def\Yjpmpo{Y_{j+m+1}^m}

\def\Yjpm{Y_{j+m}^m}

\def\Wjpmpo{W_{j+m+1}}

\def\Wjpm{W_{j+m}}
\def\Vjpmpo{V_{j+m+1}}

\def\Vjpm{V_{j+m}}
\def\Ujpmpo{U_{j+m+1}}

\def\Ujpm{U_{j+m}}

\def\lamjpm{{\sqrt{(j+m)(j+m+1)}\over\sqrt{2}}}
\def\lamjpmpo{{\sqrt{(j+m+1)(j+m+2)}\over\sqrt{2}}}

\def\delljpm{\delta_{j+m,l}}
\def\delljpmpo{\delta_{j+m+1,l}}

\def\Ylms{{_{s}}Y_{{l} m}}

\def\Ylmspo{{_{s+1}}Y_{{l} m}}
\def\Ylmsmo{{_{s-1}}Y_{{l} m}}
\def\Ylmsone{{_{s_1}}Y_{{l}_1 m_1}}
\def\Ylmstwo{{_{s_2}}Y_{{l}_2 m_2}}
\def\Ylmsthree{{_{s_3}}Y_{{l}_3 m_3}}

\def\Ylpmps{{_{s}}Y_{{l}^\prime m^\prime}}

\def\Ylamz{{_{0}}Y_{\Lambda}}

\def\Ylmz{{_{0}}Y_{{l} m}}
\def\Ylamo{{_{1}}Y_{\Lambda}}

\def\Ylammo{{_{-1}}Y_{\Lambda}}

\def\Ylamt{{_{2}}Y_{\Lambda}}

\def\Ylamtt{{_{2}}Y_{\Lambda_2}}

\def\Ylammt{{_{-2}}Y_{\Lambda}}

\def\Ylamomt{{_{-2}}Y_{\Lambda_1}}

\def\Ylamthmt{{_{-2}}Y_{\Lambda_3}}
\def\Ylamoz{{_{0}}Y_{\Lambda_1}}

\def\Ylamtz{{_{0}}Y_{\Lambda_2}}
\def\Ylamto{{_{1}}Y_{\Lambda_2}}

\def\Ylamthz{{_{0}}Y_{\Lambda_3}}
\def\Ylamtho{{_{1}}Y_{\Lambda_3}}
\def\Ylamthmo{{_{-1}}Y_{\Lambda_3}}

\def\pgtm{(+\leftrightarrow -)}

 \def\klabg{\kappa_{ABC}}

\def\sym{{\bf{\cal S}}}

\def\rhat{{\hat r}}
\def\thhat{{\hat \theta}}
\def\phhat{{\hat \varphi}}

\def\fzlam{f_0^\Lambda }

\def\gzlam{g_0^\Lambda }

\def\azlam{a_0^\Lambda }

\def\bzlam{b_0^\Lambda }
\def\fslam{f_s^\Lambda }

\def\fmslam{f_{-s}^\Lambda }

\def\fzlamo{f_0^{\Lambda_1} }
\def\fzlamt{f_0^{\Lambda_2} }
\def\fzlamth{f_0^{\Lambda_3} }

\def\gzlamo{g_0^{\Lambda_1} }

\def\gzlamt{g_0^{\Lambda_2} }

\def\gzlamth{g_0^{\Lambda_3} }

\def\azlamt{a_0^{\Lambda_2} }

\def\Gslam{G_s^\Lambda }

\def\Cslam{C_s^\Lambda }
\def\Gplam{G_+^\Lambda }

\def\Ezlam{E_0^\Lambda }

\def\Cplam{C_+^\Lambda }

\def\Gmlam{G_-^\Lambda }

\def\Cmlam{C_-^\Lambda }

\def\Dzlam{D_0^\Lambda }

\def\Fslam{F_s^\Lambda }
\def\Fplam{F_+^\Lambda }

\def\fplm{f_+^{{l} m}}
\def\fmlm{f_-^{{l} m}}
\def\fzlm{f_0^{{l} m}}
\def\Fmlam{F_-^\Lambda }

\def\Hslam{H_s^\Lambda }
\def\Hplam{H_+^\Lambda }

\def\Hmlam{H_-^\Lambda }

\def\fplam{f_+^\Lambda }

\def\fplamt{f_+^{\Lambda_2} }
\def\fplamth{f_+^{\Lambda_3} }

\def\Gplamo{G_+^{\Lambda_1} }
\def\Gplamt{G_+^{\Lambda_2} }
\def\Gplamth{G_+^{\Lambda_3} }
\def\Ezlamo{E_0^{\Lambda_1} }

\def\Cplamt{C_+^{\Lambda_2} }

\def\Fplamt{F_+^{\Lambda_2} }
\def\Fplamth{F_+^{\Lambda_3} }

\def\Hplamt{H_+^{\Lambda_2} }

\def\Dzlamo{D_0^{\Lambda_1} }

\def\Hmlamo{H_-^{\Lambda_1} }

\def\Hmlamth{H_-^{\Lambda_3} }

\def\fmlam{f_-^\Lambda }

\def\fmlamth{f_-^{\Lambda_3} }

\def\bplam{b_+^\Lambda }

\def\bmlam{b_-^\Lambda }

\def\xiumlcdm{\xi^m_{\enskip ;m}}
\def\xiumlcdl{\xi^m_{\enskip ;l}}
\def\xiumlcdmbar{\xi^m_{\enskip ;\bar m}}
\def\xiumbarlcdm{\xi^{\bar m}_{\enskip ; m}}
\def\xiumbarlcdl{\xi^{\bar m}_{\enskip ; l}}
\def\xiumbarlcdmbar{\xi^{\bar m}_{\enskip ; \bar m}}
\def\xiullcdm{\xi^l_{\enskip ;m}}
\def\xiullcdl{\xi^l_{\enskip ;l}}
\def\xiullcdmbar{\xi^l_{\enskip ;\bar m}}

\def\b{B}
\def\Bumlcdm{\b^m_{\enskip ;m}}
\def\Bumlcdl{\b^m_{\enskip ;l}}
\def\Bumlcdmbar{\b^m_{\enskip ;\bar m}}

\def\Bullcdm{\b^l_{\enskip ;m}}
\def\Bullcdl{\b^l_{\enskip ;l}}

\def\Bul{\b^l}
\def\Bum{\b^m}
\def\Bumbar{\b^{\bar m}}

\def\Bumldm{\b^m_{\enskip ,m}}
\def\Bumldl{\b^m_{\enskip ,l}}
\def\Bumldmbar{\b^m_{\enskip ,\bar m}}

\def\Bulldm{\b^l_{\enskip ,m}}
\def\Bulldl{\b^l_{\enskip ,l}}

\def\gamI{{1\over r}}
\def\ovrttr{{1\over\sqrt{2}r}}
\def\ovr{{1\over r}}
\def\gamII{{\cot\theta\over\sqrt{2} r}}
\def\mvec{{\bf m}}

\def\mbarvec{{\bf \bar m}}
\def\lvec{{\bf l}}

\def\vtz{\vartheta_0}
\def\vto{\vartheta_1}
\def\vtmo{\vartheta_{-1}}

\def\Avec{{\bf A}}
\def\bvec{{\bf B}}

\def\ofrthetaphi{(r,\theta,\varphi)}
\def\fofrthph{{(r,\pi-\theta,\varphi)}}

\def\rpar{\right )}
\def\lpar{\left (}

\def\opL{{\bf L}}

\def\rhonot{\rho}
\def\phinot{\phi}
\def\pnot{p}

\def\Thetaujli{\Theta^j_{\enskip i}}
\def\Thetauilj{\Theta^i_{\enskip j}}
\def\Xiujli{\Xi^j_{\enskip i}}
\def\Xiuilj{\Xi^i_{\enskip j}}

\def\uvec{{\bf u}}

\def\xivecofxt{\xivec\ofxt}
\def\Omvec{{\mbox{\boldmath $\Omega$}}}

\def\xivec{{\mbox{\boldmath $\xi$}}}

\def\grad{{\mbox{\boldmath $\nabla$}}}
\def\gradi{\grad_{0,i}}
\def\gradj{\grad_{0,j}}
\def\gradk{\grad_{0,k}}

\def\gradjp{\gradj^\prime}
\def\gradkp{\gradk^\prime}

\def\eperm{{\cal E}}
\def\xvec{{\bf x}}
\def\uvec{{\bf u}}
\def\xzero{\xvec_0}

\def\xzerop{\xzero^\prime}
\def\ofxt{(\xzero,t)}
\def\ofxpt{(\xzero^\prime,t)}

\def\xij{\xivec_j\ofxt}

\def\xijp{\xivec_j\ofxpt}
\def\xikp{\xivec_k\ofxpt}
\def\volx{{\rm d}^3 x\,}
\def\volzero{{\rm d}^3\xzero\,}
\def\volzerop{\volzero^\prime}
\def\Fvec{{\bf F}}
\def\Fi{\Fvec_i}
\def\det{{\rm det}}

\def\Jxxzero{J(\xvec,\xzero)}
\def\dM{{\rm d}M}

\def\dMp{\dM^\prime}

\def\ttwo{T_2}
\def\lag{{\cal L}}
\def\kin{{\cal T}}
\def\pot{{V}}

\def\Ugh{{U}}
\def\Vgh{\pot_G}

\def\fvec{{\bf f}}
\def\pivec{{\mbox{\boldmath $\pi$}}}
\def\muvec{\vec\mu}

\begin{document}
\title{Nonlinear mode coupling in rotating stars and the $r$-mode instability
in neutron stars}
\author{A. K. Schenk}
\address{Sloan Center for Theoretical Neurobiology, UCSF,
 513 Parnassus, San Francisco, CA 94143-0444}
\author{P. Arras}
\address{Canadian Institute for Theoretical Astrophysics, University of Toronto}
\author{\'{E}. \'{E}. Flanagan, S. A. Teukolsky, and I. Wasserman}
\address{Center for Radiophysics and Space Research, Cornell
         University, Ithaca, New York, 14853}
\date{\today}
\maketitle

\begin{abstract}

We develop the formalism required to study the nonlinear interaction
of modes in rotating Newtonian stars, assuming that the mode amplitudes are
only mildly nonlinear.  The formalism is simpler than previous
treatments of mode-mode interactions for spherical stars, and
simplifies and corrects previous treatments for rotating stars.
At linear order, we elucidate and extend slightly a formalism due to
Schutz, show how to decompose a general motion of a rotating star into a 
sum over modes, and obtain uncoupled equations of motion for the mode 
amplitudes under the influence of an external force. 
Nonlinear effects are added perturbatively via three-mode couplings,
which suffices for moderate amplitude modal excitations; the formalism
is easy to extend to higher order couplings. We describe a new, efficient
way to compute the modal coupling coefficients, to zeroth order in the
stellar rotation rate, using spin-weighted spherical harmonics.
The formalism is general enough to allow computation of the initial
trends in the evolution of the spin frequency and differential
rotation of the background star.

We apply this formalism to derive some properties of the coupling
coefficients relevant to the nonlinear interactions of unstable
$r$-modes in neutron stars, postponing numerical integrations of the
coupled equations of motion to a later paper.  First, we clarify some
aspects of the  
expansion in stellar rotation frequency $\Omega$ that is often used
to compute approximate mode functions.  We show that in zero-buoyancy
stars, the rotational modes (those modes whose frequencies vanish as
$\Omega \to 0$) are orthogonal to zeroth order in $\Omega$.
From an astrophysical viewpoint, the most
interesting result of this paper is that many couplings of $r-$modes
to other rotational modes are small:  either
they vanish altogether because of various selection rules,
or they vanish to lowest order in $\Omega$ or in compressibility.
In particular, in zero-buoyancy stars, 
the coupling of three $r-$modes is forbidden entirely
and the coupling of two $r$-modes to one hybrid, or $r$-$g$ rotational
mode vanishes to zeroth order in rotation frequency.
The coupling of any three rotational modes vanishes to zeroth order in
compressibility and in $\Omega$.
In nonzero-buoyancy stars, coupling of the $r$-modes to each other
vanishes to zeroth order in $\Omega$ .
Couplings to regular modes (those modes whose frequencies are finite
in the limit $\Omega \to 0$), such as $f-$modes, are not zero, 
but since the natural frequencies of these modes are relatively large
in the slow rotation limit compared to those of the $r$-modes, energy
transfer to those modes is not expected to be efficient.

\end{abstract}

\pacs{04.40.Dg, 97.10.Sj, 97.60.Jd}


\section{Introduction}
\label{sec:intro}
Various authors have proposed that $r$-modes might be linearly unstable inside
fairly rapidly rotating neutron stars, depending on a competition
between gravitational radiation reaction, which drives the
instability, and viscous effects, which inhibit it; several
other investigators have explored implications for the spin evolution of
rapidly rotating newborn and accreting neutron stars
\cite{1998ApJ...502..708A,1998ApJ...502..714F,prl...80...4843,1998ApJ...501L..89B,physrevD...58...084020,1999ApJ...516..307A,1999ApJ...510..846A,1999ApJ...517..328L,astro-ph0006028,astro-ph9911188,gr-qc9909084,1999A&A...125..193,diffrot,anderssonreview}.
An open question is how the instability saturates, and at what
perturbation amplitude. This question is not only important from
a theoretical standpoint, but also must be answered in order to
assess whether or not the gravitational radiation emitted during
the development of the instability is detectable. 
The saturation amplitude also determines the final spin rate of the
neutron star when it exits the instability region of parameter space.
While turbulence generated
in the strong shear layer at the interface between the fluid core and solid 
crust has been shown to saturate the $r$-mode \cite{astro-ph0006123}, a
crust will not be present initially for newly born neutron stars, and
other hydrodynamical saturation mechanisms are still being
investigated and may in fact be more important.

There have been, to date, two fully nonlinear numerical calculations of the
development of the instability, but neither completely settles the
question of saturation, principally because there are practical
limitations on what it is feasible to compute. The first calculation,
by Stergioulas and Font \cite{gr-qc0007086},
considered the evolution of the fluid inside a neutron star with
no buoyancy force
and fixed spacetime geometry. Their
calculations began with a large amplitude excitation in the 
mode expected to be most unstable, the ${l}=m=2$ $r$-mode, and
followed the subsequent hydrodynamics for about 20 stellar rotation
periods, approximately 25 ms for their calculations.  During this time
no substantial change in the amplitude  
of the modal excitation was seen, and no evidence for significant
nonlinear excitation of other stellar normal modes was detected.
Lindblom, Tohline and Vallisneri \cite{astro-ph0010653}
subsequently performed simulations based on the Newtonian equations
of hydrodynamics and gravitation, including an artificially enhanced
gravitational radiation force consisting of the correct force
\cite{prd...55..714,ApJ1999...525..939} multiplied by 
a factor of about 4500 \cite{Leenote}.  The linear instability growth
timescale in  
their simulations was about 13 rotation periods, rather than the
correct value of $\agt 10^4$ rotation periods.  The initial $r-$mode
amplitude chosen by Lindblom, Tohline and Vallisneri
was small (about 0.1), and, as expected, grew via the radiation-driven
instability to large amplitude, where shocks ultimately formed,
accompanied by a decrease in $r-$mode amplitude.
Both calculations show that nonlinear mode-mode coupling is comparatively
weak for $r-$modes, producing little effect even at large modal amplitude
on timescales $\sim 10-20$ stellar rotation periods.

A second, more analytical approach to this problem is possible.
One assumes, at the outset, that the modes develop only moderately
nonlinear amplitudes, and nonlinear effects are modeled perturbatively,
via $n-$mode couplings. When amplitudes are small, relatively low order
couplings, such as three-mode interactions, suffice, and
(when dissipation and radiation reaction are ignored) the equations
of motion for the fluid can be modeled as 
an infinite-dimensional Hamiltonian system with
a polynomial interaction potential (which is cubic when only three-mode
couplings are kept). This approach has the advantage of allowing
explicit calculation of the strength of the lowest order
coupling of a given $r-$mode
to other modes of the star, but has the disadvantage of failing when
mode amplitudes grow very large. Although it is possible that the
$r-$mode instability saturates only in the fully nonlinear regime,
neither of the numerical calculations establish this,
because it is possible that modal couplings act on 
timescales longer than covered by the simulations, but shorter than
the growth time ($\sim 10^4$ rotation periods) of the
gravitational-radiation instability.  The analytic approach we are
taking should apply if the instability 
saturates before attaining very large amplitudes.  Moreover, by
computing the three-mode coupling coefficients for $r-$modes,
we should be able to shed some light on why nonlinear coupling is
so inefficient for them on dynamical timescales in the simulations.

In this paper, we develop the tools needed to calculate the
nonlinear evolution of the unstable modes in the weakly nonlinear
regime. We concentrate on three-mode interactions for the most
part, although some of the results we derive could also apply
to interactions involving $n>3$ modes of the star.

We begin with a review of the linear perturbation theory in 
Sec.~\ref{sec:sec2}. A principal goal of that section of the paper
is to obtain equations of motion for modes of a rotating star acted
on by an external force that are uncoupled from one another at
linear order. The method we develop is based on a series of papers by
Schutz and collaborators \cite{1978ApJ...221..937F,1978ApJ...222..281F,1979ApJ...232..874S,1979RSLPS.368..389D,1980MNRAS.190....7S,1980MNRAS.190....21S},
but has not been 
applied consistently to $r-$modes before. The key results, for a reader
wishing to skip the details, are the mode decomposition formulae
(\ref{goodexpansion1a}) and (\ref{inverseexpansion1a})
and the equation of motion for each mode (\ref{fa1}). In Sec.~\ref{sec:sec3},
we review the slow rotation approximation to mode functions and
frequencies, with particular attention to the {\it rotational} modes,
whose frequencies are zero in the limit of zero rotation frequency,
$\Omega$.  The key new result in this section is the fact that the
rotational mode functions are orthogonal to leading order in the
stellar angular velocity.

In Sec.~\ref{sec:sec5}, 
we discuss the second order Lagrangian perturbation theory at the
heart of our calculation of the evolution of modes in the weakly
nonlinear regime.  The key results of this section are (i) the 
coupled nonlinear equations of motion (\ref{fanl3}) for the mode
coefficients, and (ii) the formula (\ref{eq:23}) 
for the three-mode coupling coefficients that can be
evaluated once the modes have been found for a given background star
model. In  
Sec.~\ref{sec:sec6}, we discuss properties of the coupling coefficients.
We show how general parity arguments and selection rules can be used to
explain why some modes do not couple to one another, and also show
that there are other couplings that are suppressed,
in that they may vanish to lowest order in
$\Omega$, or to lowest order in compressibility, but are nonzero more
generally. In Sec.~\ref{sec:sec7}, 
we develop a new and efficient way of computing coupling coefficients
that employs spin-weighted spherical harmonics \cite{newmanpenrose}.
The method greatly simplifies the calculation of the coupling
coefficients for nonrotating stars as well as rotating stars, and
should be useful for other astrophysical applications (see, e.g.,
Refs.\ \cite{Thesis:Wu,1996ApJ...466..946K}).

This paper concentrates on formal issues primarily, postponing
numerical integration of the time-dependent evolution of a network
of weakly interacting modes to a subsequent publication.
From a formal viewpoint, the most substantial result of this paper
is a formulation of the equations of motion that, at linear order,
separates the responses of individual modes to an external force,
and, at the first nonlinear order, involves coupling coefficients
that we can calculate relatively economically.
From an astrophysical viewpoint, the most interesting result is
that many interactions of $r-$modes with themselves can be shown
to be either forbidden entirely by selection rules, or 
suppressed to lowest order in some parameter, such as
rotation frequency or compressibility. By itself, this
is not quite enough to explain the smallness of the coupling
seen in numerical simulations, but we may already be
seeing a hint of why the transfer of energy from $r-$modes to other
modes is so inefficient on dynamical timescales.

\section{Linear Perturbation Theory for Rotating Stars:  The Schutz 
Formalism}
\label{sec:sec2}

\subsection{Overview}
\label{sec:Linear-overview}

In this section we develop the formalism we need in the linear
Lagrangian perturbation theory of rotating stars.  
There are two aspects to the perturbation theory that will be
essential for our purposes:

(i) Solving for the frequencies $\omega_\alpha$ and mode functions
$\xivec_\alpha(\xvec)$.  This aspect is well understood and highly
developed; see for example Ch. VI of Unno {\it et al} \cite{unnoetal},
as well as Refs.\
\cite{1990ApJ...355..226I,1996ApJ...460..827B,astro-ph9909009}.   

(ii) Decomposing general motions in a star into sums over modes, and 
obtaining equations of motion for the mode coefficients that are
uncoupled at linear order when an external force acts on the star. 

The second aspect of the theory --- deriving equations of motion for
the mode coefficients --- has not been well-understood in the
past.  The necessary theory is implicit in a
series of papers by Schutz and collaborators 
\cite{1978ApJ...221..937F,1978ApJ...222..281F,1979ApJ...232..874S,1979RSLPS.368..389D,1980MNRAS.190....7S,1980MNRAS.190....21S},
but the formalism has not been applied widely in astrophysics.
For example, a classic computation for which modal equations of motion
would be useful is the excitation of the modes of a rotating star by the
tidal field of a binary companion.  While there have been many papers
on this topic, we are not aware of any that have used the correct uncoupled
equations of motion. For example, Refs.\ \cite{1987A&A...175...81R} and
\cite{1981MNRAS.196..371P} use
a type of mode expansion [Eq.~(\ref{badexpansion1}) below] for
which the equations of motion of the various mode amplitudes are coupled,
and
Refs.\ \cite{1995ApJ...449..294K}, \cite{1997ApJ...490..847L} and \cite
{1999MNRAS.308..153H} use the same mode expansion but simply drop
the coupling terms between the different modes.  
Some authors have resorted to numerically solving the
linearized hydrodynamic equations in the frequency domain, after
factoring out a factor of $e^{i m \varphi - i \omega t}$, rather
than using a mode expansion \cite{1995MNRAS.277..471S,1997MNRAS.291..633S,1999A&A...341..842W}.

This section of the paper is devoted to presenting and explaining the
equations of motion that result from Schutz's
formalism when one makes 
a number of simplifying assumptions that usually are valid in
applications.  We have attempted to make the presentation transparent
so as to be accessible to a wide audience.  
All the derivations are relegated to an appendix.  We start in
Sec.~\ref{sec:eqn-foundations} by summarizing the governing  
equations.  In Sec. \ref{sec:schutz-formalism} we describe the 
form of the mode expansion and equations of motion for the mode
coefficients. 

\subsection{Governing equations}
\label{sec:eqn-foundations}

We assume that the unperturbed background star is uniformly rotating
with angular velocity ${\bf \Omega}$.  The fluid equations of motion
in the corotating frame are 
\begin{equation}
{\partial \rho \over \partial t} + {\bf \nabla} \cdot ( \rho {\bf u})
=0
\label{continuity0}
\end{equation}
and 
\begin{equation}
{\partial {\bf u} \over \partial t} + ({\bf u} \cdot {\bf \nabla})
{\bf u} + 2 {\bf \Omega} \times {\bf u}
+ {\bf \Omega} \times \left( {\bf \Omega} \times {\bf x} \right) = 
- {{\bf \nabla} p \over
\rho} - {\bf \nabla} \phi + {\bf a}_{\rm ext},
\label{euler0}
\end{equation}
where $\rho$ is the density, $p$ the pressure, ${\bf u}$ the
velocity and ${\bf a}_{\rm ext}$ is any acceleration due to external forces.
The gravitational potential $\phi$ is given by 
${\bf \nabla}^2 \phi = 4 \pi G \rho$. 
In the background solution, ${\bf u}$ vanishes and $\rho$ is time
independent.

Now consider linearized perturbations characterized by Eulerian perturbations 
$\delta {\bf u}$, $\delta \rho$ and $\delta p$.  
The first two of these are related to the linearized Lagrangian
displacement $\xivec(\xvec,t)$ by 
\beq
\delta \rho = - {\bf \nabla} \cdot( \rho \xivec)
\label{deltarhodef}
\endeq
 and $\delta {\bf u} =  {\dot \xivec}$ (since ${\bf u}=0$).  
We assume that the perturbations are characterized by an adiabatic index 
$\Gamma_1$, so that
\begin{equation}
{\Delta p \over p} = 
\Gamma_1\, { \Delta \rho \over \rho},
\label{Gamma1def}
\end{equation}
where $\Delta \rho = \delta \rho + (\xivec \cdot \bfnabla ) \rho$ and
$\Delta p = \delta p + (\xivec \cdot \bfnabla) p$ are the Lagrangian
perturbations of density and pressure\footnote{If the background star
and the perturbations both obey the same one-parameter  
equation of state $p = p(\rho)$, then $\Gamma_1 = (\rho/p) dp/ d\rho$
depends only on the background density $\rho$.  More generally, we can
regard $\Gamma_1 = \Gamma_1({\bf x})$ as a function of position
determined by the structure of the background star.  A very general
definition of the adiabatic index $\Gamma_1 = \Gamma_1({\bf x})$ that
is consistent with the phenomenological relation (\ref{Gamma1def}) is
given in Eq.\ (\ref{gammaonedef}) below.}.
By combining these relations with linearized versions of the
continuity and Euler equations 
(\ref{continuity0}) and (\ref{euler0}),
\begin{equation}
{\delta {\dot \rho}} + {\bf \nabla} \cdot ( \rho \delta {\bf u}) =0,
\label{continuity0l}
\end{equation}
and
\begin{equation}
\delta {\dot {\bf u}} + 2 {\bf \Omega} \times  \delta {\bf u} = - {{\bf
\nabla} \delta p \over \rho} + { {\bf \nabla} p \over \rho^2} \delta
\rho - {\bf \nabla} \delta \phi + {\bf a}_{\rm ext},
\label{euler0l}
\end{equation}
where ${\bf \nabla}^2 \delta \phi = 4 \pi G \delta \rho$, we obtain
the equations of motion for linearized perturbations in the form
\cite{lbo}
\begin{eqnarray}
{\ddot {\bfxi}} + {\bf B} \cdot {\dot {\bfxi}} + {\bf C} \cdot
{\bfxi}= {\bf a}_{\rm ext}(\xvec,t).
\label{basic0}
\end{eqnarray}
Here the operator ${\bf B}$ is given by
\begin{eqnarray}
{\bf B} \cdot {\bfxi} \equiv 2 {\bf \Omega} \times \bfxi,
\label{Bdef}
\end{eqnarray}
and the operator ${\bf C}$ is given by
\cite{lbo}
\begin{eqnarray}
\rho \left({\bf C} \cdot \bfxi\right)_i &=&  -
\nabla_i(\Gamma_1 p \nabla_j \xi^j) + \nabla_i p \, \nabla_j \xi^j
- \nabla_j p \, \nabla_i \xi^j \nonumber \\
\mbox{} && + \rho \xi^j \nabla_j \nabla_i \phi 
  + \rho \xi^j \nabla_j \nabla_i \phi_{\rm rot} +
  \rho \nabla_i \delta \phi,
\label{Cdefa}
\end{eqnarray}
where
\beq
\phi_{\rm rot}({\bf x}) = - {1 \over 2} ({\bf \Omega} \times {\bf
x})^2.
\label{phirotdef}
\endeq
The ${\bf C}$ operator can be broken up into two pieces 
\baray
{\bf C}={\bf C}_a
+ {\bf C}_b
\label{Cdef}
\earay
where ${\bf C}_a$ is proportional to the vectorial Schwarzschild
discriminant 
\beq
{\bf {A}} = { \bfnabla \rho \over \rho} - { 1 \over \Gamma_1}
{\bfnabla p \over p},
\endeq
and ${\bf C}_b$ is independent of ${\bf A}$.  We find
\beq
{\bf C}_a \cdot {\bfxi} =  \grad 
\left( \frac{\Gamma_1 p}{\rho} \frac{\delta \rho}{\rho} + \delta \phi \right)
\label{Cdef-simple}
\endeq
and 
\baray
{\bf C}_b \cdot {\bfxi} & = & \grad \left( \frac{\Gamma_1 p}{\rho}
\xibf \cdot \Avec \right) 
+ \frac{\delta p}{\rho} \Avec + \frac{\grad p}{\rho} \xibf \cdot \Avec
\\ &=& { p \over \rho^2} \delta \rho \Gamma_1 \Avec + {1 \over \rho} {\bf \nabla} \left( \Gamma_1 p \xivec \cdot \Avec \right).
\label{Cdef-b}
\earay
The buoyancy force described by the operator ${\bf C}_b$
gives rise to $g$ modes whose frequency scale is set by 
the Brunt-V\"{a}is\"{a}l\"{a} frequency $N$ (Brunt for short),
defined by
\begin{equation}
N^2 = {{\bf \nabla} p \over \rho} \cdot {\bf A}.
\end{equation}

For much of this paper we will specialize to perturbations of stars
without buoyancy, i.e., situations where ${\bf {A}}=0$\footnote{For
such stars $dp \wedge d\rho =0$ and so the background star satisfies
$p = p(\rho)$ for some equation of state.  Furthermore the
perturbations obey the same equation of state, so the star is barotropic.  
[A special case of this is a star of uniform specific entropy, i.e.,
an isentropic star.]  For more general stars $dp$ and $d\rho$ need not
be proportional.}.  For 
example, 
if the background star and the perturbations obey one parameter
equations of state with adiabatic indices $\Gamma$ and $\Gamma_1$,
respectively, then ${\bf {A}} \propto \Gamma-\Gamma_1$, and so the
zero-buoyancy case is $\Gamma=\Gamma_1$.  
For ordinary stars, the Brunt frequency in the convection zone is
approximately zero when the superadiabatic gradient is small.
For neutron stars, ${\bf
A}=0$ means a cold, zero-entropy gas in which the composition adjusts
instantly to changes in density.   In real neutron stars the
adjustment speed is limited by beta and inverse-beta decays, which
gives rise to $g$ modes \cite{gmoderefs} (see Appendix
\ref{variational} for a related discussion).
Our analysis of $r$ modes will make the zero-buoyancy
approximation ${\bf A}=0$, which is expected   
to be good in the regime $N \ll \Omega$.  However, this inequality
will only be marginally satisfied in newly born neutron stars, so 
it will be important for future analyses to analyze nonlinear
couplings of $r$ modes in nonzero-buoyancy stars.  We choose to focus
on the zero-buoyancy case for simplicity.

The dynamical equation (\ref{basic0}) forms the starting point for our
discussions.  We can find a large class of solutions 
for the case ${\bf a}_{\rm ext}=0$ of no forcing if we 
make the ansatz 
\begin{eqnarray}
\bfxi({\bf x},t) = e^{- i \omega t} \bfxi({\bf x}),
\end{eqnarray}
where $\omega$ is the rotating-frame frequency.  This yields 
\begin{eqnarray}
\left[ - \omega^2  - i  \omega {\bf B}  + {\bf C} \right] \cdot
{\bfxi}=0.
\label{basic1}
\end{eqnarray}
The quadratic eigenvalue equation (\ref{basic1}) is the standard
equation that one solves to  
obtain the eigenfunctions $\xivec({\bf x})$ and eigenfrequencies
$\omega$ for rotating stars.  A mode of the star will consist of a
pair $(\xivec,\omega)$ that satisfy Eq.\ (\ref{basic1}).

\subsection{Mode decomposition formalism}
\label{sec:schutz-formalism}

\subsubsection{Nonrotating stars}
\label{sec:nonrot}

We start by recalling the standard mode decomposition formalism for
non-rotating stars for which ${\bf B}=0$, in order to contrast it with
the rotating case below.  For non-rotating stars Eq.\ 
(\ref{basic1}) reduces to 
\begin{equation}
{\bf C} \cdot \xivec = \omega^2 \xivec,
\label{basic-nr}
\endeq
which represents a standard eigenvalue problem for $\omega^2$.
We define the inner product on the space ${\cal H}$ of
complex vector functions 
$\bfxi = \bfxi({\bf x})$ by
\begin{eqnarray}
\left< \bfxi \, , \, \bfxi^\prime \right> = \int d^3x \,
\rho({\bf x}) \, \bfxi({\bf x})^*  \cdot  \bfxi^\prime({\bf x}).
\label{innerproduct}
\end{eqnarray}
The operator ${\bf C}$ is Hermitian with respect to this inner
product\footnote{Throughout this paper we assume that the space ${\cal H}$ is
finite dimensional, as it will be in any numerical computations, 
to avoid discussing subtleties related to infinite dimensional spaces.}:
for any elements $\bfxi$ and 
$\bfxi^\prime$ of ${\cal H}$, 
\begin{eqnarray}
\left< \bfxi \, , \, {\bf C} \cdot \bfxi^\prime \right> = 
\left< {\bf C} \cdot \bfxi \, , \, \bfxi^\prime \right>,
\end{eqnarray}
while ${\bf B}$ is anti-Hermitian
\begin{eqnarray}
\left< \bfxi \, , \, {\bf B} \cdot \bfxi^\prime \right> = 
- \left< {\bf B} \cdot \bfxi \, , \, \bfxi^\prime \right>.
\label{eq:antiHermitian}
\end{eqnarray}
It follows that one can find a set $\{ \xivec_\alpha \}$ of
eigenvectors of ${\bf C}$ with eigenvalues $\omega_\alpha^2$
which are orthonormal
\beq
\left< \xivec_\alpha \, , \, \xivec_\beta \right> = 
\delta_{\alpha\beta}
\label{orthonormal0}
\endeq
and which also form a basis for the vector space ${\cal H}$.  Hence we can
decompose any Lagrangian displacement $\xivec(\xvec,t)$ as
\footnote{
The sum over $\alpha$ in Eq.\ (\ref{nonrot-decompos}) is over all
distinct solutions $\xivec_\alpha$ to Eq.\ (\ref{basic-nr}), not over
the larger set of distinct pairs $(\xivec,\omega)$, as there is
only one term in the sum (\ref{nonrot-decompos}) corresponding to the
two solutions $(\xivec_\alpha,\omega_\alpha)$ and
$(\xivec_\alpha,-\omega_\alpha)$.  This will be important below.}
\beq
\xivec(\xvec,t) = \sum_\alpha \, q_\alpha(t) \, \xivec_\alpha(\xvec),
\label{nonrot-decompos}
\endeq
where
\beq
q_\alpha(t) = \left< \xivec_\alpha \, , \, \xivec(t) \right>.
\label{qalphadef}
\endeq
Finally, by inserting the expansion (\ref{nonrot-decompos}) into the
dynamical equation (\ref{basic0}) with ${\bf B}=0$ and by contracting
with $\xivec_\alpha$, one obtains the equation of motion
\beq
{\ddot q}_\alpha(t) + \omega_\alpha^2 q_\alpha(t) = 
 \left<
\xivec_\alpha \, , \, {\bf a}_{\rm ext}(t) \right>,
\label{eenr}
\endeq
which is just the forced harmonic oscillator equation.

\subsubsection{Rotating stars}
\label{sec:linear_rot}

Turn, now, to the corresponding formalism for rotating stars.
We make a number of simplifying assumptions:
\begin{itemize}
\item The background star is stable, so all frequencies are real.
\item All the modes are non-degenerate: for each frequency $\omega$
the dimension of the space of eigenmodes is one.  This assumption will
be relaxed in Sec.\ \ref{sec:degeneracy} below.  
\end{itemize}
A more general treatment, without these assumptions, is given in
Appendix \ref{sec:proofs}.

The eigenvalue equation (\ref{basic1}) can be reformulated as a
standard eigenvalue equation [Eq.\ (\ref{rtevec2}) below] for $\omega$
involving a non-Hermitian operator.  Non-Hermitian operators in
general do not posess sufficiently many right eigenvectors 
to form a basis. However, for such operators there is a standard
procedure described in linear algebra textbooks to obtain a 
basis by adding to the eigenvectors certain additional vectors,
sometimes called generalized eigenvectors or Jordan chain vectors.
Each actual eigenvector may have one or more generalized eigenvectors
associated with it, forming a so-called Jordan chain. 

There are two classes of modes $(\xivec,\omega)$ in a rotating star:
modes associated with non-trivial Jordan chains
\cite{1980MNRAS.190....7S}, which we shall call Jordan-chain modes,
and modes not associated with such Jordan chains (i.e., the length
of the chain is zero; see Appendix
\ref{sec:proofs} for more details). 
A mode that is just at the point of becoming unstable is always
a Jordan-chain mode \cite{1980MNRAS.190....7S}. 
However, even for stable rotating stars, Jordan-chain
modes of zero frequency are always present.  One set of such modes
corresponds to displacements of the star to nearby equilibria
with slightly different angular velocities
\beq
\Omega \to \Omega + \delta \Omega(r_\perp),
\endeq
where $r_\perp$ is the distance from the rotation axis (see Appendix
\ref{sec:proofs3} for a proof of this in the zero-buoyancy
case)\footnote{There are also Jordan chain modes corresponding to
tilting the axis of rotation of the star.}.  Therefore, one cannot
assume that no Jordan chains    
occur.  In this section we shall nevertheless, for simplicity,
describe the formalism that would apply if there were no Jordan-chain
modes.  This formalism should be useful, for example, in describing
situations in which the Jordan-chain modes are 
unimportant dynamically. In the remainder of this paper we shall
assume that the 
Jordan chain modes are unimportant for the process of saturation of
$r$-modes\footnote{However, note that several studies have found
that the gravitational radiation reaction and/or hydrodynamic
nonlinearities induce 
significant differential rotation in the star
\protect{\cite{diffrot,gr-qc0007086,astro-ph0010653}}.  A set of coupled
equations for the growth of the $r$-mode 
as well as for the differential rotation can be formulated using the results
of Appendix \ref{sec:proofs}.  Since the differential rotation corresponds
to a set of Jordan chains of length one (Appendix \ref{sec:proofs3}),
Eqs.\ (\ref{ee3}) and (\ref{ee4}) 
can be used to evolve the differential rotation in the linear regime;
however the linear approximation breaks down after a time $\sim
1/\delta \Omega$.}.
A more complete treatment of the formalism, allowing
Jordan-chain modes, is detailed in Appendix \ref{sec:proofs}.   

Given these assumptions, there are two obstacles to obtaining a
mode decomposition formalism similar to that for 
non-rotating stars.  First,
distinct modes $(\xivec,\omega)$ and $(\xivec^\prime,\omega^\prime)$
satisfying Eq.\ (\ref{basic1}) with $\omega \ne \omega^\prime$ need
not be orthogonal
with respect to the inner product (\ref{innerproduct}) .  Although 
the operator
\beq
{\bf L}(\omega) = - \omega^2 - i \omega {\bf B} + {\bf C}
\label{ldef}
\endeq
is Hermitian for real $\omega$ \cite{lbo}, the vectors
$\xivec$ and $\xivec^\prime$ are eigenvectors of the two
different operators ${\bf L}(\omega)$ and ${\bf
L}(\omega^\prime)$ and so need not be orthogonal.   
Nevertheless, it is possible to find a set $\{ \xivec_\alpha \}$ of
solutions to Eq.\ (\ref{basic1}) that form a basis of ${\cal H}$, so
that every Lagrangian displacement can be uniquely
decomposed as a superposition of the form 
\beq
\xivec(\xvec,t) = \sum_\alpha \, q_\alpha(t) \, \xivec_\alpha(\xvec).
\label{badexpansion}
\endeq
The second obstacle is the following: {\it There is in general no
choice of 
basis} $\{ \xivec_\alpha \}$ {\it of eigenvectors for which the
equations of motion of the coefficients} $q_\alpha(t)$ {\it defined by
Eq.~(\ref{badexpansion}) are uncoupled from one another}.  

To circumvent this obstacle, it is necessary to use a phase space mode
expansion instead of a configuration space mode expansion, as pointed
out by Dyson and Schutz \cite{1979RSLPS.368..389D}.  Let's label
the distinct non-Jordan-chain solutions\footnote{Here we define two solutions
$(\xivec,\omega)$ and $(\xivec^\prime,\omega^\prime)$ to be 
distinct if $\omega \ne \omega^\prime$ or if $\xivec$ and
$\xivec^\prime$ are linearly independent.} $(\xivec,\omega)$ to
Eq.~(\ref{basic1}) as 
$(\xivec_A,\omega_A)$, so that
\beq
\left[ -\omega_A^2 - i \omega_A {\bf B} + {\bf C} \right] \cdot
\xivec_A =0.
\label{sdf}
\endeq
The set of vectors
\begin{eqnarray}
 \left[ \begin{array}{c}  
	\xivec_A\\
	- i \omega_A \xivec_A
	\end{array} \right]
\label{basis2}
\end{eqnarray}   
forms a basis for the space ${\cal H} \oplus {\cal
H}$ of pairs of vectors $[ \bfxi, \bfxi^\prime]$.  
We deliberately use a different type of index
here --- capital Roman indices rather than lower case Greek indices ---
as a reminder that the number of distinct $A$'s is twice the
dimension of ${\cal H}$, whereas the number of distinct $\alpha$'s in
the sums (\ref{nonrot-decompos}) and (\ref{badexpansion}) is 
just the dimension of ${\cal H}$.

The mode functions $\xivec_A$ are not orthogonal in general, that is,
$
\left< \xivec_A \, , \, \xivec_B \right>
$
need not vanish when $A \ne B$.
We give an explicit example demonstrating the non-orthogonality in
Sec.~\ref{sec:sec3} below.  However, as shown by Ref.\
\cite{1978ApJ...222..281F}, 
the modes $(\xivec_A,\omega_A)$ do
obey a modified type of orthogonality relation, which is
\footnote{The left hand side of Eq.\ (\ref{orthonormal1}) is
proportional to the symplectic product $W(\xivec_1,\xivec_2)$ defined by 
Ref.\ \cite{1978ApJ...222..281F}, where $\xivec_1({\bf x},t) = \xivec_A({\bf
x}) \exp[- i \omega_A t]$ and $\xivec_2({\bf x},t) = \xivec_B({\bf x})
\exp[- i \omega_B t]$.  An alternative proof of the orthogonality
relation (\ref{orthonormal1}) from a phase space construction was
given by Ref.\ \cite{1980MNRAS.190....7S}, which we review in Appendix
\ref{sec:proofs}.}
\beq
\left< \xivec_A \, , \, i {\bf B} \cdot \xivec_B \right> + ( \omega_A
+ \omega_B ) \left< \xivec_A \, , \xivec_B \right> = 0
\label{orthonormal1}
\endeq
for $A \ne B$.  

Now, at any time $t$, we can form the vector 
\begin{eqnarray}
 \left[ \begin{array}{c}  
	\xivec(t)\\
	{\dot \xivec}(t)
	\end{array} \right]
\label{zetadef0}
\end{eqnarray}   
from the Lagrangian displacement $\xivec(\xvec,t)$ and its time
derivative ${\dot \xivec}(\xvec,t)$.  We can expand this vector in the basis
(\ref{basis2}) as\footnote{Note that it follows from Eq.\ (\ref{goodexpansion})
that
$$
\sum_A \, ( {\dot c}_A + i \omega_A c_A ) \xivec_A =0.
$$
However, it does not follow that ${\dot c}_A + i \omega_A c_A =0$ for
each $A$ since the set $\{ \xivec_A \}$ is {\it not} a basis for
${\cal H}$; it has twice too many basis elements.}  
\begin{eqnarray}
 \left[ \begin{array}{c}  
	\xivec(t)\\
	{\dot \xivec}(t)
	\end{array} \right] = 
\sum_A c_A(t) \left[ \begin{array}{c}  
	\xivec_A\\
	- i \omega_A \xivec_A
	\end{array} \right]. 
\label{goodexpansion}
\end{eqnarray}   
Using the orthogonality relation (\ref{orthonormal1}), the expansion
(\ref{goodexpansion}) can be inverted (see Appendix \ref{sec:proofs})
to obtain 
\beq
c_A(t) = {1 \over b_A} \, \left< \xivec_A \, , \, \omega_A \xivec(t)
+ i {\dot \xivec}(t) + i {\bf B} \cdot \xivec(t)\right>,
\label{inverseexpansion}
\endeq
where the constant $b_A$ is given by
\footnote{Note that adjusting the normalization of $\xivec_A$ will
change the magnitude but not the sign of $b_A$.  Hence we cannot find
a basis that achieves $b_A=1$ for all $A$, since some of the $b_A$'s
are negative and some are positive.}
\beq
b_A = \left< \xivec_A \, , \, i {\bf B} \cdot \xivec_A \right> + 2 \omega_A
 \left< \xivec_A \, , \xivec_A \right>.
\label{normalizationdef}
\endeq
Since the operator $i {\bf B}$ is Hermitian, $b_A$ is real although it
need not be positive.
In Appendix \ref{sec:proofs} we show that for $\omega_A \ne 0$, the
constant $b_A$ is related to the rotating-frame mode energy $\varepsilon_A$
at unit amplitude by $b_A = \varepsilon_A / \omega_A$.
Finally, the equation of motion for the mode coefficient $c_A(t)$ is
the first-order-in-time equation
\beq
{\dot c}_A(t) + i \omega_A c_A(t) = {i \over b_A} \left< \xivec_A \, ,
\, {\bf a}_{\rm ext}(t) \right>.
\label{fa}
\endeq
Using the mode decomposition (\ref{goodexpansion}), its inverse
(\ref{inverseexpansion}), and the equation of motion for each mode
(\ref{fa}), it is straightforward to compute the response of the star
to any externally applied acceleration ${\bf a}_{\rm ext}({\bf x},t)$.
See Appendix \ref{sec:proofs} for justification of all the claims and
formulae of this subsection.

\subsubsection{Degeneracy}
\label{sec:degeneracy}

Consider next the situation where there is degeneracy, that is, where
distinct modes $(\xivec_A,\omega_A)$ and $(\xivec_B,\omega_B)$ have
identical frequencies $\omega_A = \omega_B$.  We introduce the
following index notation: we write the distinct eigenfrequencies as
$\omega_a$, and we let the associated eigenvectors be $\xivec_{a,k}$
for $1 \le k \le n_a$, where $n_a$ is the degeneracy associated with
$\omega_a$.  Thus, we identify the index $A$ with the pair of indices
$(a,k)$, and we make the identifications
\begin{eqnarray}
\omega_A & = & \omega_{a,k} = \omega_a, \nonumber \\
\xivec_A &=& \xivec_{a,k}.
\end{eqnarray}
We can also rewrite sums over $A$ as
\beq
\sum_A = \sum_a \ \sum_{k=1}^{n_a}.
\end{equation}
Consider now the matrix 
\beq
{\cal M}_{AB} \equiv \left< \xivec_A \, , \, i {\bf B} \cdot \xivec_B
\right> + ( \omega_A + \omega_B ) \left< \xivec_A \, , \xivec_B \right>.
\label{calMdef}
\end{equation}
Using the notation $A = (a,k)$ and $B = (b,l)$, we can write the
matrix as ${\cal M}_{AB} = {\cal M}_{ak,bl}$.  In Appendix
\ref{sec:proofs} we show that the matrix ${\cal M}$ is always block
diagonal in the sense that ${\cal M}_{ak,bl} =0$ for $a \ne b$.  In
other words, for $\omega_A \ne \omega_B$ we have ${\cal M}_{AB}=0$,
which is a generalization of the orthogonality relation
(\ref{orthonormal1}) above. 

Within a given degenerate subspace, however, we have
\beq
{\cal M}_{ak,al}  = \left< \xivec_{a,k} \, , \, i {\bf B} \cdot \xivec_{a,l}
\right> + 2 \omega_a \left< \xivec_{a,k} \, , \xivec_{a,l} \right>,
\label{calMdef1}
\end{equation}
which is an $n_a \times n_a$ matrix that depends on the choice of
basis $\xivec_{a,k}$.  Since we are assuming in this section that all the
eigenfrequencies are real, the matrix (\ref{calMdef1})
is Hermitian and can be diagonalized by an appropriate chance
of basis $\xivec_{a,k} \to T_k^{\ l} \xivec_{a,l}$ within the
degenerate subspace.  A choice of basis $\xivec_{a,k}$ that
diagonalizes the matrix (\ref{calMdef1}) is the analog for rotating
stars of orthonormal bases for non-rotating stars.  
For such bases, the formulae of Sec.\ \ref{sec:linear_rot} above are
valid in the degenerate as well as the non-degenerate cases, namely
the mode decomposition (\ref{goodexpansion}), its inverse
(\ref{inverseexpansion}), the definition (\ref{normalizationdef}) of
the constant $b_A$, and the equation of motion (\ref{fa}).

\subsubsection{Reality conditions}
\label{sec:reality}

So far, the formalism is valid for complex Lagrangian displacements
$\xivec({\bf x},t)$.  We now describe simplifications that occur in
the physically relevant case of real $\xivec({\bf x},t)$.

If $(\xivec_A,\omega_A)$ is a solution of the quadratic eigenvalue equation
(\ref{basic1}), then another solution $(\xivec_A^*,-\omega_A)$ can be
obtained by complex-conjugating the mode function and reversing the
sign of the frequency \cite{lbo}.   
Under this transformation, the normalization constant 
$b_A$ flips
sign, from Eq.\ (\ref{normalizationdef}) \footnote{The inertial-frame
frequency $\omega + m \Omega$, where $m$ is the azimuthal quantum
number, also flips sign under this transformation.}.  
In Appendix \ref{sec:proofs} we show that all non Jordan-chain
modes have $b_A \ne 0$.   Hence, all the distinct non Jordan-chain modes
occur in pairs $(\xivec,\omega)$ and $(\xivec^*,-\omega)$.
\footnote{If a mode satisfies (i) $\xivec_A$ is purely real, and (ii)
$\omega_A=0$, then from Eqs.\ (\ref{normalizationdef}) and
(\ref{eq:antiHermitian}) that mode also has $b_A=0$ and hence is a
Jordan-chain mode.}

Let us now focus attention on the set of modes with 
$b_A > 0$ \footnote{For modes with $\omega_A \ne 0$, we have
$$
b_A = \omega_A \left< \xivec_A\, , \, \xivec_A \right> + {1 \over
\omega_A} \left< \xivec_A \, , \, {\bf C} \cdot \xivec_A \right>,
$$
from Eqs.\ (\ref{basic1}) and (\ref{normalizationdef}).  Therefore
positive frequency 
$\omega_A>0$ would correspond to $b_A > 0$ if the operator ${\bf C}$
were positive.  However, ${\bf C}$ need not always be a positive
operator.  Nevertheless we suspect that for modes of nonzero frequency,
positive frequency always corresponds to $b_A > 0$,
although we have been unable to prove this.}.
We introduce the following index notation: we write the distinct
modes with 
$b_A > 0$ as $(\xivec_\alpha,\omega_\alpha)$.
We identify the index $A$ with the pair of indices
$(\alpha,\epsilon)$, where $\epsilon$ takes on the values $\epsilon=+$
and $\epsilon=-$, and we make the identifications
$\xivec_A = \xivec_{\alpha,\epsilon}$, $\omega_A =
\omega_{\alpha,\epsilon}$, and $b_A = b_{\alpha,\epsilon}$ with 
\begin{eqnarray}
\omega_{\alpha,+} & = & \omega_\alpha \nonumber \\
\omega_{\alpha,-} & = & - \omega_\alpha \nonumber \\
\xivec_{\alpha,+} &=& \xivec_\alpha \nonumber \\
\xivec_{\alpha,-} &=& \xivec_\alpha^* \nonumber \\
b_{\alpha,+} &=& b_\alpha \nonumber \\
b_{\alpha,-} &=& - b_\alpha.
\label{gnotations}
\end{eqnarray}
We can also rewrite sums over $A$ as
\beq
\sum_A = \sum_\alpha \ \sum_{\epsilon=\pm}.
\end{equation}

Using these notations, the mode expansion (\ref{goodexpansion}) can be
written as
\begin{eqnarray}
 \left[ \begin{array}{c}  
	\xivec(t)\\
	{\dot \xivec}(t)
	\end{array} \right] = 
\sum_\alpha c_{\alpha,+}(t) \left[ \begin{array}{c}  
	\xivec_\alpha\\
	- i \omega_\alpha \xivec_\alpha
	\end{array} \right] + 
c_{\alpha,-}(t) \left[ \begin{array}{c}  
	\xivec_\alpha^*\\
	 i \omega_\alpha \xivec_\alpha^*
	\end{array} \right],
\label{goodexpansion1}
\end{eqnarray}
and its inverse (\ref{inverseexpansion}) can be written as
\beq
c_{\alpha,+}(t) = {1 \over b_\alpha} \, \left< \xivec_\alpha \, , \,
\omega_\alpha \xivec(t) + i {\dot \xivec}(t) + i {\bf B} \cdot
\xivec(t)\right>, 
\label{inverseexpansion1}
\endeq
and
\beq
c_{\alpha,-}(t) = {1 \over b_\alpha} \, \left< \xivec_\alpha^* \, , \,
\omega_\alpha \xivec(t) - i {\dot \xivec}(t) - i {\bf B} \cdot
\xivec(t)\right>.
\label{inverseexpansion2}
\endeq
It follows that 
$\xivec(t)$ will be real if and only if $c_{\alpha,-}(t) =
c_{\alpha,+}(t)^*$ for all $\alpha$.  In this case, writing $c_{\alpha} \equiv
c_{\alpha,+}$, 
the mode expansion (\ref{goodexpansion1}) becomes
\begin{eqnarray}
 \left[ \begin{array}{c}  
	\xivec(t)\\
	{\dot \xivec}(t)
	\end{array} \right] = 
\sum_\alpha c_{\alpha}(t) \left[ \begin{array}{c}  
	\xivec_\alpha\\
	- i \omega_\alpha \xivec_\alpha
	\end{array} \right] + 
c_{\alpha}(t)^* \left[ \begin{array}{c}  
	\xivec_\alpha^*\\
	 i \omega_\alpha \xivec_\alpha^*
	\end{array} \right],
\label{goodexpansion1a}
\end{eqnarray}
the inverse mode expansion (\ref{inverseexpansion1}) becomes
\beq
c_{\alpha}(t) = {1 \over b_\alpha} \, \left< \xivec_\alpha \, , \,
\omega_\alpha \xivec(t) + i {\dot \xivec}(t) + i {\bf B} \cdot
\xivec(t)\right>, 
\label{inverseexpansion1a}
\endeq
and the equation of motion (\ref{fa}) becomes
\beq
{\dot c}_\alpha(t) + i \omega_\alpha c_\alpha(t) = {i \over b_\alpha}
\left< \xivec_\alpha \, , \, {\bf a}_{\rm ext}(t) \right>.
\label{fa1}
\endeq

Note that the number of distinct $\xivec_\alpha$'s is the dimension of
${\cal H}$.  In Appendix \ref{sec:proofs1} we show that the
$\xivec_\alpha$'s are linearly independent and thus form a basis of
${\cal H}$ \footnote{More precisely, they would form a basis for
${\cal H}$ if there were no Jordan-chain modes.}.  Hence, one {\it
could} perform a configuration-space 
mode expansion of the form 
\beq
\xivec(\xvec,t) = \sum_\alpha \, q_\alpha(t) \, \xivec_\alpha(\xvec),
\label{badexpansion1}
\endeq
instead of the phase space mode expansion (\ref{goodexpansion1a}).
However, the coefficients  
$q_\alpha(t)$ defined by Eq.\ (\ref{badexpansion1}) are not 
related in any simple way to the coefficients $c_{\alpha}(t)$ 
\footnote{For the special case of a non-rotating star and real
$\xivec({\bf x},t)$, there is a simple relation between the two sets
of coefficients, namely 
$2 c_{\alpha,+} = q_\alpha + i {\dot q}_\alpha / \omega_\alpha$
and $2 c_{\alpha,-} = q_\alpha^* - i {\dot q}^*_\alpha /
\omega_\alpha$.}, and 
furthermore the equations of motion for the coefficients $q_\alpha$
will not be uncoupled from one another.

In Appendix \ref{sec:proofs2} we give an alternative form of the
equations of motion, valid for arbitrary complex $\xivec({\bf x},t)$,
which combines the two first order equations for $c_{\alpha,+}(t)$ and
$c_{\alpha,-}(t)$ in a single second order equation.  That alternative
form, when specialized to a non-rotating star and to a real basis of
eigenmodes $\xivec_\alpha({\bf x})$, coincides with the standard
formalism (\ref{eenr}) for a non-rotating star.

Finally, in Appendix \ref{sec:eam} we give expressions for the energy
and angular momentum of perturbations in terms of the mode amplitudes.

\section{The Slow Rotation Expansion}
\label{sec:sec3}

\subsection{Overview}
\label{sec:slow_rot1}

In this section we discuss the approximation method of solving for the 
mode functions $\xivec_A({\bf x})$ and frequencies $\omega_A$ by
using a perturbative expansion in the star's angular velocity
$\Omega$.  The main result of this section is that the leading
order (in $\Omega$) mode functions for the inertial or hybrid modes in
a slowly rotating zero-buoyancy
star are orthogonal with respect to the
inner product (\ref{innerproduct}) 
of the non-rotating star.

There are two general classes of modes in rotating stars
\cite{Thesis:Lockitch,1999ApJ...521..764L}: 
\begin{itemize}
\item Modes for which the frequency goes to zero in the non-rotating
limit.  These modes can be called {\it rotational} since they 
have non-zero frequency only 
in the presence of rotation.
For zero-buoyancy stars, these modes are the hybrid modes of
Ref. \cite{1999ApJ...521..764L}, also called inertial modes.  For
nonzero-buoyancy stars, they are purely axial. 
\item Modes with a finite frequency 
in the non-rotating limit, such as $f$ and $p$ modes, and $g$ modes in
nonzero-buoyancy stars.
\end{itemize}
There is a substantial literature on computing mode functions and
frequencies perturbatively in powers of $\Omega$.  
For nonzero-buoyancy stars,
the rotational or Rossby modes have been
computed to leading order in $\Omega$ by Refs.\ \cite{1980SSRv...27..653S},
\cite{1981A&A....94..126P} and \cite{1982ApJ...256..717S}, and to
all orders in $\Omega$ by Ref.\ \cite{yoshidaleenonisentropic}.
Non-rotational modes in the nonzero-buoyancy case 
have been computed by Ref.\ \cite{1987AcA....37..313D}, and by Ref.\
\cite{1998A&A...334..911S} beyond the leading order in $\Omega$.

For zero-buoyancy stars,
there are two classes of rotational modes.  The
so-called pure or classical $r$-modes \cite{1978MNRAS.182..423P} exist
only for ${l}=|m|$
and are purely axial.  
The remaining rotational modes
have both polar and axial pieces and have been called generalized
$r$-modes, hybrid modes or inertial modes.  They have been  
obtained to leading order in $\Omega$ by Ref.\ \cite{1999ApJ...521..764L}
and by Ref.\ \cite{PhysrevD...59...044009} for Maclaurin spheroids, to higher
order in $\Omega$ by Refs.\ \cite{gr-qc9902052} and
\cite{yoshidaleeisentropic}, to all orders in $\Omega$ by Ref.\
\cite{astro-ph9901532}, 
and in relativistic stars by Ref.\ \cite{gr-qc0008019}.
The hybrid modes consist of mixtures of the $r$-modes 
of nonzero-buoyancy stars 
together with the zero-buoyancy limit of
$g$-modes \cite{1999ApJ...521..764L}; see also 
Ref.\ \cite{1980A&A....89..314S} for hints in this direction.

In this section we will re-derive the equations obtained by
Ref.\ \cite{1999ApJ...521..764L} that describe the leading order mode
functions of the hybrid modes [Eqs.\ (\ref{df1})--(\ref{df2}) below],
by using a formal perturbation theory expansion of the quadratic
eigenvalue equation (\ref{basic1}).  The main benefit of the analysis 
is that it shows that all of the rotational modes are orthogonal with
respect to the inner product of the non-rotating star.  In addition,
it clarifies why, in the analysis of Ref.\ \cite{1999ApJ...521..764L}, it is
sufficient to consider the curl of the perturbed Euler equation and to
neglect its longitudinal part.

We also note that Ref.\ \cite{1980MNRAS.190....21S} shows that if a
background star has Jordan chains of length $p$, and is subject to a
perturbation of order $\varepsilon$, then generically the leading order
change in mode frequencies scales as $\varepsilon^{1/(p+1)}$.  Now,
non-rotating stars always have Jordan-chain modes of length 1 (see
Appendix \ref{sec:proofs3}), and the leading order effects of rotation
on the mode dynamics is linear in the star's angular velocity
$\Omega$.  Hence, one might suspect the existence
of modes in rotating stars whose frequencies scale as $\sqrt{\Omega}$
as $\Omega \to 0$
\footnote{Modes whose frequencies $\omega$ are proportional to
$\sqrt{\Omega}$ at large $\Omega$ have been found by
\protect{Ref.\ \cite{1996ApJ...460..827B}}, but for these modes $\omega
\to \omega_0 \ne 0$ as $\Omega \to 0$.}.
We show explicitly in Appendix \ref{sec:noexist} that no such modes
exist, and explain why the argument of Ref.\ \cite{1980MNRAS.190....21S}
does not apply in Appendix \ref{sec:proofs3}.

\subsection{Normalization convention}

The formalism of Sec.\ \ref{sec:sec2} is invariant under changes of
normalization of the basis functions $\xivec_A$; changes of the
normalization of the modes are compensated for by changes of the
normalization of the mode coefficients $c_A(t)$.  In general one can
choose any convenient normalization convention.  However, in the
context of the slow rotation expansion, statements about the behavior
of the mode functions $\xivec_A$ as $\Omega \to 0$ are dependent on
the normalization convention\footnote{For example, if one chooses to
normalize an $r$ mode in such a way that an amplitude $c_A$ of order unity
corresponds to a mode energy of order the stellar binding energy, then
$\xivec_A \propto 1/\Omega$ as $\Omega \to 0$.}.  Throughout the rest
of this paper we 
shall adopt the convention 
\beq
\left< \xivec_A \, , \xivec_A \right> = M R^2,
\label{normc}
\endeq
where $M$ is the stellar mass and $R$ the stellar radius.  Then the
mode coefficients are dimensionless, and order unity amplitudes $c_A
\sim 1$ correspond to Lagrangian displacements of order size of the
star.  With this convention, all of the mode functions $\xivec_A$ will
have finite limits as $\Omega \to 0$.

\subsection{Perturbation expansion}

We denote by ${\cal H}_0$ the subspace of the ${\cal H}$ of Lagrangian
displacements $\xivec({\bf x})$ spanned by modes of the non-rotating
star of zero frequency. In the 
zero-buoyancy case, 
the space ${\cal H}_0$ consists of those perturbations $\xivec({\bf
x})$ for which $\delta \rho = - \nabla \cdot (\rho \xivec)$ vanishes,
and in the nonzero-buoyancy
case it consists of purely axial vectors
(under reasonable assumptions).  These characterizations of the space
of zero-frequency modes have been 
proved for fully relativistic stars by Ref.\ \cite{gr-qc0008019}.  
In Appendix \ref{sec:zero-freq} we give Newtonian versions of the
proofs of Ref.\ \cite{gr-qc0008019}, and also derive some other properties
of the space ${\cal H}_0$.

The operators ${\bf B}$ and ${\bf C}$ can be expanded as power series
in the angular velocity $\Omega$ as \footnote{More precisely, the
expansion parameter is the dimensionless quantity $\Omega
\sqrt{R^3/(G M)}$, where $M$ and $R$ are the stellar mass and radius.}
\beq
{\bf B} = \Omega {\bf B}^{(1)} 
\label{Bexpansion}
\endeq
and
\beq
{\bf C} = {\bf C}^{(0)} + \Omega^2 {\bf C}^{(2)} + O(\Omega^4).
\label{Cexpansion}
\endeq
Also the inner product (\ref{innerproduct}) depends on $\Omega$
through the background density $\rho$ and can be expanded as
\beq
\left< \,\xivec \, ,  \xivec^\prime \,\right> = 
\left< \,\xivec \, ,  \xivec^\prime \,\right>_0
+ \Omega^2 \left< \,\xivec \, ,  \xivec^\prime \,\right>_2 + O(\Omega^4).
\end{equation}
Since the operators ${\bf C}$ and $i {\bf B}$ are Hermitian with
respect to $\left< \ , \ \right>$, it follows that ${\bf C}^{(0)}$ and
$i {\bf B}^{(1)}$ (although not ${\bf C}^{(2)}$) are Hermitian with
respect to $\left< \ , \ \right>_0$.

The operator ${\bf C}^{(0)}$ governs the modes of the non-rotating star.  
Let $\omega_A^{(0)}$, ${\hat \bfxi}_A$, for $A = 1,2,3 \ldots$,
be a complete set of solutions of the eigenvalue equation for
the spherical star:
\begin{eqnarray}
{\bf C}^{(0)} \cdot {\hat \bfxi}_A =  \omega_A^{(0)\,2} \,
{\hat \bfxi}_A.
\label{eig0a}
\end{eqnarray}
As in Sec.\ \ref{sec:reality} above we can take $A = (\alpha,\epsilon)$ with
$\epsilon = +$ or $\epsilon=-$, and with ${\hat \bfxi}_{\alpha,+} = {\hat
\bfxi}_{\alpha,-} \equiv {\hat \bfxi}_\alpha$,
and $\omega_{\alpha,+}^{(0)} = - \omega_{\alpha,-}^{(0)} \equiv
\omega_\alpha^{(0)} \ge 0$.  Since ${\bf C}^{(0)}$ is a Hermitian operator,
the basis
$\{ {\hat \bfxi}_\alpha \}$ is a 
complete, orthonormal basis of ${\cal H}$.  The basis ${\hat
\bfxi}_\alpha$ can be chosen arbitrarily within each  
degenerate subspace of the operator ${\bf C}^{(0)}$.  For example, the
basis can be chosen arbitrarily within the subspace ${\cal H}_0$.
Below we will introduce a different, preferred basis $\bfxi_\alpha^{(0)}$
of eigenvectors of ${\bf C}^{(0)}$ that are the $\Omega \to 0$ limit
of modes of the rotating star.

Consider now trying to solve for the one parameter family or families of modes 
$\xivec_\alpha(\Omega), \omega_\alpha(\Omega)$ of the rotating star
for which $\omega_\alpha(\Omega) \to \omega_\alpha^{(0)}$ as $\Omega
\to 0$
\footnote{It suffices to consider the case $\epsilon=+$ and
to seek a one-parameter family of modes whose limiting frequency is
$\omega^{(0)}_{\alpha,+}$, since a one parameter family whose limiting
frequency is $\omega^{(0)}_{\alpha,-}$ can then be obtained using the
transformation $(\xivec,\omega) \to (\xivec^*,-\omega)$; see Sec.\
\ref{sec:reality} above.}. 
As explained above, there are two types of such modes, rotational
modes for which $\omega_\alpha^{(0)} =0$, and non-rotational modes for
which $\omega_\alpha^{(0)} \ne 0$.
For both types of modes, we make the ansatz that
the frequency and mode function can be expanded as
\begin{eqnarray}
\omega_\alpha(\Omega) = \omega_\alpha^{(0)} + \Omega \,
\omega_\alpha^{(1)} + \Omega^2 \, \omega_\alpha^{(2)} + O(\Omega^3),
\label{lambdaexpansion}
\end{eqnarray}
and 
\begin{eqnarray}
\bfxi_\alpha(\Omega) = \bfxi_\alpha^{(0)} + \Omega \,
\bfxi_\alpha^{(1)} + \Omega^2 \bfxi_\alpha^{(2)} + O(\Omega^3).
\label{xiexpansion}
\end{eqnarray}
If we substitute the expansions
(\ref{lambdaexpansion})--(\ref{xiexpansion}) and also the expansions  
(\ref{Bexpansion})--(\ref{Cexpansion}) for the operators ${\bf B}$ and
${\bf C}$ into the quadratic eigenvalue equation (\ref{basic1}) we get
a series of equations.  First, at order $O(\Omega^0)$ we get
\begin{eqnarray}
\left[ - \omega_\alpha^{(0)\,2} + {\bf C}^{(0)} \right] \cdot
\bfxi_\alpha^{(0)} =0, 
\label{eig0}
\end{eqnarray}
which is the usual eigenvalue equation for spherical stars.  At order
$O(\Omega)$ we get  
\begin{eqnarray}
\left[ - \omega_\alpha^{(0)\,2} + {\bf C}^{(0)} \right] \cdot
\bfxi_\alpha^{(1)} - \omega_\alpha^{(0)} \left[ 2 \omega_\alpha^{(1)}
+ i {\bf B}^{(1)} \right] \cdot 
\bfxi_\alpha^{(0)} =0.
\label{eig1}
\end{eqnarray}
Finally, at order $O(\Omega^2)$, we get the equation
\begin{eqnarray}
&& \left[ - \omega_\alpha^{(0)\,2} + {\bf C}^{(0)} \right] \cdot
\bfxi_\alpha^{(2)} - \omega_\alpha^{(0)} \left[ 2 \omega_\alpha^{(1)}
+ i  {\bf B}^{(1)} \right] \cdot 
\bfxi_\alpha^{(1)}   
\nonumber \\ \mbox{} &&
+ \left[ - 2 \omega_\alpha^{(0)} \omega_\alpha^{(2)} -
\omega_\alpha^{(1)\,2} - i \omega_\alpha^{(1)} {\bf B}^{(1)} + {\bf
C}^{(  2)} \right] \cdot \bfxi_\alpha^{(0)} 
=0.
\label{eig2}
\end{eqnarray}

We next express these perturbation equations in the basis of modes of
the non-rotating star.
We expand the
mode functions 
$\bfxi_\alpha^{(0)}$, $\bfxi_\alpha^{(1)}$ and $\bfxi_\alpha^{(2)}$ in
terms of the eigenbasis $\{ {\hat \bfxi}_\alpha \}$ of the
non-rotating star as
\begin{eqnarray}
\bfxi_\alpha^{(0)} = \sum_\beta c^{(0)}_{\alpha\beta} \, \, {\hat \bfxi}_\beta,
\label{xi0ex}
\end{eqnarray}
\begin{eqnarray}
\bfxi_\alpha^{(1)} = \sum_\beta  c^{(1)}_{\alpha\beta} \, \, {\hat
\bfxi}_\beta, 
\label{xi1ex}
\end{eqnarray}
and
\begin{eqnarray}
\bfxi_\alpha^{(2)} = \sum_\beta  c^{(2)}_{\alpha\beta} \, \, {\hat
\bfxi}_\beta. 
\label{xi2ex}
\end{eqnarray}
Note that we cannot assume that $c^{(0)}_{\alpha\beta} =
\delta_{\alpha\beta}$ because 
of the possibility of degeneracies.  Just as in degenerate
perturbation theory in quantum mechanics, the modes of the rotating
star define a preferred basis $\bfxi_\alpha^{(0)}$ of each degenerate 
subspace.  Since we do not know this basis before we
solve for the modes of the rotating star, we must start our
computation with some arbitrary choice of basis ${\hat \bfxi}_\alpha$
in each degenerate subspace.  

Next, we substitute the expansion (\ref{xi0ex}) of $\bfxi^{(0)}_\alpha$ into
the zeroth-order eigenvalue equation (\ref{eig0}).  This gives, using
Eq.\ (\ref{eig0a}), 
\begin{eqnarray}
\sum_\beta c_{\alpha\beta}^{(0)} \left[ \omega_\alpha^{(0)\,2} -
\omega_\beta^{(0)\,2} \right] \, {\hat \bfxi}_\beta =0.
\end{eqnarray}
It follows that $c_{\alpha\beta}^{(0)}=0$ when $\omega_\alpha^{(0)} \ne
\omega_\beta^{(0)}$.  In other words, the matrix
$c_{\alpha\beta}^{(0)}=0$  is block 
diagonal, with each block corresponding to a degenerate subspace.
Hence the zeroth order mode function can be written as, from
Eq.~(\ref{xi0ex}), 
\begin{eqnarray}
\bfxi_\alpha^{(0)} = \sum_{\beta \ {\rm with}\
\omega_\beta^{(0)}=\omega_\alpha^{(0)}} \, 
c^{(0)}_{\alpha\beta} \, \, {\hat \bfxi}_\beta,
\label{pr}
\end{eqnarray}
and thus lies within the degenerate subspace corresponding to the eigenvalue
$\omega_\alpha^{(0)}$.  In particular, for the modes with
$\omega_\alpha^{(0)}=0$, the zeroth order 
mode function $\bfxi_\alpha^{(0)}$ lies inside the space ${\cal H}_0$.  

The analysis now divides into two cases for the two different types of
mode discussed above.

\subsection{Modes with finite frequency in the non-rotating limit}

If we substitute the expansions (\ref{xi0ex}) and (\ref{xi1ex}) for
$\bfxi_\alpha^{(0)}$ and $\bfxi_\alpha^{(1)}$ into the first-order eigenvalue
equation (\ref{eig1}), and then take the zeroth order inner product 
$\left< \ , \ \right>_0$ with
${\hat \bfxi}_\beta$, we get
\begin{eqnarray}
\left[ ( \omega_\alpha^{(0)})^2 - (\omega_\beta^{(0)})^2 \right]
c_{\alpha\beta}^{(1)} && + 2 \omega_\alpha^{(0)} \omega_\alpha^{(1)}
\, c_{\alpha\beta}^{(0)}  \nonumber \\ \mbox{} &&
+ i \omega_\alpha^{(0)} \sum_\gamma B^{(1)}_{\beta\gamma}
\, c^{(0)}_{\alpha\gamma} =0,
\label{regular}
\end{eqnarray}
where 
\begin{eqnarray}
B^{(1)}_{\beta\gamma} \equiv \left< {\hat \bfxi}_\beta \, , \, {\bf
B}^{(1)} \cdot 
{\hat \bfxi}_\gamma \right>_0. 
\end{eqnarray}
If we specialize Eq.\ (\ref{regular}) to values of $\beta$ for which
$\omega_\beta^{(0)} = \omega_\alpha^{(0)}$, and divide across by
$\omega_\alpha^{(0)}$ we get 
\begin{eqnarray}
2 \omega_\alpha^{(1)} \, c_{\alpha\beta}^{(0)} +i \sum_\gamma
B^{(1)}_{\beta\gamma} \, c^{(0)}_{\alpha\gamma} =0, 
\label{regular1}
\end{eqnarray}
which is a standard eigenvalue equation.  Thus, the zeroth order modes are
given by diagonalizing the operator $B^{(1)}_{\beta\gamma}$ within the
degenerate subspace of modes $\beta$ with $\omega_\beta^{(0)} =
\omega_\alpha^{(0)}$.  The 
eigenvalues of $B^{(1)}_{\beta\gamma}$ give the first order changes
$\omega_\alpha^{(1)}$ to the eigenfrequency, just 
as in degenerate perturbation theory in quantum mechanics.  

Suppose now we switch to the basis that diagonalizes
$B^{(1)}_{\alpha\beta}$ within each degenerate subspace, so that
$c^{(0)}_{\alpha\beta} = \delta_{\alpha\beta}$.  Then, 
applying Eq.\ 
(\ref{regular}) for values of $\beta$ with $\omega_\beta^{(0)} \ne
\omega_\alpha^{(0)}$ 
gives for the first order change in the mode functions
\begin{eqnarray}
c^{(1)}_{\alpha\beta} =  { - i \omega_\beta^{(0)} \over 
\omega_\alpha^{(0)\,2} - \omega_\beta^{(0)\,2} } \ B^{(1)}_{\beta\alpha}.
\label{c1ans3}
\end{eqnarray}
Using the expansions (\ref{xiexpansion}), (\ref{xi0ex}), and (\ref{xi1ex})
together with Eq.\ (\ref{c1ans3}) shows that the inner product of two
mode functions with $\omega_\alpha^{(0)} \ne \omega_\beta^{(0)}$ is
\beq
\left< \xivec_\alpha(\Omega) \, , \,  \xivec_\beta(\Omega) \right> = \
{ - i \Omega \over \omega_\alpha^{(0)} + \omega_\beta^{(0)} } \,
B^{(1)}_{\alpha\beta} + O(\Omega^2).
\label{nonorthogeg}
\end{equation}
This explicitly demonstrates that modes in a rotating star are not
orthogonal in general, since we can find pairs of modes $\alpha$,
$\beta$ of the non-rotating star for which the matrix element
$B_{\alpha\beta}^{(1)}$ is non vanishing.

\subsection{Modes with vanishing frequency in the non-rotating limit}

Things work somewhat differently for the second class of modes.  If we
substitute $\omega_\alpha^{(0)} = 0$ into Eq.\ (\ref{eig1}) we
get
\begin{eqnarray}
{\bf C}^{(0)} \cdot \bfxi^{(1)}_\alpha = 0.
\end{eqnarray}
Thus, the first order equation (\ref{eig1}) does not determine
$\bfxi^{(1)}_\alpha$, except to dictate that $\bfxi^{(1)}_\alpha$ lie
in the space 
${\cal H}_0$ of zero-frequency modes of the non-rotating star.  In
addition, the ambiguity in the choice of basis of $\bfxi_\alpha^{(0)}$ of
${\cal H}_0$ is not resolved at this linear order.

Hence we must use the second order equation
(\ref{eig2}).  For $\omega_\alpha^{(0)}=0$, this equation
reduces to 
\begin{eqnarray}
{\bf C}^{(0)} \cdot \bfxi^{(2)}_\alpha + 
\left[ - \omega_\alpha^{(1)\,2} - i 
\omega_\alpha^{(1)} {\bf B}^{(1)} + {\bf C}^{(2)} \right] \cdot
\bfxi_\alpha^{(0)} =0.
\label{eig2a}
\end{eqnarray}
If we substitute the expansions (\ref{xi0ex}) and (\ref{xi2ex}) into
Eq.\ (\ref{eig2a}) and 
take the zeroth order inner product 
$\left< \ , \ \right>_0$ with
${\hat \bfxi}_\beta$, we get
\begin{eqnarray}
- \omega_\alpha^{(1)\,2} \, c^{(0)}_{\alpha\beta} 
+  \omega_\beta^{(0)\,2} \,  c_{\alpha\beta}^{(2)} 
+ \sum_\gamma \left[ - i \omega_\alpha^{(1)}
B^{(1)}_{\beta\gamma} \, + C^{(2)}_{\beta\gamma} \right]
c_{\alpha\gamma}^{(0)}  =0,
\label{hybrid}
\end{eqnarray}
where
\begin{eqnarray}
C^{(2)}_{\beta\gamma} \equiv \left< {\hat \bfxi}_\beta \, , \,
{\bf C}^{(2)} \cdot 
{\hat \bfxi}_\gamma \right>_0. 
\end{eqnarray}
We now specialize Eq.\ (\ref{hybrid}) to values of $\beta$ for which
$\omega_\beta^{(0)}=0$, i.e., project both sides of the equation into
${\cal H}_0$
\footnote{The component of Eq.\ (\ref{hybrid}) orthogonal
to ${\cal H}_0$ is not needed to determine the leading order
quantities $\xivec_\alpha^{(0)}$ and $\omega_\alpha^{(1)}$.  It
determines the higher order correction $\xivec_\alpha^{(2)}$ to the
mode function.}.  This gives 
\begin{eqnarray}
- \omega_\alpha^{(1)\,2} \, c^{(0)}_{\alpha\beta} 
+ \sum_\gamma \left[ - i \omega_\alpha^{(1)}
B^{(1)}_{\beta\gamma} \, + C^{(2)}_{\beta\gamma} \right]
c_{\alpha\gamma}^{(0)}  =0,
\label{hybrid1}
\end{eqnarray}
which is a quadratic eigenvalue equation that determines both
$\omega_\alpha^{(1)}$ and $c^{(0)}_{\alpha\beta}$.  Equation
(\ref{hybrid1}) can be 
rewritten as 
\begin{eqnarray}
{\bf P}_0 \cdot \left[ - \omega_\alpha^{(1)\,2} - i \omega_\alpha^{(1)} {\bf
B}^{(1)} + {\bf C}^{(2)} \right] \cdot {\bf P}_0 \cdot \bfxi_\alpha^{(0)} = 0,
\label{hybrid2}
\end{eqnarray}
where ${\bf P}_0$ is the orthogonal projection operator (with respect
to $\left< \ , \ \right>_0$) that projects
into ${\cal H}_0$ (see appendix \ref{sec:zero-freq}).

It is at this point in the analysis that the difference between
stars with zero and non-zero buoyancy
enters.  We treat the two
cases separately.  

\subsubsection{Zero-buoyancy stars}

When $\Avec = 0$, 
the term ${\bf P}_0 \cdot {\bf C}^{(2)} \cdot {\bf P}_0 \cdot
\bfxi_\alpha^{(0)}$ in Eq.\ (\ref{hybrid2}) vanishes.  To see 
this, let
\begin{eqnarray}
\bfkappa = {\bf C} \cdot \bfxi_\alpha^{(0)} = {\bf C} \cdot {\bf P}_0
\cdot \bfxi_\alpha^{(0)}.
\label{kappadef}
\end{eqnarray}
It follows from the formula (\ref{Cdef-simple}) for the operator ${\bf
C}$ that $\bfkappa({\bf x})$ must be a pure gradient, so that ${\bf
\nabla} \times 
\bfkappa=0$.  However, Eq.\ (\ref{ans4}) of Appendix \ref{sec:zero-freq}
then implies that ${\bf P}_0 \cdot \bfkappa=0$, and expanding this
last equation to second order in $\Omega$ gives the desired result.

Equation  (\ref{hybrid2}) now simplifies to
\begin{eqnarray}
\left[ {\bf P}_0 \cdot i {\bf B}^{(1)} \cdot {\bf P}_0 
 \right] \cdot \bfxi_\alpha^{(0)} = - \omega_\alpha^{(1)} \,
 \bfxi_\alpha^{(0)},  
\label{hybrid3}
\end{eqnarray}
which is a standard eigenvalue equation in ${\cal H}_0$.
Thus, the zeroth order mode functions $\bfxi^{(0)}_\alpha$ for the
hybrid modes are simply the
eigenvectors in ${\cal H}_0$ of the Hermitian operator $i {\bf P}_0
\cdot {\bf B}^{(1)} \cdot {\bf P}_0$.  It follows that these zeroth order mode
functions are orthogonal to each other with respect the inner product
$\left< \ , \ \right>_0$ of the non-rotating star.  They are also
orthogonal to the zeroth order 
mode functions of the modes with $\omega_\alpha^{(0)}\ne 0$, since
those mode functions $\bfxi_\alpha^{(0)}$ lie completely inside degenerate
subspaces that are orthogonal to ${\cal H}_0$, by Eq.\ (\ref{pr}).

We now rewrite the equations defining the hybrid modes
$\xivec_\alpha^{(0)}({\bf x})$ in a more
accessible notation.  First, from Eq.\ (\ref{pr}) we know that
$\xivec_\alpha^{(0)}$ lies in ${\cal H}_0$, which from
Appendix \ref{sec:zero-freq} is equivalent to 
\begin{eqnarray}
{\bf \nabla} \cdot ( \rho \bfxi_\alpha^{(0)}) =0.
\label{df1}
\end{eqnarray}
Second, Eq.\ (\ref{hybrid3}) can be rewritten using the definition
(\ref{Bdef}) of ${\bf B}$ and Eq.\ (\ref{ans4}) as
\begin{eqnarray}
{\bf \nabla} \times \left[ - i \omega_\alpha^{(1)} \bfxi_\alpha^{(0)}+
2 {\bf \Omega} \times \bfxi_\alpha^{(0)} \right]=0.
\label{df2}
\end{eqnarray}
Equations (\ref{df1}) and (\ref{df2}) are the equations that
are solved to obtain the hybrid modes in Ref.\ \cite{1999ApJ...521..764L}.

Finally we note that going from Eq.\ (\ref{hybrid2}) to Eq.\ (\ref{hybrid3})
entailed dividing by $\omega_\alpha^{(1)}$.  There is another solution
to Eq.\ (\ref{hybrid2}) with $\omega_\alpha^{(1)}=0$.  Examining the
perturbation expansion to higher order for this solution shows that
$\omega_{\alpha}^{(2)}=0$ also.  We suspect that this solution is a
zero-frequency solution to all orders in $\Omega$.  These modes are
not Jordan-chain modes, but they are pure gauge in the sense of 
Ref.\ \cite{1978ApJ...221..937F}, at least to linear order in the mode
amplitude.  We note that it is impractical to include these modes (and
also the similar modes of Appendix \ref{sec:proofs3}) in nonlinear mode
evolution calculations since the perturbation expansion in mode
amplitude for these modes typically breaks down on a timescale $\sim
1/\Omega$.

\subsubsection{Stars with buoyancy}

For $\Avec \neq 0$ (stars for which there is a nonzero buoyancy force), 
the space ${\cal H}_0$ is the space of
purely axial vectors (see Appendix \ref{sec:zero-freq}).  In this case
the quadratic eigenvalue equation (\ref{hybrid2}) for the mode
functions can be written as
\beq
{\bf P}_0 \cdot \left[ - \omega_\alpha^{(1)\,2} \xivec_\alpha^{(0)} -
2 i \omega_\alpha^{(1)} {\bf \Omega} \times \xivec_\alpha^{(0)} 
+ {\bf C}_a^{(2)} \cdot \xivec_\alpha^{(0)} 
+ {\bf C}_b^{(2)} \cdot \xivec_\alpha^{(0)} \right] = 0,
\label{hybrid2a}
\end{equation}
where ${\bf P}_0 \cdot {\bf v}$ is now the axial part of ${\bf v}$,
and ${\bf C}_a^{(2)}$ and ${\bf C}_b^{(2)}$ are the second order in
$\Omega$ pieces of the operators (\ref{Cdef-simple}) and (\ref{Cdef-b}).
The term ${\bf C}^{(2)}_a \cdot \xivec_\alpha^{(0)}$ drops
out since it is a gradient and thus has no axial part, and the term 
${\bf C}^{(2)}_b \cdot \xivec_\alpha^{(0)}$ drops out since it depends
only on ${\bf \nabla} \cdot \xivec_\alpha^{(0)}$
and $\xivec_\alpha^{(0)} \cdot {\bf e}_r$ which vanishes as
$\xivec_\alpha^{(0)}$ is purely axial.  Using Eq.\ (\ref{ans44}), the
equation now reduces to
\begin{eqnarray}
{\bf r} \cdot {\bf \nabla} \times \left[ - i \omega_\alpha^{(1)}
\bfxi_\alpha^{(0)}+ 2 {\bf \Omega} \times \bfxi_\alpha^{(0)} \right]=0.
\label{df2a}
\end{eqnarray}
In Ref.\ \cite{1981A&A....94..126P} it is shown that Eq.\ (\ref{df2a}) determines the
frequency $\omega_\alpha^{(1)}$ but not the eigenfunction
$\xivec_\alpha^{(0)}$, and that one must go to higher order in the
$\Omega$ expansion in order to determine the eigenfunction; see also 
Refs.\ \cite{1980SSRv...27..653S} and \cite{1982ApJ...256..717S}.

\section{Second Order Lagrangian Perturbation Theory in Rotating Stars}
\label{sec:sec5}

In this section we develop a second order Lagrangian perturbation
theory for rotating stars, extending previous analyses of non-rotating
stars (see, e.g., Refs.\
\cite{1994A&A...286..879V,1989ApJ...342..558K,Thesis:Wu}).  The main 
new result is that the expression for the three-mode coupling
coefficient in terms of the mode functions [Eq.\ (\ref{eq:23}) below] in
the same for rotating stars as for non-rotating stars.  
Although we restrict attention in this paper to second order
perturbation theory, we lay the foundations in Appendix
\ref{variational} for computing mode coupling coefficients at third
and higher orders.

\subsection{Second order equation of motion}
\label{sec:Lagr-Pert-Theory}

To describe the fluid perturbation we use the rotating-frame 
Lagrangian displacement $\xivec(\xvec,t)$.  The definition of this
quantity is the same in a nonlinear context as in linear theory: 
a fluid element at rotating-frame location ${\bf x}$ 
in the unperturbed star is moved to rotating-frame location
$\xvec + \xivec(\xvec,t)$ by the perturbation.
We assume that the pressure $p$ in the background star is
a function $p(\rho,\muvec)$ of the 
the background mass density $\rho$ and of a vector $\muvec$ of other
fluid variables such as entropy or composition, which can vary with
position.  We also assume that the fluid variables $\muvec$ of each
fluid element are preserved by the perturbation.

Under these assumptions, the equation of motion for $\xivec$ is
derived in Appendix \ref{app:nvpd} 
to second order in $\xivec$ via direct perturbation of the Euler equation,
and in Appendix \ref{variational} to any order in $\xivec$ from a variational
principle.  The result has the form
\begin{eqnarray}
{\ddot {\bfxi}} + {\bf B} \cdot {\dot {\bfxi}} + {\bf C} \cdot
{\bfxi}= {\bf a}[\xivec],
\label{basicnonlin0}
\end{eqnarray}
where the left hand side contains all the linear terms
[cf.~Eq.~(\ref{basic0}) above] and the right hand side ${\bf a}[\xivec]$
is a nonlinear acceleration.
That acceleration can be written as ${\bf a} = {\bf a}_P + {\bf a}_G$,
where ${\bf a}_P$ and ${\bf a}_G$ are the contributions from pressure
gradients and gravity; the corresponding force densities are ${\bf f}
= \rho {\bf a}$, ${\bf f}_P = \rho {\bf a}_P$, and ${\bf f}_G = \rho
{\bf a}_G$.

We expand the acceleration as 
\beq
{\bf a}[\xivec] = {\bf a}^{(2)}[\xivec,\xivec] + O(\xivec^3)
={\bf a}^{(2)}_P[\xivec,\xivec] + {\bf a}^{(2)}_G[\xivec,\xivec] +
O(\xivec^3),
\label{basicnonlin1}
\endeq
where ${\bf a}^{(2)}[\xivec,\xivec^\prime]$, 
${\bf a}^{(2)}_P[\xivec,\xivec^\prime]$, and
${\bf a}^{(2)}_G[\xivec,\xivec^\prime]$ are symmetric bilinear
functions of their arguments.
In order to give the explicit results for the second order pieces
${\bf a}^{(2)}_P[\xivec,\xivec]$ and ${\bf a}^{(2)}_G[\xivec,\xivec]$
of the pressure 
gradient and 
gravitational accelerations, we need to introduce some notations.  
The first and second order Eulerian perturbations 
$\delta^{(1)}\phi$ and $\delta^{(2)}\phi$
to the Newtonian potential are defined as functionals of $\xivec$ by
the equations [see Eq.\ (\ref{phi3}) below] 
\beq
{1 \over 4 \pi G} \nabla^2 \, \delta^{(1)}\phi = - {\bf \nabla} \cdot
(\rho \xivec)
\label{delta1phidef}
\label{eq:18}
\endeq
and
\beq
{1 \over 4 \pi G} \nabla^2 \, \delta^{(2)}\phi = {1 \over 2} \nabla_i
\nabla_j (\rho \xi^i \xi^j). 
\label{eq:18aa}
\endeq
We define the tensors $\Thetauilj$, $\Xiuilj$ and $\chi^i_{\enskip j}$
by
\begin{eqnarray}
	\Thetauilj&\equiv&\xiuilcdj\xiuklcdk,\\
	 \Xiuilj&\equiv&\xiuilcdk\xiuklcdj,
		\label{thetaxipsidef}
\end{eqnarray}
and
\beq
\chi^i_{\enskip j} \equiv \xiuilcdl\xiullcdk\xiuklcdj.
\endeq
Here semicolons denote covariant derivatives, for example $\xi^i_{\
;j} \equiv \nabla_j \xi^i$.  Finally, we denote by $\Gamma_1$ the
generalized adiabatic index governing the perturbations, defined by
Eq.~(\ref{eq:6}) or Eq.~(\ref{gammaonedef}).

The second order gravitational acceleration is [Eqs.~(\ref{sauleq}) and
(\ref{iraeq}) below]
\beq
a_{G\,j}^{(2)}[\xivec,\xivec] = 
- \nabla_j \delta^{(2)}\phi - \xi^k \nabla_k \nabla_j \delta^{(1)}\phi 
- {1 \over 2} \xi^k \xi^l \nabla_k \nabla_l \nabla_j \phi.
\label{eq:17}
\endeq
The second order pressure term is [Eqs.~(\ref{sauleq}), (\ref{ugheq})
and (\ref{forces0}) below] 
\beq
a_{P\,i}^{(2)}[\xivec,\xivec] = - {1 \over \rho} \nabla_j \left[ p
(\Gamma_1 -1) \Thetaujli + p \Xiujli + \Psi \delta^j_i \right],
\label{eq:16}
\endeq
where
\beq
\Psi = {1 \over 2} p \Theta \left[ (\Gamma_1 -1)^2 + {\partial
\Gamma_1 \over \partial \ln \rho} \right] + {1 \over 2} p (\Gamma_1 -1)
\Xi,
\endeq
and $\Theta$ and $\Xi$ are the traces of $\Thetaujli$ and $\Xiujli$.
More general expressions that are valid to all orders in $\xivec$ are
given in Eqs.\ (\ref{forces}) below.

A key feature of the equation of motion (\ref{basicnonlin0}) is that there
is no dependence on the angular velocity $\Omega$ of the rotating frame
in the nonlinear terms on the right hand side.  The dependence on
$\Omega$ occurs only in the linear terms on the left hand side.  This
can be understood most easily from the variational principle that
underlies the hydrodynamical equations.   As explained in Appendix 
\ref{variational}, the only term in the action that is not
explicitly invariant under a change of rotational frame is the kinetic
energy term, and that kinetic energy term is quadratic in the 
Lagrangian displacement $\xivec$ and its time derivative ${\dot
\xivec}$.  

Finally, we note that in Lagrangian perturbation theory only pressure and
gravity can contribute to the nonlinear force. In spite of the fact that
the nonlinear advection term ${\bf v} \cdot \grad {\bf v}$ appears
in the Euler equation, no nonlinear advection terms are present in the
Lagrangian framework, since the kinetic energy is explicitly quadratic
in $\xivec$.

\subsection{Coupled equations of motion for the mode coefficients}
\label{sec:Deriv-Mode-Eqns}

We now use the equation of motion given by Eqs.~(\ref{basicnonlin0}),
(\ref{basicnonlin1}), (\ref{eq:17}) and (\ref{eq:16}) to derive the
coupled, nonlinear ordinary differential equations satisfied by the
mode expansion coefficients $c_A(t)$.  The basic equation
(\ref{basic0}) of Sec.\ \ref{sec:sec2} can be applied with the
externally applied acceleration ${\bf a}_{\rm ext}({\bf x},t)$
replaced by the nonlinear acceleration ${\bf a}[\xivec]$.  
Therefore we can directly carry over the equation of motion (\ref{fa}),
which gives
\beq
{\dot c}_A(t) + i \omega_A c_A(t) = {i \over b_A} \left< \xivec_A \, ,
\, {\bf a}^{(2)}[\xivec,\xivec] \right> + O(\xivec^3).
\label{fanl}
\endeq
Next, we take the complex conjugate of the phase space mode expansion
(\ref{goodexpansion}) and use the fact that $\xivec(\xvec,t)$ is real
to obtain
\beq
\xivec(\xvec,t) = \sum_A \, c_A(t)^* \, \xivec_A({\bf x})^*.
\endeq
Substituting this expansion into the right hand side of Eq.\
(\ref{fanl}) gives
\beq
{\dot c}_A(t) + i \omega_A c_A(t) = {i \over b_A} \sum_{B,C} 
\kappa_{ABC}^* \, c_B(t)^* c_C(t)^* + O(c^3).
\label{fanl1}
\endeq
where the three-mode coupling coefficient is
\beq
\kappa_{ABC} = 
\left< \xivec_A^* \, ,\, {\bf a}^{(2)}[\xivec_B,\xivec_C] \right>
\label{kappaabcdef}
\endeq
which is completely symmetric in the indices $A$, $B$ and $C$.
For $\omega_A \ne 0$, this equation can also be written as
\beq
{\dot c}_A(t) + i \omega_A c_A(t) = {i \omega_A} \sum_{B,C} 
{ \kappa_{ABC}^* \over \varepsilon_A} \, c_B(t)^* c_C(t)^* + O(c^3),
\label{fanl1a}
\endeq
where we have used Eq.\ (\ref{varepsilondef}).
This form of the equation has a simple physical significance,
since $\kappa_{ABC} / \varepsilon_A$ is the ratio 
of the nonlinear interaction energy at unit amplitude 
to the energy of the mode at unit amplitude.

The equation of motion (\ref{fanl1}) is valid for arbitrary complex
$\xivec(\xvec,t)$.  However, for the physically relevant case of real
$\xivec(\xvec,t)$, the coefficients $c_A(t)$ will not all be
independent, as explained in Sec.\ \ref{sec:reality} above.  
The modes will occur in complex conjugate pairs $(\xivec,\omega)$ and
$(\xivec^*,-\omega)$, and it is convenient to use the set of modes
$\xivec_\alpha$ defined in Sec.\ \ref{sec:reality} consisting of one
mode from each complex-conjugate pair.  For this set of modes, the
mode expansion is given by Eq.\ (\ref{goodexpansion1a}):
\beq
\xivec(t) = \sum_\alpha \left[ c_{\alpha}(t) \xivec_\alpha + c_{\alpha}(t)^*
\xivec_\alpha^* \right], 
\label{goodexpansion1b}
\endeq
and the inverse mode expansion by Eq.\ (\ref{inverseexpansion1a}).
For this expansion, the variables 
$c_\alpha(t)$ are all independent.  If we now use the 
expansion (\ref{goodexpansion1b}) in the equation of motion
(\ref{fa1}) we find as a replacement for Eq.\ (\ref{fanl1})
the equation
\beq
{\dot c}_\alpha + i \omega_\alpha c_\alpha = {i \over b_\alpha}
\sum_{\beta,\gamma} \left[ \kappa_{{\bar \alpha}\beta\gamma} c_\beta
c_\gamma + \kappa_{{\bar \alpha}{\bar \beta}\gamma} c_\beta^* c_\gamma
+ \kappa_{{\bar \alpha}\beta{\bar \gamma}} c_\beta c_\gamma^* + 
\kappa_{{\bar \alpha}{\bar \beta}{\bar \gamma}} c_\beta^* c_\gamma^*
\right],
\label{fanl3}
\endeq
where
\beq
\kappa_{\alpha\beta\gamma} = 
\left< \xivec_\alpha^* \, ,\, {\bf
a}^{(2)}[\xivec_\beta,\xivec_\gamma] \right> 
\endeq
which is symmetric in $\alpha$, $\beta$, and $\gamma$.  In Eq.\
(\ref{fanl3}) we have used a notational convention for the coupling
coefficients $\kappa$ where a bar over an index means that the
corresponding mode function is to be complex conjugated in the
expression for the coupling coefficient.  For example, we have
\beq
\kappa_{\alpha\beta{\bar \gamma}} = 
\left< \xivec_\alpha^* \, ,\, {\bf
a}^{(2)}[\xivec_\beta,\xivec_\gamma^*] \right>.
\endeq
Equation (\ref{fanl3}) is the final equation of motion.  
To model the saturation of $r$-modes, it will have to be supplemented
by viscous damping and gravitational radiation reaction terms.
We now turn
to evaluating the coupling coefficients $\kappa_{ABC}$ to insert into this
equation, with the understanding that each capital Roman index can be
either a unbarred or a barred Greek lower case index, for example $A =
\alpha$ or $A = {\bar \alpha}$.  In the notation of Sec.\
\ref{sec:reality}, $(\alpha,+)$ corresponds to $\alpha$ and
$(\alpha,-)$ corresponds to ${\bar \alpha}$.  

\subsection{Explicit expression for the three-mode coupling coefficient}
\label{sec:Deriv-Coupl-Coeff}

We can obtain an explicit expression for the coupling coefficient
$\kappa_{ABC}$ in terms of the mode functions $\xivec_A(\xvec)$,
$\xivec_B(\xvec)$ and $\xivec_C(\xvec)$ by substituting the formulae
(\ref{basicnonlin1}), (\ref{eq:17}) and (\ref{eq:16}) for the
nonlinear acceleration into the definition (\ref{kappaabcdef}) of
$\kappa_{ABC}$, using the definition (\ref{innerproduct}) of the
inner product, and performing several integrations by parts.
The computation can be simplified by (i) computing
the functional $\left< \xivec^* \, ,\, {\bf a}^{(2)}[\xivec,\xivec]
\right>$ of $\xivec$, (ii) equating this functional to
$\kappa(\xivec,\xivec,\xivec)$ where
$\kappa(\xivec,\xivec^\prime,\xivec^{\prime\prime})$ is a symmetric
trilinear functional of its arguments, in order to determine the
functional $\kappa(\xivec,\xivec^\prime,\xivec^{\prime\prime})$, and
(iii) evaluating $\kappa(\xivec_A,\xivec_B,\xivec_C)$ to obtain the coupling
coefficient $\kappa_{ABC}$.  For the gravitational terms, the domain
of the spatial 
integral can be 
taken to be all of space, if one includes the $\delta$-function
contributions in Eq. (\ref{phi3}).  The gravitational terms can then
be integrated by parts, and the boundary terms at infinity which are
generated vanish.  The domain of integration for the pressure terms
can be taken to be the interior of the star; the boundary terms
generated by the 
integration by parts vanish since $p=0$ on the stellar surface.  The
final result is
\begin{eqnarray}
\kappa_{ABC}
& = & \frac{1}{2} \int d^3x  \, p
\Biggl[
(\Gamma_1-1)\left( \Xi_{AB} \grad \cdot \xivec_C + 
\Xi_{BC} \grad \cdot \xivec_A + \Xi_{CA} \grad \cdot \xivec_B \right)
\nonumber \\ & + & 
\left\{  (\Gamma_1-1)^2 
+ \frac{\partial \Gamma_1}{\partial \ln \rho }\right\} 
\grad \cdot \xivec_A \grad \cdot \xivec_B  \grad \cdot \xivec_C 
+ \chi_{ABC} + \chi_{ACB} \Biggr]
\nonumber \\ & - & 
\frac{1}{2} \int d^3x \, \rho \left[ 
 \xi^i_A \xi^j_B \delta^{(1)} \phi_{C;ij} 
+ \xi^i_B \xi^j_C \delta^{(1)} \phi_{A;ij}
+ \xi^i_C \xi^j_A \delta^{(1)} \phi_{B;ij}  
\nonumber \right. \\ & + & \left.
\xi^i_A \xi^j_B \xi^k_C \phi_{;ijk}
\right].
\label{eq:23}
\end{eqnarray}
The notations here are as follows.  The quantity $\delta^{(1)} \phi_A$
is defined by Eq.\ (\ref{delta1phidef}) with $\xivec$ on the right
hand side replaced by $\xivec_A$.  
We define the functions 
\begin{eqnarray}
	\Thetauilj[\xivec,\xivec^\prime]&\equiv&\xiuilcdj \xi^{\prime\,k}_{\ ;k},\\
	 \Xiuilj[\xivec,\xivec^\prime]&\equiv&\xiuilcdk
	\xi^{\prime\,k}_{\ ;j},
		\label{thetaxip sidef1}
\end{eqnarray}
and
\beq
\chi^i_{\enskip j}[\xivec,\xivec^\prime,\xivec^{\prime\prime}] \equiv
\xiuilcdl \xi^{\prime\,l}_{\ ;k} \xi^{\prime\prime\,k}_{\ \ ;j}.
\endeq
The tensors $\Thetauilj$, $\Xiuilj$ and $\chi^i_{\enskip j}$ which we
defined previously in Eq.\ (\ref{thetaxipsidef}) are obtained from
these expressions evaluated at $\xivec = \xivec^\prime =
\xivec^{\prime\prime}$.  Finally we define the traces
\beq
\Xi_{AB} = \delta^j_i \, \Xiuilj[\xivec_A,\xivec_B]
\label{Xiabdef}
\endeq
and
\beq
\chi_{ABC} = \delta^j_i \, \chi^i_{\enskip j}[\xivec_A,\xivec_B,\xivec_C].
\label{eq:chiabcdef}
\endeq
The domain of integration for the gravitational term in Eq.\
(\ref{eq:23}) is all of space.
The domain of integration
for the pressure terms can be taken to be either the interior of the
star, or all of space.

Note that the expression (\ref{eq:23}) for the three mode coupling
coefficient is valid for arbitrary angular velocities $\Omega$ of the
background star.  The expression has no explicit dependence on
$\Omega$, however, and therefore is also valid for non-rotating stars.   
In fact the expression (\ref{eq:23}) agrees with previous 
results \cite{1994A&A...286..879V,1989ApJ...342..558K,Thesis:Wu}
for non-rotating stars, except for an overall factor of $-3$ due to
differing conventions for the form of the equation of motion.
Some of the previous analyses omitted the pressure force term
containing the factor $\partial \Gamma_1/\partial \ln \rho$,
and some used the Cowling approximation and thus omitted the
gravitational terms\footnote{Ref.\ \cite{1989ApJ...342..558K} makes a
slight error in calculating the gravitational terms.  However, since
they use the Cowling approximation in which $\delta\phi=0$, this
error does not affect their results.}.

For the special case of constant $\Gamma_1$, it is possible to
obtain an alternative form of the pressure terms in the coupling
coefficient by (i) taking the domain of integration in both the
gravitational and pressure terms to be all of space, and (ii)
performing further integrations by parts \cite{Wuref}.   
The result can be expressed as
\be
\kappa_{ABC} = \kappa(\xivec_A,\xivec_B,\xivec_C)
\ee
where
\begin{eqnarray}
\kappa(\xivec,\xivec,\xivec) 
& = & - \frac{1}{2} \int d^3x  \,
\Biggl[ \xi^i \xi^j \xi^k p_{;ijk} + 3 (\grad \cdot \xivec) \xi^i
\xi^j p_{;ij} - \Gamma_1 (\Gamma_1+1) p (\grad \cdot \xivec)^3
\nonumber \\ & + & 
3 \xi^i \xi^j (\delta p)_{;ij} + 6 (\grad \cdot \xivec) \xi^i (\delta
p)_{;i} + 3 \xi^i \xi^j \rho (\delta \phi)_{;ij} + \xi^i \xi^j \xi^k
\rho \phi_{;ijk} \Biggr],
\label{eq:23aa}
\end{eqnarray}
and $\delta p = -\Gamma_1\pnot\delxi -\xivec\cdot\grad \pnot$ is the
Eulerian pressure perturbation.  The domain of integration here is all
of space, and therefore one should include the $\delta$-functions at
the stellar surface in the factors $p_{;ijk}$, $p_{;ij}$ and possibly
$\phi_{;ijk}$.  However, one can show that all of the contributions
from the $\delta$ functions cancel in general, and therefore one can
take the domain of integration in the expression (\ref{eq:23aa}) to
be the interior of the star.

\subsection{Application to Rossby Modes}

Although the expression for the three-mode coupling coefficient is
the same for rotating stars as for non-rotating stars, it is much
more difficult to evaluate the coupling coefficient for rotating
stars, for two reasons.  First, including centrifugal flattening of
the background pressure and density profiles makes the integrals much
more difficult to perform.  Second, the mode functions
$\xivec_A(\xvec)$ for rotating stars typically consist of a sum of
terms with different values of the quantum number $l$, which
complicates the calculation.

We shall avoid the first of these difficulties by using the slow
rotation expansion.  First, we will compute the mode functions to
leading order in $\Omega$, as explained in Sec.\ \ref{sec:sec3} above.  
Second, in evaluating the coupling coefficient expression
(\ref{eq:23}), we will use the background density, pressure and
potential profiles of a spherical star, i.e., we will use the $\Omega
\to 0$ limits of the variables of the background star.  This
procedure will give the zeroth order piece of the coupling
coefficient $\kappa_{ABC}$ in an expansion in powers of $\Omega$.  In
cases where this zeroth order piece vanishes, the coupling coefficient
might be nonzero at higher orders in $\Omega$.

Before discussing in more detail how to compute the coefficients,
we describe some useful selection rules.

\section{Selection Rules for the Coupling Coefficients}
\label{sec:sec6}

\subsection{Terminology}

In this section we derive a number of selection rules that apply to
the three-mode coupling coefficient.  We start by reviewing some
terminology to describe modes.  Any vector field $\xivec(\xvec)$ can be
expanded in vector spherical harmonics as
\beq
\xivec(r,\theta,\varphi) = \sum_{lm} \bigg[ A_{lm}(r)
Y_{lm}(\theta,\varphi) {\bf e}_{\hat r} + B_{lm}(r) {\bfnabla}
Y_{lm}(\theta,\varphi) + C_{lm}(r) {\bf r} {\bf \times} {\bf \nabla}
Y_{lm}(\theta,\varphi) \bigg],
\label{lagpert} 
\end{equation}
where $(r,\theta,\varphi)$ are spherical polar coordinates and ${\bf
e}_{\hat r}$, ${\bf e}_{\hat \theta}$ and ${\bf e}_{\hat \varphi}$ is
the associated orthonormal basis.  Modes for which $A_{lm} = B_{lm}
=0$ for all $l,m$ are called {\it axial} modes.
We define $h$ to be the parity map $(x,y,z) \to (-x,-y,-z)$ and $f$
be the map $(x,y,z) \to (x,y,-z)$ 
or $(r,\theta,\varphi) \to (r,\pi - \theta,\varphi)$
which we call $z$-parity.  Rotating
stars, both Newtonian and relativistic, are invariant under these
maps, and therefore we can always choose to use a basis of modes for
which all the modes have definite parity and $z$-parity transformation
properties; see Refs.\ \cite{yoshidaleeisentropic,Graff}.  Specifically, we
define the pullback $f_* \xivec$ of 
$\xivec$ under $f$ by \cite{bishopgoldberg}
\begin{eqnarray}
	f_*\xivec &=& f_*\left[\xi^{\hat{r}}\ofrthetaphi\unit_{\rhat} 
		+\xi^{\thhat}\ofrthetaphi\unit_\thhat +
		\xi^\phhat\ofrthetaphi\unit_\phhat\right]
		\nonumber\\
	&=& \xi^{\hat{r}}\fofrthph \unit_{\rhat} -\xi^{\thhat}\fofrthph\unit_\thhat +
		\xi^\phhat\fofrthph\unit_\phhat.
		\label{fstaronvec}
\end{eqnarray}
Then we can always find a basis for which all modes are either
{\it z-parity even}, satisfying
\beq
f_* \xivec = \xivec,
\endeq
or {\it z-parity odd}, satisfying
\beq
f_* \xivec = -\xivec.
\endeq
Note that $f_* Y_{lm} = (-1)^{l+m} Y_{lm}$.  A similar statement is
true for the regular parity map $h$; we can take all modes to be
either parity even or parity odd.  Finally we can take all modes to
have definite values of $m$, so that if $z$ is the mapping
$(r,\theta,\varphi) \to (r,\theta,\varphi + \Delta \varphi)$, then
\beq
z_* \xivec = \exp[i m \Delta \varphi] \xivec.
\label{mqndef}
\endeq
Recall also 
that all modes can be classified as either {\it rotational} ($\omega
\to 0$ as $\Omega \to 0$) or {\it regular} ($\omega \to $ finite as
$\Omega \to 0$), as discussed in Sec.\ \ref{sec:slow_rot1}
above.

\subsection{Classes of modes}

We next review the various different classes of modes to which we
shall apply the selection rules.  In zero-buoyancy stars, we
distinguish three different classes of modes:

\begin{itemize}

\item The rotational modes which are axial to zeroth order in
$\Omega$.  These modes are characterized by single
values of $l$ and of $m$.  If one is working with all the modes
$\xivec_A$, then one can have $m=l$ or $m=-l$.  If one is working only
with the set of modes $\xi_\alpha$ with positive rotating-frame
frequency $\omega$ (the remaining modes being complex conjugates of
these), then only $m=-l$ is allowed.  These modes are all $z$-parity odd. 
We shall call these modes the {\it pure r-modes}.

\item The rotational modes whose $\Omega \to 0$ limits are not axial.
These have fixed $m$ but do not have a fixed value of $l$.  To zeroth
order in $\Omega$ 
they can be expanded as \cite{1999ApJ...521..764L}
\begin{eqnarray}
	\xivec(\xvec) = 
		\sumjinf\left[{\Wjpmpo\ofr\over r}
	\Yjpmpo\ofr \unit_{\hat r} +
		\Vjpmpo\ofr\grad\Yjpmpo-{\Ujpm\ofr\over r}\opL\Yjpm\right],
       \label{axialgen}
\end{eqnarray}
in the $z$-parity odd case,
and as
\begin{eqnarray}
	\xivec(\xvec) =  \sumjinf\left[
		{\Wjpm\ofr\over r}\Yjpm\unit_{\hat r} +
		\Vjpm\ofr\grad\Yjpm-{\Ujpmpo\ofr\over r}\opL\Yjpmpo\right]
	\label{polargen}
\end{eqnarray}
in the $z-$parity even case,
where $\opL = -i( {\bf r}\times\grad)$ and $j$ runs over $0,2,4,
\ldots$.  Following Ref.\ \cite{gr-qc0008019} we shall call these the
{\it rotational hybrid} 
modes, although they have also been called inertial modes or
generalized $r$-modes.

\item The regular modes, which can be $z$-parity even or $z$-parity
odd, and consist of $f$ and $p$ modes.

\end{itemize}

Similarly, in stars with buoyancy we shall distinguish the following
classes of modes:

\begin{itemize}

\item The rotational modes.  All of these modes are axial to zeroth
order in $\Omega$ (Appendix \ref{sec:zero-freq}).  They are also
characterized by single values of $l$ and $m$, but any values of $l$,
$m$ with $l \ge |m|$ are allowed.  They have $z$-parity
$(-1)^{l+m+1}$ and parity $(-1)^{l+1}$.  We shall call these the
$r$-modes.  [Unlike the pure 
$r$-modes above, their radial eigenfunctions are not polynomial in
$r$.]

\item The regular modes, which may be $z$-parity even or $z$-parity
odd, and consist of $f$, $p$ and $g$ modes.

\end{itemize}

\subsection{Selection rules}

We can write the three mode coupling coefficient as a symmetric
function of the three mode functions $\xivec_A$, $\xivec_B$,
$\xivec_C$:
\beq
\kappa_{ABC} = \kappa(\xivec_A,\xivec_B,\xivec_C).
\label{sr1}
\endeq
Now apply the pullback $f_*$ to this equation.  Since the left hand
side is a pure number, it is invariant.  On the right hand side, we
can pull the $f_*$ operator inside the function $\kappa$ since pullback
commutes with geometrical operations like taking covariant
derivatives.  Thus
\beq
\kappa_{ABC} = \kappa(f_* \xivec_A, f_* \xivec_B, f_* \xivec_C).
\label{sr2}
\endeq
If we denote the $z$-parity of mode $A$ by $\epsilon_A = \pm 1$, so that $f_*
\xivec_A = \epsilon_A \xivec_A$, then it follows from Eqs.\ (\ref{sr1})
and (\ref{sr2}) that 
\beq
\left[ 1 - \epsilon_A \epsilon_B \epsilon_C \right] \kappa_{ABC} =0.
\label{sr3}
\endeq
Therefore we get the selection rule
\beq
\FL
{\text z-parity\ selection\ rule:\ \ \ \ }{\text {\rm\ odd\ number\ of\
z-parity\ odd\ modes\ \ \ }} \Rightarrow \kappa_{ABC} =0.
\label{zpsel}
\endeq
For example, if all three modes are $z$-parity odd, then the coupling
coefficient vanishes.  An identical argument gives the selection rule
for regular parity
\beq
\FL
{\text \,parity\ selection\ rule:\ \ \ \ }{\rm\ odd\ number\ of\ parity\
odd\ modes\ \ \ } \Rightarrow \kappa_{ABC} =0.
\label{psel}
\endeq
Finally, applying Eq.\ (\ref{sr3}) with $f$ replaced by the mapping
$\varphi \to \varphi + \Delta \varphi$ and with $\epsilon_A$ replaced
by $\exp[i m_A \Delta \varphi]$ [cf.\ Eq.\ (\ref{mqndef}) above]
yields
\beq
\FL
m \ {\text selection\ rule:\ \ \ \ \ } m_A + m_B + m_C \ne 0
\ \ \ \ \Rightarrow \kappa_{ABC} =0.
\label{mselection}
\endeq
These selection rules are valid in rapidly rotating stars as well as
in the slow rotation limit, and also retain their validity for
relativistic stars.  The arguments can also be generalized to
four-mode coupling coefficients $\kappa_{ABCD}$.

The selection rules (\ref{zpsel}), (\ref{psel}), (\ref{mselection})
are based on 
symmetries of the background star and thus apply to all orders
in $\Omega$.  Therefore, when they are applicable, they restrict not
only the piece of $\kappa_{ABC}$ which is zeroth order in $\Omega$
(which is all we 
compute in this paper), but also all correction terms that are higher
order in $\Omega$.  By contrast, in Appendix \ref{sec:vanishing} we
derive two selection rules which are valid to zeroth order in $\Omega$
only.  First, we show the coupling coefficient for the coupling of three
axial modes is $O(\Omega)$.  
Thus we have
\beq
\FL
{\text \,Axial\ selection\ rule\ 1:\ \ \ \ }
\xivec_A, \xivec_B, \xivec_C = {\rm axial\ } + O(\Omega) \ \ 
\Rightarrow \kappa_{ABC} =O(\Omega).
\label{axialsel1}
\endeq
We also prove in Appendix \ref{sec:vanishing} the following second
selection rule.
If (i) two of the modes are axial, (ii) the third has
vanishing Eulerian density perturbation, and (iii) the background star 
and perturbations obey the same one-parameter equation of state (i.e., the
zero-buoyancy or barotropic case), then $\kappa_{ABC}= O(\Omega)$.  
We can write this selection rule as
\beq
\FL
{\text \,Axial\ selection\ rule\ 2:\ \ \ \ }
\delta \rho_A = O(\Omega),
\ \xivec_B, \xivec_C = {\rm axial\ } + O(\Omega),
\ {\rm zero\ buoyancy} \ \ \ \Rightarrow \kappa_{ABC} =O(\Omega).
\label{axialsel2}
\endeq
Note that axial modes automatically have vanishing density perturbation
to zeroth order in $\Omega$, from Eq.\ (\ref{deltarhodef}) and using
${\bf \nabla} \cdot \xivec = \xi^r = 0$ and the fact that the
background density is a function of $r$ only to leading order in $\Omega$.   
The selection rules (\ref{axialsel1}) and (\ref{axialsel2}) 
might fail for relativistic stars.

In incompressible stars, we prove in Appendix \ref{sec:vanishing} that
the coupling between any three rotational modes vanishes to zeroth
order in $\Omega$.  A precise statement of this selection rule is 
\beq
\FL
{\text \,Incompressible\ selection\ rule:\ \ \ \ }
\delta \rho_A, \delta p_A, \delta \rho_B, \delta p_B, \delta \rho_C,
\delta p_C = O(1 / \Gamma_1) +O(\Omega) 
\ \ \ \Rightarrow \kappa_{ABC} =O(1/\Gamma_1) + O(\Omega),
\label{incompsel}
\endeq
where $\Gamma_1$ is the adiabatic index of the perturbations.
Note that the regular modes ($f$-modes) of an incompressible star have
$\delta \rho \to {\rm const}$ in the limit $\Gamma_1 \to \infty$, and
do not satisfy the hypothesis of this selection rule, whereas the
rotational modes have exactly vanishing $\delta \rho$ and $\delta p$
(to zeroth order in $\Omega$) and so do satisfy the hypothesis.

The $\Omega$ dependence of the coupling coefficients $\kappa_{ABC}$
arises in two ways:  (i) through the dependence of the expression
(\ref{eq:23}) on the background star, for which the leading order
corrections arise at $O(\Omega^2)$, and (ii) through the dependence of
the mode functions themselves on $\Omega$, for which the leading order
corrections are $O(\Omega^2)$ for pure $r$-modes
\cite{gr-qc9902052,yoshidaleeisentropic}, but can be
$O(\Omega)$ for regular modes.  Therefore, when the zeroth order
coupling coefficient vanishes, the leading order corrections 
may be $O(\Omega)$ or $O(\Omega^2)$, depending on the modes involved.

\subsection{Applications of selection rules}

We now discuss some applications of these rules.
Consider first the case of zero-buoyancy stars.   The most interesting
coupling coefficients are those in which an unstable pure $r$-mode
appears twice, since if the amplitudes of all 
the other modes are zero initially, it is via these coupling
coefficients that energy can leak out of the pure $r$ mode.  
In particular we are interested in the most unstable mode, the $l=m=2$
mode.  If $\xivec_A =
\xivec_B$ is an unstable pure $r$-mode, then the third mode $\xivec_C$
must be a $z$-parity even mode from rule (\ref{zpsel}).
In particular, the third mode cannot
be a pure $r$-mode as pure $r$ modes have odd $z$ parity.  Coupling
coefficients between any three pure $r$-modes vanish. 
Nonzero coefficients can be obtained by taking the third mode to be
a $z$-parity even regular mode.  However, if the third mode is taken
to be a $z$-parity even rotational hybrid, then the second axial selection
rule (\ref{axialsel2}) applies due to Eq.\ (\ref{df1}), and implies that
the coupling coefficient vanishes to  
zeroth order in $\Omega$.  It may be nonvanishing to higher order in
$\Omega$ \cite{sharonprivate}.
If one considers one pure $r$-mode and two rotational hybrid modes,
the corresponding coupling vanishes in an incompressible star by the
incompressible selection rule, but may be nonvanishing at
$O(1/\Gamma_1)$.

In Table~\ref{tab:cdCCsum} we summarize which classes of
modes can and cannot couple to zeroth order in $\Omega$ in 
in zero-buoyancy stars.  Stars indicate non-zero 
coupling coefficients.   Since $\kappa_{ABC}$ is 
symmetric, Table~\ref{tab:cdCCsum} lists all possible types of coupling
coefficient except one, the coupling coefficient between one pure
$r$-mode, one regular mode and one hybrid rotational mode; this
coupling coefficient is non-zero.   The table shows the kinds of coupling
coefficients that can be non-zero when $z-$parity restrictions are met.
For example, the table indicates that three hybrid rotational
modes can couple together, but this can only happen when the coupling
involves an even number of $z-$parity odd modes.

For nonzero-buoyancy stars, consider the coupling coefficient between
three $r$-modes.  If the three modes have quantum numbers $l_A$,
$m_A$, $l_B$, $m_B$ and $l_C$, $m_C$, then since each mode has
$z$-parity $(-1)^{l+m+1}$, Eq.\ (\ref{sr3}) gives
\beq
\left[1 - (-1)^{3 + m_A + m_B + m_C + l_A + l_B + l_C} \right]
\kappa_{ABC}=0.
\endeq
Invoking the $m$ selection rule (\ref{mselection}) implies that $l_A +
l_B + l_C$ must be odd for $\kappa_{ABC}$ to be non-zero.  
If all three modes are of the $|m|=l$ variety, this contradicts $m_A +
m_B + m_C =0$, so the coupling of three $l=|m|$ modes is zero.
If some of the three modes do not satisfy $l = |m|$, then the
$z$-parity and parity selection rules do not apply.  However, the
axial selection rule (\ref{axialsel1}) does apply and therefore
the coupling coefficient vanishes to zeroth order in $\Omega$.  
To higher order in $\Omega$ however, these coupling coefficients
with $l_A + l_B + l_C$ odd and $m_A + m_B + m_C =0$ need not vanish.
In fact, Morsink \cite{sharonprivate} has calculated coupling
coefficients of this type to $O(\Omega^2)$ and has found them to be
nonzero.

\section{Computational method for Coupling Coefficients}
\label{sec:sec7}

In this section we describe an efficient computational method
for computing coupling coefficients in rotating stars.  Specific
coupling coefficients relevant to $r$-mode saturation will be computed
numerically in a subsequent paper.  

\subsection{Overview}

Direct calculation of the 
expression (\ref{eq:23}) for the coupling coefficients is algebraically
intensive even in the case where all three modes have no sums over ${l}$.
If one proceeds in spherical coordinates to directly 
calculate all the covariant 
derivatives needed one encounters many terms that scale as 
${1/\sin^n\theta}$ with $n\ge 2$.  These terms  are divergent as 
$\theta\rightarrow 0$ or $\pi$.  Since the full coupling coefficient is 
obviously finite at the poles ($\theta=0,\pi$), these terms must cancel
with other terms of the same order in $1/\sin \theta$.  However, the
amount of algebra  
necessary to cancel these terms by hand is unmanageable.
Calculation of the coupling coefficients in Cartesian coordinates is also 
daunting. Even though all covariant derivatives are highly simplified
in Cartesian coordinates,
the total number of terms is greater than in spherical coordinates 
and one must still integrate over
the sphere in the end, which is awkward.  
Cartesian coordinates also obscure selection rules that might lead to 
insight into which modes couple and which don't. Methods based on 
integration by parts simplify the algebra for non-rotating
stars \cite{Thesis:Wu,1989ApJ...342..558K},
but, for rotating stars, modes 
acquire toroidal pieces that severely complicate the 
angular integrations even after integration by parts.  

To circumvent
these difficulties, we have found that using the spin-weighted spherical 
harmonics of Ref.\ \cite{newmanpenrose} as angular basis functions is
ideal.  The use of these basis functions
allows one to immediately reduce the angular integrals 
to Wigner $3-j$ symbols.  It also 
reduces dramatically the number of terms that must be calculated.   
For example, for modes that have only {\it one} value of ${l}$ each,
(\ref{eq:23}) requires
explicit calculation of $\sim 100 $ terms, including many that
require computing angular 
derivatives of the mode functions.  
When modes involve sums over $n$ values of 
${l}$, the number of terms to calculate scales as $n^3$.
Once this is all done, one must then integrate 
the resulting complicated expression over angles.  
However,  in the spin-weighted
formalism, the calculation involves just $6 n^3$ terms,
none of which requires computing angular derivatives.  Moreover, those
$6 n^3$ terms are completely spherically symmetric, all angular
dependence being already 
integrated out in terms of Wigner $3-j$ symbols.  
The coupling coefficients
are then built from these spherically symmetric terms, their
corresponding $3-j$ symbols, and their complex conjugates  
and symmetrizations.
The method is easy to automate, including the various sums
over the different ${l}$'s of the modes, leaving only $6$ terms to actually
code up.

This section is organized as follows.  In
Sec.~\ref{sec:An-Appr-Choice} we define the spin-weighted basis  
and express the covariant derivatives needed for the coupling
coefficients in terms of components on that basis.  Some details are
relegated to Appendix ~\ref{app:ccdswsh}.  
Section \ref{sec:Integr-Coupl-Coeff} derives the 
explicit expression (\ref{eq:30}) 
for the coupling coefficient in terms of spherically
symmetric functions and Wigner $3-j$ symbols.

The method described here is specialized to slowly rotating stars and
is valid only to zeroth order in $\Omega$, since we take the 
background density and pressure to be functions of $r$ only.
However, it is straightforward to extend the method to higher orders
in $\Omega$.  For example, in calculating coupling coefficients to
second order in $\Omega$, contributions of order $O(\Omega^2)$ will
arise both from corrections to the background stellar model
(centrifugal flattening), and from corrections to the modes.
Including the additional angular
dependence of the background density and pressure would require
computing angular integrals of products of four spin-weighted
spherical harmonics instead of three.

\subsection{An appropriate choice of basis}\label{sec:An-Appr-Choice}

Consider a given mode function $\xivec({\bf x})$.  Since any
mode has a fixed value of $m$, we can assume an expansion of the mode
function on a basis of vector spherical harmonics of the form
\begin{eqnarray}
	\xivec = \sum_{{l}=\vert m\vert}^\infty 
	\left[ {W_{{l} m}\ofr\over r}Y_{{l} m}
	 \unit_{\hat r} +
		{V_{{l} m}\ofr}
	\grad Y_{{l} m}-
		{U_{{l} m}\ofr\over r}\opL Y_{{l} m}\right],
	\label{genformmodes}
\end{eqnarray}
where $\opL = -i( {\bf r}\times\grad)$.
We define a set of {\it spin-weighted} basis vectors consisting of 
two complex null 3-vectors, $\mvec$ and $\mbarvec$ given by
\be
	\mvec = \rtti\left( \eveclthhat +i\eveclphhat\right)
	\label{m}
\ee
and
\be
	\mbarvec = \rtti\left( \eveclthhat -i\eveclphhat\right)
	\label{mbar}
\ee
and a vector orthogonal to those two,
\be
	\lvec = \eveclrhat.\label{l}
\ee
The components of the metric on this non-coordinate basis are
\begin{eqnarray}
	&\gull& = \glll=1,\\
	&\gummbar& = \glmmbar = 1,\label{lmmbarmet}
\end{eqnarray}
with all other components zero.  
The components of the mode function $\xivec$ on this basis are given by
\begin{eqnarray}
\xi_l &\equiv& \vartheta_0 = \lvec\cdot \xivec \nonumber\\
\xi_m &\equiv& \vartheta_1 = \mvec\cdot \xivec \nonumber\\
\xi_{\bar m} &\equiv& \vartheta_{-1} = \mbarvec \cdot \xivec,
\label{xicomps}
\end{eqnarray}
where the subscript $0,1$ or $-1$ on the scalar function 
$\vartheta\ofrthph$ represents
the spin weight\footnote{For a definition of spin-weight, see 
Appendix~\ref{app:ccdswsh}.} of the component. 
Note that 
\begin{eqnarray}
 \xi_l &=& \xi^l = \vartheta_0 \nonumber \\
 \xi_m &=& \xi^{\bar m} = \vartheta_1 \nonumber \\
 \xi_{\bar m} &=& \xi^{m} = \vartheta_{-1},
\label{raiselower}
\end{eqnarray}
and that the mode function can be reconstructed as
\beq
\xivec = \xi^l {\bf l} + \xi^{{\bar m}} {\bf {\bar m}} +
\xi^{{m}} {\bf {m}}.
\endeq

Spin-weighted spherical harmonics are defined in terms of the matrix
representations of the rotation group, $D^{l}_{-sm}$, by
\begin{equation}
	\Ylms\ofthph = \sqrt{(2{l}+1)/4\pi}D^{l}_{-sm}
			(\phi,\theta,0),\label{eq:25}
\end{equation}
where $s$ is the spin weight and $\Ylmz$ are the ordinary spherical
harmonics $\Ylm$.  See  
Ref.\ \cite{edmonds} for a definition and a detailed list of the
properties of the quantities $D^{l}_{-sm}$.
For each value of the spin weight $s$, these spin-weighted spherical harmonics
form a complete orthonormal set, i.e.
\begin{equation}
	\int {\Ylms}^*\Ylpmps\, d\Omega = \delta_{{l}^\prime {l}}
			\delta_{m^\prime m}.\label{eq:26}
\end{equation}
Thus we can expand $\vtz,\vto,\vtmo$ on the
spin-weighted spherical harmonics bases as
\begin{eqnarray}
	\vtz\ofrthph &=& \sum_\Lambda \fzlam\ofr\,\,\Ylamz\ofthph\nonumber\\
	\vto\ofrthph &=& \sum_\Lambda\fplam\ofr\,\,\Ylamo\ofthph\nonumber\\
	\vtmo\ofrthph &=& \sum_\Lambda\fmlam\ofr\,\,\Ylammo\ofthph,
		\label{eq:28}
\end{eqnarray}
where the subscripts $+$ and $-$ are understood to mean $+1$ and $-1$
respectively, and $\Lambda$ denotes $(lm)$.  Therefore the expansion
of the mode function can be written as
\be
\xivec = \sum_{lm} \left[ 
f_0^{lm}(r) \, {}_0 Y_{lm} \, {\bf l} \ \ 
+ f_+^{lm}(r) \, {}_{1} Y_{lm} \, {\bar {\bf m} } \ \ 
+ f_-^{lm}(r) \, {}_{-1} Y_{lm} \, {\bf m}
\right].
\ee
In Appendix \ref{app:ccdswsh} we relate the coefficients $f_s^{lm}(r)$
appearing in this expansion to the coefficients $U_{lm}(r)$,
$V_{lm}(r)$, and $W_{lm}(r)$ appearing in the expansion
(\ref{genformmodes}) above [see Eqs.\ (\ref{axialfs}) and (\ref{polarfs})].

We will be using a nonstandard notational convention involving the
symbol $\Lambda$ which we now explain.
In rotating stars, ${l}$ is generally not a good
``quantum number'' while $m$ is.  In general there will be other good
quantum numbers 
that will allow us to enumerate the modes.  What is meant by the symbol
$\Lambda$ is a set of numbers representing both the 
good and bad quantum numbers of the object in question. For example, when 
$\Lambda$ is used on a $\Ylms$ we mean the string of quantum numbers $({l},
m)$.   However, when $\Lambda$ is used on the mode function $\fplam$ it 
may mean a string of numbers that includes,  but is not limited
to ${l}$ and $m$.
When we sum over $\Lambda$ it is implied that we sum only over bad quantum
numbers.  Thus we might label the above component $\vtz$ with another 
index 
($A$ say) representing only the good quantum numbers that are
left after the bad ones have been summed over, e.g.
\begin{equation}
         \vtz^A\ofrthph = \sum_\Lambda \fzlam\ofr\,\,\Ylamz\ofthph.
        \label{eq:29}
\end{equation}
And finally, when we use the symbol $\Lambda$ {\it not} in a sub- or super-
script it is defined to be $\Lambda\equiv\sqrt{{l}({l}+1)}$.

We now express the covariant derivatives of the mode functions that
appear in the coupling  
coefficient (\ref{eq:23}) in terms of our new expansion coefficients
$f_s^\Lambda(r)$.  The details of
the calculation are given in Appendix~\ref{app:ccdswsh}.
We define the functionals $\Gslam[\fslam\ofr]$, $\Fslam[\fslam\ofr]$, 
and $\Hslam[\fslam\ofr]$ to be
\begin{eqnarray}
	\Gslam[\fslam\ofr]&\equiv& \ovr\left(\fzlam\ofr 
		-{s\Lambda\over\rttwo}
		\fmslam\ofr\right)\nonumber\\
	\Fslam[\fslam\ofr]&\equiv&-\ovr\left(\fslam\ofr 
		+{s\Lambda\over\rttwo}\fzlam\ofr
		\right)\nonumber\\
	\Hslam[\fslam\ofr]&\equiv& 
		-{s\over\rttwo r}\fslam\ofr\sqrt{\Lambda^2-2}.
		\label{gfhdefs}
\end{eqnarray}
Then the components in the basis $(\lvec,\mvec,\mbarvec)$ of the tensor
$\xi^i_{;j}$ can be written as
\begin{eqnarray}
	\xiullcdl &=& \sum_\Lambda{\fzlam}_{,r}\ofr\,\,\Ylamz\nonumber\\
	\xiumbarlcdl &=& \sum_\Lambda{\fplam}_{,r}\ofr\,\,\Ylamo,\quad\quad
	     \xiumlcdl = \sum_\Lambda{\fmlam}_{,r}\ofr\,\,\Ylammo\nonumber\\
	\xiumlcdm &=& \sumlam\Gplam[f]\,\,\Ylamz,\quad\quad
		\xiumbarlcdmbar = \sumlam\Gmlam[f]\,\,\Ylamz,\nonumber\\
	\xiumbarlcdm &=& \sumlam\Hplam[f]\,\,\Ylamt,\quad\quad
		\xiumlcdmbar = \sumlam\Hmlam[f]\,\,\Ylammt,\nonumber\\
	\xiullcdm &=& \sumlam\Fplam[f]\,\,\Ylamo,\quad\quad
		\xiullcdmbar = \sumlam\Fmlam[f]\,\,\Ylammo,
		\label{covdivsimp}
\end{eqnarray}
where we abbreviate $\Gslam[\fslam\ofr]$ as $\Gslam[f]$, etc.
Note that this formalism allows us to write down forms
for the covariant derivatives in terms of functions of $r
$ only. The angular derivatives have all been computed.
Thus the formalism avoids the large curvature
terms that can arise from angular covariant derivatives in spherical 
coordinates.  Another advantage is that the formalism allows us to 
integrate the coupling coefficients over angles {\it immediately}; no
integrations by parts are needed.  To see how this works we note that
the integral over angles of three of the spin-weighted spherical 
harmonics can be written simply in terms of Wigner $3-j$ symbols 
\cite{edmonds}, which are easily calculated:
\begin{eqnarray} 
	\int \Ylmsone\Ylmstwo\Ylmsthree {d\Omega}
	&\equiv&\langle\Ylmsone\Ylmstwo\Ylmsthree\rangle\nonumber\\
	&=& 4\pi\sqrt{(2{l}_1+1)/4\pi}\sqrt{(2{l}_2+1)/4\pi}
		\sqrt{(2{l}_3+1)/4\pi}\nonumber\\
	& &\times\int D^{{l}_1}_{-s_1 m_1}\ofphthz
		   D^{{l}_2}_{-s_2 m_2}\ofphthz 
			D^{{l}_3}_{-s_3 m_3}\ofphthz
			{d\Omega}\nonumber\\
	&=& 4\pi\sqrt{(2{l}_1+1)/4\pi}\sqrt{(2{l}_2+1)/4\pi}
		\sqrt{(2{l}_3+1)/4\pi}\nonumber\\
	& &\times\left(\begin{array}{ccc}
		{l}_1&{l}_2&{l}_3\\
		-s_1&-s_2&-s_3
		\end{array}\right)	
	     \left(\begin{array}{ccc}
		 {l}_1&{l}_2&{l}_3\\
		m_1&m_2&m_3
		\end{array}\right),
		\label{int3ylms}
\end{eqnarray}
Note that this integral vanishes unless $m_1+m_2+m_3=0$,
{\it cf}. the $m$ selection rule (\ref{mselection}).

\subsection{Integrating the coupling coefficients over the sphere}
\label{sec:Integr-Coupl-Coeff}
To begin integrating the expression (\ref{eq:23}) over angles we write
\begin{eqnarray}
\kappa_{ABC} \equiv
		\int r^2dr \,\,
\kappa_{ABC} 
		\ofr,\label{kappadefb2}
\end{eqnarray}
and note that in calculating $\klabg$ we can ignore mode indices
as long as we symmetrize at the end of the computation.  To this end
define an operator 
$\sym$ that symmetrizes over mode indices:
\begin{eqnarray}
	\sym L_{\Lambda_1\Lambda_2\Lambda_3}\equiv 
		L_{(\Lambda_1\Lambda_2\Lambda_3)}&\equiv&
		{1\over 6}\left(L_{\Lambda_1\Lambda_2\Lambda_3}
	+L_{\Lambda_2\Lambda_1\Lambda_3}\right.\nonumber\\
	&&+\left.L_{\Lambda_3\Lambda_2\Lambda_1}
	+L_{\Lambda_1\Lambda_3\Lambda_2}
		+L_{\Lambda_3\Lambda_1\Lambda_2}+
		L_{\Lambda_2\Lambda_3\Lambda_1}\right).\label{symdef}
\end{eqnarray}
Now expand the scalar $ \grad \cdot \xivec_A$ as
\begin{eqnarray}
\grad \cdot \xivec_A \equiv \sumlam \gzlam\ofr\,\,\Ylamz,\label{gdef}
\end{eqnarray}
and the scalar $\delta\phi_A$ as 
\begin{eqnarray}
       a\equiv\delta\phi_A \equiv \sumlam
               \azlam\ofr\,\,\Ylamz.\label{adef1}
\end{eqnarray}
The second covariant derivatives of this scalar are then
given by
\begin{eqnarray}
	a_{;mm} &=&\sumlam\Ezlam[a]\,\,\Ylamt,\quad\quad 
		a_{;\bar m\bar m} =\sumlam\Ezlam[a]\,\,\Ylammt\nonumber\\
		a_{;lm}&=&a_{;ml} = \sumlam\Cplam[a]\,\,\Ylamo,\quad\quad
		a_{;l\bar m}=a_{;\bar ml} 
			= \sumlam\Cmlam[a]\,\,\Ylammo\nonumber\\
	a_{;m\bar m} &=& a_{;\bar m m} = \sumlam\Dzlam[a]\,\,\Ylamz
	,\quad\quad a_{;{l}{l}} = \sumlam {\azlam}_{,rr}\,\,
		\Ylamz,\label{adef2}
\end{eqnarray}
where the functionals $\Ezlam[a\ofr],\Cslam[a\ofr]$ and
$\Dzlam[a\ofr]$ are
\begin{eqnarray}
	\Ezlam[a\ofr] &\equiv& {\Lambda\over 2 r^2}\sqrt{\Lambda^2 -2}
		\, \azlam
		\ofr,\nonumber\\
	\Dzlam[a\ofr] &\equiv&{1\over 2r^2}\left (2r\,{\azlam}_{,r} -
		\azlam \Lambda^2\right),\nonumber\\
	\Cslam[a\ofr] &\equiv& {s\Lambda\over \rttwo r}\left(
		\ovr\,\azlam -{\azlam}_{,r}\right).\label{ddcds}
\end{eqnarray}
We are now in a position to write
down $\klabg\ofr$:
\begin{eqnarray}
	\klabg\ofr &=& \half \sumlamotth\Bigl\{
		\Lo\langle\Ylamoz\Ylamtz\Ylamthz\rangle
			\nonumber\\
	& &
		\quad\quad+6\sym\left[
			\Lt\langle\Ylamoz\Ylamto\Ylamthmo\rangle
			\right]
				\nonumber\\
	& &
		\quad\quad+6\sym\left[
			\Lth\langle\Ylamoz\Ylamtt\Ylamthmt
			\rangle
			\right]
				\nonumber\\
	& &
		\quad\quad+6\sym\left[
			\Lf\langle\Ylamomt\Ylamto\Ylamtho\rangle
			\right]\nonumber\\
	& &	\quad\quad+\pgtm\Bigr\},\label{eq:30}
\end{eqnarray}
where the $\pgtm$ symbol means repeat all terms in the above equation
replacing $s\rightarrow -s$ in the integrals and interchanging $+$ and
$-$ subscripts in the functions (detailed below). The
functions $\Li$ are given by 
\begin{eqnarray}
	\Lo &=&\half\pnot 
\left\{ (\Gamma_1-1)^2 + \frac{\partial \Gamma_1}{\partial \ln \rho_0} \right\} 
\gzlamo\gzlamt
			\gzlamth-
		\half\rhonot{d^3\phinot\over dr^3}\fzlamo\fzlamt\fzlamth
				\nonumber\\
	& &+\sym\left\{\pnot(\Gamma_1-1)3\left[\half\gzlamo{\fzlamt}_{,r}
			{\fzlamth}_{,r} +\gzlamo\Gplamt[f]\Gplamth[f]
			\right]  \right.\nonumber\\
		& &+ 2\pnot\left[\half{\fzlamo}_{,r}{\fzlamt}_{,r}
		{\fzlamth}_{,r} + \Gplamo[f]\Gplamt[f]\Gplamth[f]
			 \right]\nonumber\\
		& &\left .-\rhonot {3\over 2}\fzlamo{\azlamt}_{,rr}
			\fzlamth\right\},
		\label{eq:31}\\
	\Lt &=& \pnot(\Gamma_1-1)\gzlamo\Fplamt[f]{\fmlamth}_{,r}
		\nonumber\\
	& &-\rhonot \left(\half\Dzlamo[a]\fplamt\fmlamth +\fzlamo\Cplamt[a]
		\fmlamth\right)\nonumber\\
	& &+\pnot\left[{\fzlamo}_{,r}
		+\Gplamo[f]\right]\Fplamt[f]{\fmlamth}_{,r}
			\nonumber\\
	& &-\half\rhonot{d\over dr}\left({1\over r}{d\phinot
		\over dr}\right)\fzlamo\fplamt\fmlamth,\label{ltdef}\\
	\Lth &=& \half\pnot(\Gamma_1-1)\gzlamo\Hplamt[f]\Hmlamth[f]
		+\pnot\Gplamo[f]\Hplamt[f]\Hmlamth[f], \label{eq:34}\\
	\Lf &=& \pnot\Hmlamo[f]{\fplamt}_{,r}\Fplamth[f]
		-\rhonot\half\Ezlamo[a]\fplamt\fplamth,\label{lfvdef}
\end{eqnarray}
where a comma indicates ordinary differentiation.

To summarize, given a set of three modes $\xivec_A$, $\xivec_B$,
$\xivec_C$, the computational method is to (i) Use Eqs.\
(\ref{axialfs}) and (\ref{polarfs}) to compute 
the expansion coefficients $f_s^{lm}(r)$ for each mode in terms of the
more standard expansion coefficients $U_{lm}(r)$, $V_{lm}(r)$, and
$W_{lm}(r)$ appearing in the expansion (\ref{genformmodes}), and (ii) Use
the expressions (\ref{int3ylms}), (\ref{kappadefb2}) and
(\ref{eq:30}) -- (\ref{lfvdef}) to compute the coupling coefficient
$\kappa_{ABC}$ from the functions $f_s^{lm}(r)$.

\section{Summary}
\label{sec:sec8}

In this paper, we have formulated a perturbative approach to the nonlinear
interactions of unstable $r-$modes in a neutron star.  Our formalism presumes
that mode growth saturates at moderately low amplitudes, so that we can
model the modal interactions via three-mode couplings. By developing 
further a previous perturbation theory for rotating stars, we have
found equations of motion for the mode amplitudes of rotating stars
that are uncoupled at linear order when acted upon by an external
force.  This feature of our formalism is essential for following the
cascade of energy from one mode to another when lowest order nonlinear
couplings are included. 

The important astrophysical question is what determines the saturation
amplitude for unstable $r-$modes.  Fundamentally, if mode-mode coupling is
the dominant damping, we expect the saturation
amplitude to be set by a competition between the growth rate of the 
instability, and the amplitude-dependent rate of drainage of energy from
an unstable mode to other stellar modes.  Our formalism gives explicit
formulae for the lowest-order coupling coefficients among stellar modes.
The numerical simulations by 
Stergioulas and Font \cite{gr-qc0007086},
and Lindblom, Tohline and Vallisneri \cite{astro-ph0010653},
have found the coupling of the unstable ${l}=\vert m\vert=2$
$r-$mode to other stellar modes
to be surprisingly weak, resulting in a timescale for energy
transfer to other modes that is at least $\sim 20$ rotation periods even
when the modal amplitude is substantial. 

Although we shall present detailed numerical solutions of our equations
of motion for modal amplitudes in a subsequent publication, there are
already hints, in the formal developments presented here, of an explanation
for the apparent weakness of the coupling of the ${l}=\vert m\vert=2$ $r-$mode
to other stellar modes. 
Strong coupling requires either near-resonance or a large coupling
coefficient, or both.
Although the rotational modes have frequencies comparable to the $r$-mode,
we have found that the possible couplings to such modes may be
limited.  For example, in zero-buoyancy stars 
parity arguments prevent the coupling of three
$r-$modes, and the couplings involving two $r$ modes and one hybrid rotational
mode vanish to zeroth order in the stellar angular velocity.
In nonzero-buoyancy stars, the coupling between $r$-modes is again
vanishing to zeroth order in $\Omega$, but is nonzero when correction
terms of order $O(\Omega^2)$ are included \cite{sharonprivate}.
Thus, the couplings among $r-$modes are small, generically.
The coupling of two $r-$modes to a $f-$mode need not be small.
However, since the eigenfrequencies of $f-$modes are $\sim(GM/R^3)^{1/2}$,
which is much larger than $\Omega$ for a slowly rotating star, the excitation
of an $f-$mode from small amplitude by an unstable $r-$mode is likely to
be suppressed, by factors $\sim\Omega^2R^3/GM$.

Finally, as explained in the introduction, the possibility of the
$r$-modes saturating at small amplitude is not necessarily
incompatible with the numerical simulations to date, or with 
the apparent weakness of the nonlinear couplings of
the $r$-modes.  If the modes do saturate at small or moderate
amplitude, then the formalism developed in this paper may be
sufficient to explore the saturation process. 
On the other hand, it is possible that the formalism developed here
may have to be developed further in order to understand the saturation.
Some possibilities are

\begin{itemize}

\item It may be that the $r$-modes stop growing only in the very
nonlinear regime, where shocks develop, as in the simulations of
Lindblom, Tohline and Vallisneri \cite{astro-ph0010653}.  If this is
the case then the perturbative formalism developed here will not be
useful.  However, as explained in the introduction, the simulations to
date do not show that the strongly nonlinear regime is reached. 

\item It may be that an analysis with only three-mode couplings will
be insufficient, as cubic potentials generically have
instabilities at large amplitude \cite{Thesis:Wu}, but that including
four-mode couplings will be sufficient to allow an exploration of the
saturation process.

\item This paper has only studied the coupling coefficients to zeroth
order in the stellar angular velocity.  However, it may be that the
dominant energy-transfer channels involve coupling coefficients that
are nonzero only at $O(\Omega^2)$.  

\item We have focused attention in this paper on zero-buoyancy stars,
for simplicity.  However the quantitative details of the saturation
process in real neutron stars may be altered by the presence of
buoyancy forces.

\item Finally, we have used Newtonian gravity throughout.  It is
conceivable that important energy transfer channels could come about
via coupling coefficients which vanish at Newtonian order (to zeroth
order in $\Omega$), but which are non-vanishing when post-1-Newtonian
corrections are included.  This will depend on the relative sizes of
the two dimensionless parameters $G M/(c^2 R)$ and $R^3 \Omega^2 / (G M)$.  
The fact that the gravitational radiation reaction force on the
$l=m=2$ $r$-mode is dominated not by the ``Newtonian'' quadrupole
coupling but instead by the ``post-Newtonian'' gravitomagnetic
coupling is a hint in this direction.

\end{itemize}

\begin{acknowledgements}
This work was supported in part by NSF grants
PHY-9900672 and PHY-9722189 and NASA grants NAG5-7264 
and NAG5-8356 to Cornell University.  PA received support from the
Jeffrey L. Bishop Fellowship and EF from the Alfred P. Sloan foundation. 
PA wishes to thank Yanqin Wu and Chris Matzner for many useful conversations.
We thank Larry Kidder, Dong Lai, Sharon Morsink and Mark Scheel for
useful conversations, and Sharon Morsink for detailed and helpful
suggestions on the manuscript.  

\end{acknowledgements}
 

\appendix

\section{Hamiltonian Analysis of Linearized Perturbations}
\label{sec:proofs}
The purpose of this appendix is to derive the mode decomposition formalism
described in Sec.\ \ref{sec:sec2} of the body of the paper.
As discussed in Sec.\ \ref{sec:sec2}, most of that
formalism is contained in a series of papers by Schutz and
collaborators
\cite{1979ApJ...232..874S,1979RSLPS.368..389D,1980MNRAS.190....7S,1980MNRAS.190....21S}.
Schutz introduces the phase space mode expansion, shows 
that the set of vectors (\ref{basis2}) form a basis for ${\cal H}
\times {\cal H}$, and gives an extensive discussion of Jordan chains.  
The main new feature that we introduce is the explicit computation of
left eigenvectors [defined in Eq.\ (\ref{leftemode}) below] in terms of right
eigenvectors, and the subsequent derivation of the equations of motion in the
explicit form (\ref{fa}) above.  
For completeness, we sketch in this Appendix derivations of all the
building blocks of the formalism.  Secs.\ \ref{sec:phase} --
\ref{sec:Specialization-to-no} detail the formalism for non-Jordan
chain modes.  However, evolving the differential rotation and/or total
spin of the star requires an extension of the formalism to compute the
equations of motion for Jordan chain modes.  That extension is given
in Sec.\ \ref{sec:generalcase}. 

\subsection{Phase space equations of motion}
\label{sec:phase}

We start by reformulating the equation of motion (\ref{basic1}) as a
pair of first order equations rather than a single second order
equation.  Our treatment closely follows \cite{1979RSLPS.368..389D},
except that we use canonically conjugate variables instead of $\xivec$
and ${\dot \xivec}$.  

The equation of motion (\ref{basic1}) can be derived from the
Lagrangian density
\beq
{\cal L} = {1 \over 2} {\dot \xivec} \cdot {\dot \xivec} + 
{1 \over 2} {\dot \xivec} \cdot {\bf B} \cdot \xivec 
- {1 \over 2} \xivec \cdot {\bf C} \cdot \xivec  + {\bf a}_{\rm ext}(t) \cdot
\xivec.
\end{equation}
The momentum canonically conjugate to $\xivec$ is
\beq
{\bfpi} = {\partial {\cal L} \over \partial \dot \xivec} = {\dot \xivec}
+ {1 \over 2} {\bf B} \cdot \xivec,
\label{pidef}
\end{equation}
and the associated Hamiltonian density is
\beq
{\cal H} = { 1 \over 2} \left( {\bfpi} - {1 \over 2} {\bf B} \cdot
\xivec \right)^2 
+ {1 \over 2} \xivec \cdot {\bf C} \cdot \xivec  - {\bf a}_{\rm ext}(t) \cdot
\xivec.
\label{eq:hamiltoniandensity}
\end{equation}
The Hamiltonian equations of motion can be written as
\begin{eqnarray}
{\dot {\bfzeta}} = {\bf T} \cdot \bfzeta + {\bf F}(t), 
\label{dynamical}
\end{eqnarray}
where 
\begin{eqnarray}
\bfzeta(t,{\bf x}) \equiv \left[ \begin{array}{c}  
	\bfxi(t,{\bf x})\\
	\bfpi(t,{\bf x}) 
	\end{array} \right],
\label{zetadef}
\end{eqnarray}   
the operator ${\bf T}$ is
\begin{eqnarray}
{\bf T} = \left[ \begin{array}{cc}  
	- {1 \over 2} {\bf B} & {\bf 1} \\
	- {\bf C} + {1 \over 4} {\bf B}^2 & - {1 \over 2} {\bf B}
	\end{array} \right],
\label{Tdef}
\end{eqnarray}
and where 
\begin{eqnarray}
{\bf F}(t) = \left[ \begin{array}{c}  
	0 \\
	{\bf a}_{\rm ext}(t)
	\end{array} \right].
\label{Fdef}
\end{eqnarray}   
If we now specialize to the case of no forcing term, ${\bf a}_{\rm ext}(t)=0$,
and assume a solution of the form  
\begin{eqnarray}
\bfzeta(t,{\bf x}) = e^{-i \omega t} \bfzeta({\bf x}),
\end{eqnarray}
then we get from Eq.\ (\ref{dynamical}) the eigenvalue equation 
\begin{eqnarray}
\left[ {\bf T} +  i \omega \right] \cdot \bfzeta({\bf x}) =0.
\label{rtevec2}
\end{eqnarray}
It is straightforward to show using Eqs.\ (\ref{pidef}) and
(\ref{zetadef}) that Eq.\ (\ref{dynamical}) is 
equivalent to the configuration space equation of motion
(\ref{basic1}), and that the eigenvalue equation (\ref{rtevec2}) is
equivalent to the quadratic eigenvalue equation (\ref{basic1}).

\subsection{Right and left eigenvectors and Jordan chains}
\label{sec:jordan}

Label the distinct right eigenvectors of ${\bf T}$ as
$\bfzeta_A$, and the associated eigenfrequencies as $\omega_A$, so
that 
\begin{eqnarray}
\left[{\bf T} + i  \omega_A \right] \cdot \bfzeta_A = 0.
\label{rtevec}
\end{eqnarray}
Since the operator ${\bf T}$ is not Hermitian, we
will have also left eigenvectors $\bfchi_A$ (distinct from the right
eigenvectors) that satisfy 
\begin{eqnarray}
\left[ {\bf T}^\dagger -i \omega_A^* \right] \cdot \bfchi_A =0.
\label{leftevec}
\end{eqnarray}
Here 
\begin{eqnarray}
{\bf T}^\dagger = \left[ \begin{array}{cc}  
	 {1 \over 2} {\bf B} &\;\; - {\bf C} + {1 \over 4} {\bf B}^2 \\
	{\bf 1} &  {1 \over 2} {\bf B}
	\end{array} \right],
\label{Tdaggerdef}
\end{eqnarray}
is the Hermitian conjugate of ${\bf T}$, since ${\bf C}^\dagger = {\bf
C}$ and ${\bf B}^\dagger = - {\bf B}$.

Since the operator ${\bf T}$ is not Hermitian, the set of its
right eigenvectors will not in general be a complete basis.  However,
one can obtain a complete basis if one includes all Jordan chains 
\cite{1980MNRAS.190....7S}.  What this means is as follows.   
For a given eigenvalue $- i \omega_A$, let ${\cal V}_A$ be the
subspace of ${\cal H}_2 \equiv {\cal H} \oplus {\cal H}$ consisting of
vectors $\bfzeta$ 
that satisfy 
\begin{eqnarray}
\left[ {\bf T} + i \omega_A \right]^m \cdot \bfzeta =0
\end{eqnarray}
for some integer $m \ge 1$.  Clearly the space ${\cal V}_A$
contains all the right eigenvectors associated with $\omega_A$.  
Now it can be shown that the direct sum of all the subspaces ${\cal
V}_A$ (one for each distinct eigenvalue) gives the entire space
${\cal H}_2$.  Hence, to obtain a basis for ${\cal H}_2$, it suffices 
to find a basis for each space ${\cal V}_A$.  One can always find such
a basis of the form $\{ \bfzeta_{A,\sigma} \}$ where $0 \le \sigma \le
p_A$, $\bfzeta_{A,0} = \bfzeta_A$ is the right eigenvector,
$\bfzeta_{A,1}, \ldots, \bfzeta_{A,p_A}$ are a set vectors in ${\cal
V}_A$ that satisfy
\begin{eqnarray}
\left[ {\bf T} + i \omega_A \right] \cdot \bfzeta_{A,\sigma} =
\bfzeta_{A,\sigma-1},
\label{rtjordan}
\end{eqnarray}
for $1 \le \sigma \le p_A$, and that form the Jordan chain of length
$p_A$ associated with $\bfzeta_A$.  For each eigenvector
$\bfzeta_A$, either $p_A=0$ and there is no associated Jordan chain,
or there is a chain of length $p_A \ge 1$.
The set $\left\{ \bfzeta_{A,\sigma} \right| \ A = 1,2,3,
\ldots, \ 0 \le \sigma \le p_A \}$ of Jordan chains, including
the right eigenvectors, forms a basis of ${\cal H}_2$.

These Jordan chains are right Jordan chains.  
We also have for each $A$ a left eigenvector
$\bfchi_A = \bfchi_{A,0}$ as discussed above, 
and left Jordan chains consisting of vectors $\bfchi_{A,\sigma}$ for $1 \le
\sigma \le p_A$ satisfying
\begin{eqnarray}
\left[ {\bf T}^\dagger -i \omega_A^* \right] 
\cdot \bfchi_{A,\sigma} = \bfchi_{A,\sigma-1},
\label{leftjordan}
\end{eqnarray}
for $1 \le A \le p_A$.  Note that the length $p_A$ of the left Jordan
chain must be the same as that of the right Jordan chain, for each
$A$.  The basis $\{ \bfchi_{A,\sigma} \}$ of left Jordan chains
can be chosen to be dual to the basis of right Jordan chains in the
sense that
\begin{eqnarray}
\left< \bfchi_{A,\sigma} \, , \,  \bfzeta_{B,\lambda}  \right> \ = \
\delta_{AB} \, \delta_{p_A \, , \, \sigma + \lambda},
\label{orthog1}
\end{eqnarray}
for all $A,B$ and for $0 \le \sigma \le p_A$ and $0 \le \lambda \le
p_B$ \cite{1980MNRAS.190....7S}.  The inner product here is defined
in the obvious way as
\beq
\left< \left[ \xivec,\bfpi \right] \, , \, 
\left[ \xivec^\prime,\bfpi^\prime \right] \right> \equiv
\left< \xivec \, , \, \xivec^\prime \right> +
\left< \bfpi \, , \, \bfpi^\prime \right>,
\endeq
where the inner products on the right hand side are given by the
definition (\ref{innerproduct}).

\subsection{General mode expansion and equations of motion for mode
coefficients} 

We expand $\bfzeta(t,{\bf x})$ on the basis of right Jordan chains as
\begin{eqnarray}
\bfzeta(t,{\bf x}) = \sum_A \, \sum_{\sigma=0}^{p_A} \, c_{A,\sigma}(t)
\bfzeta_{A  ,\sigma}({\bf x}),
\label{zetaexpand}
\end{eqnarray}
where from the orthogonality relation (\ref{orthog1}) we have
\be
c_{A,\sigma} = \left< \bfchi_{A,p_A-\sigma} \, , \, \bfzeta \right>.
\label{cAsigmaeqn}
\ee
Substituting the expansion (\ref{zetaexpand}) into the dynamical equation
(\ref{dynamical}), and using the defining properties (\ref{rtevec}) and
(\ref{rtjordan}) of right Jordan chains yields
\begin{eqnarray}
\sum_A \, \sum_{\sigma=0}^{p_A} \, 
{\dot c}_{A,\sigma}(t) \, \bfzeta_{A,\sigma} \ &=& \ \sum_A \,
\sum_{\sigma=0}^{p_A} \,  
c_{A,\sigma}(t) \left[ - i \omega_A \bfzeta_{A,\sigma} +
\bfzeta_{A,\sigma-1} \right]
+ {\bf F}(t),
\end{eqnarray}
where we define $\bfzeta_{A,-1} \equiv 0$.  Now multiplying on the
left by $\bfchi_{B,\lambda}$, using the orthogonality relation
(\ref{orthog1}), and then relabeling the indices yields
the equations
\begin{eqnarray}
{\dot c}_{A,\sigma} + i  \omega_A c_{A,\sigma} - c_{A,\sigma+1} = \left<
\bfchi_{A,p_A-\sigma} \, , \, {\bf F}(t) \right>,
\label{ee3}
\end{eqnarray}
for $0 \le \sigma \le p_A -1$, and
\begin{eqnarray}
{\dot c}_{A,p_A} +  i \omega_A c_{A,p_A} = \left<
\bfchi_{A,0} \, , \, {\bf F}(t) \right>.
\label{ee4}
\end{eqnarray}
Thus, the equations for the mode coefficients $c_{A,0}(t)$,
$c_{A,1}(t)$, 
$\ldots$, $c_{A,p_A}(t)$ are coupled together for each fixed
$A$, but 
the different $A$'s are uncoupled.  The general solution of Eqs.\
(\ref{ee3}) and (\ref{ee4}) in the case of no forcing terms is
a polynomial in time multiplied by the usual complex exponential:
\begin{eqnarray}
c_{A,\sigma}(t) = e^{- i \omega_A t} \, \sum_{\lambda=0}^{p_A-\sigma} \,
\gamma_{A,\sigma+\lambda} 
\, t^\lambda,
\label{gensoln}
\end{eqnarray}
where $\gamma_{A,0}, \gamma_{A,1}, \ldots,
\gamma_{A,p_A}$ are constants of integration.

\subsection{Specialization to no Jordan chains}
\label{sec:Specialization-to-no}

We now specialize to the situation where
there are no Jordan chains.  In this case the orthonormality relation
(\ref{orthog1}) reduces to 
\begin{eqnarray}
\left< \bfchi_{A} \, , \,  \bfzeta_{B}  \right> \ = \
\delta_{AB},
\label{orthog2}
\end{eqnarray}
and the mode expansion (\ref{zetaexpand}) becomes 
\begin{eqnarray}
\bfzeta(t,{\bf x}) = \sum_A  \, c_{A}(t)
\bfzeta_{A}({\bf x}).
\label{zetaexpand1}
\end{eqnarray}
The inverse of this mode expansion is
\beq
c_A(t) = \left< \bfchi_A \, , \, \bfzeta(t) \right>,
\label{inverseexpansion4}
\end{equation}
and the equation of motions (\ref{ee3}) and (\ref{ee4}) for the mode
coefficients reduce to 
\begin{eqnarray}
{\dot c}_{A} +  i \omega_A c_{A} = \left<
\bfchi_{A} \, , \, {\bf F}(t) \right>.
\label{ee5}
\end{eqnarray}

\subsubsection{Translation of results to configuration space variables}

Our goal now is to write these results for no Jordan chains entirely
in terms of the 
Lagrangian displacement $\xivec$ and its time derivative ${\dot
\xivec}$, and the modes $(\xivec_A,\omega_A)$ of the configuration space
formalism.  To do this we proceed as follows.
We write the right eigenvector $\bfzeta_A$ as
\begin{eqnarray}
\bfzeta_{A} = \left[ \begin{array}{c}  
	\bfxi_{A} \\
	\bfpi_{A}
	\end{array} \right],
\label{jk1a}
\end{eqnarray}   
and similarly write the left eigenvector $\bfchi_A$ as
\begin{eqnarray}
\bfchi_{A} = \left[ \begin{array}{c}  
	\bfsigma_{A} \\
	\bftau_{A}
	\end{array} \right].
\label{jk2a}
\end{eqnarray}   
Then the definitions (\ref{rtevec})
--(\ref{leftevec}) of right and left eigenvectors together
with the formula (\ref{Tdef}) for the operator ${\bf T}$ shows that
$\xivec_A$ and $\bftau_A$ are right and left eigenvectors of the
operator ${\bf L}(\omega_A) = - \omega_A^2 -i \omega_A {\bf B} + {\bf
C}$ [cf.\ Eqs.\ (\ref{ldef}) and (\ref{sdf}) above]
\begin{eqnarray}
{\bf L}(\omega_A) \cdot \xivec_A &=&0  \\
{\bf L}(\omega_A)^\dagger \cdot \bftau_A &=&0.
\label{leftemode}
\end{eqnarray}
In addition we obtain the formulae 
\be
\bfpi_A = - i \omega_A \xivec_A + {\bf B} \cdot \xivec_A/2
\label{lid1}
\ee
and
\be
\bfsigma_A = i \omega_A^* \bftau_A - {\bf B} \cdot \bftau_A /2,
\label{lid2}
\ee
which using Eqs.\ (\ref{jk1}) and (\ref{jk2})
allow us to write the phase space right 
and left eigenvectors $\bfzeta_A$ and $\bfchi_A$ entirely in terms of
the configuration space left and right eigenvectors $\xivec_A$ and
$\bftau_A$:
\begin{eqnarray}
\bfzeta_{A} = \left[ \begin{array}{c}  
	\bfxi_{A} \\
	- i \omega_A \xivec_A + {1 \over 2} {\bf B} \cdot \xivec_A
	\end{array} \right]
\label{jk1}
\end{eqnarray}   
and
\begin{eqnarray}
\bfchi_{A} = \left[ \begin{array}{c}  
	i \omega_A^* \bftau_A - {1 \over 2}{\bf B} \cdot \bftau_A   \\
	\bftau_{A}
	\end{array} \right].
\label{jk2}
\end{eqnarray}

Now using the definitions (\ref{pidef}) and (\ref{zetadef}) and the
formula (\ref{jk1}) we can rewrite the mode expansion
(\ref{zetaexpand1}) as 
\begin{eqnarray}
 \left[ \begin{array}{c}  
	\xivec(t)\\
	{\dot \xivec}(t) + {1 \over 2} {\bf B} \cdot \xivec(t)
	\end{array} \right] = 
\sum_A c_A(t) \left[ \begin{array}{c}  
	\xivec_A\\
	- i \omega_A \xivec_A + {1 \over 2} {\bf B} \cdot \xivec_A
	\end{array} \right],
\label{goodexpansion3}
\end{eqnarray}   
which is equivalent to the expansion (\ref{goodexpansion}) quoted in
the body of the paper.  
Note that there is a one-to-one correspondence between
right eigenvectors solutions $\bfzeta_A$ of Eq.\ (\ref{rtevec})
and solutions $(\xivec_A,\omega_A)$ of Eq.\ (\ref{basic1}), which
allows us to identify the sums over $A$ that appear in the
expansions (\ref{goodexpansion}) and (\ref{goodexpansion3}).

Next, we can use the formulae (\ref{jk1}) and (\ref{jk2}) to rewrite
the orthogonality relation (\ref{orthog2}), the inverse mode expansion
(\ref{inverseexpansion4}) and the equation of motion (\ref{ee5}).  The
results are
\beq
\delta_{AB} = \left< \bftau_A \, , \, {\bf B} \cdot \xivec_B \right> 
- i (\omega_A + \omega_B) \left< \bftau_A \, , \, \xivec_B \right>,
\label{orthog6}
\end{equation}
\beq
c_A(t) = - i \left< \bftau_A \, , \, \omega_A \xivec(t) + i {\bf B}
\cdot \xivec(t) + i {\dot \xivec}(t) \right>.
\label{inverseexpansion7}
\end{equation}
and 
\begin{eqnarray}
{\dot c}_{A} +  i \omega_A c_{A} = \left<
\bftau_{A} \, , \, {\bf a}_{\rm ext}(t) \right>,
\label{ee6}
\end{eqnarray}
where we have also used Eqs.~(\ref{eq:antiHermitian}), (\ref{pidef}),
(\ref{zetadef}) and (\ref{Fdef}).  

\subsubsection{Computation of the left eigenmodes}
\label{sec:explicit}

The last step in the construction is the explicit computation of the
left eigenvectors $\bftau_A$ in terms of the right eigenvectors
$\xivec_A$.  
For a given eigenfrequency $\omega$, let ${\cal V}_R(\omega)$ denote
the space of associated right eigenvectors $\xivec$
satisfying ${\bf L}(\omega) \cdot \xivec=0$
[cf.\ Eq.\ (\ref{ldef})], and let ${\cal V}_L(\omega)$ denote
the space of associated left eigenvectors $\bftau$ satisfying ${\bf
L}(\omega)^\dagger \cdot \bftau=0$.  The spaces ${\cal V}_R(\omega_A)$
and ${\cal V}_L(\omega_A)$ have the same dimension, which is the
degeneracy associated with the eigenfrequency $\omega_A$.  However,
${\cal V}_R(\omega_A)$ and ${\cal V}_L(\omega_A)$ will not in general
coincide for complex $\omega_A$.  
We now show that for real $\omega_A$, the spaces ${\cal
V}_L(\omega_A)$ and ${\cal V}_R(\omega_A)$ do coincide, which means
that the left eigenvectors associated with a given eigenfrequency can
be expressed as linear combinations of the corresponding right
eigenvectors.  This is the key result that we use to compute the left
eigenvectors. 

We shall need the following properties of the spaces ${\cal
V}_R(\omega)$ and ${\cal V}_L(\omega)$ of right and left eigenvectors
discussed by Schutz \cite{1980MNRAS.190....7S}.
Suppose that $\xivec$ lies in ${\cal V}_R(\omega)$.  Then

\begin{itemize}
\item The function $\bfxi$ also lies in ${\cal V}_L(\omega^*)$, ie, is
a left eigenvector with eigenfrequency
$\omega^*$. This can be derived by taking the Hermitian conjugate of
Eq.~(\ref{basic1}), and by using the fact that ${\bf B}$ is anti-Hermitian and
${\bf C}$ is Hermitian, which yields ${\bf L}(\omega^*)^\dagger \cdot
\xivec=0$.  

\item The function $\bfxi({\bf x})^*$ lies in ${\cal V}_R(-\omega^*)$.
This follows directly from the fact that the equation of motion
(\ref{basic0}) is real. 

\item Let $g$ be the mapping that takes $(r,\theta,\varphi)$ to
$(r,\theta,-\varphi)$ in spherical polar coordinates, or equivalently
$(x,y,z)  \to (x,-y,z)$ in Cartesian coordinates, i.e., reflection in 
the $xz$ plane.  We define $g_* \bfxi$ be the
pullback of $\bfxi$ under $g$ \cite{bishopgoldberg}, so that if
\begin{eqnarray}
\bfxi = \xi^r(r,\theta,\varphi) {\partial \over \partial r} +
\xi^\theta(r,\theta,\varphi) {\partial \over \partial \theta} +
\xi^\varphi(r,\theta,\varphi) {\partial \over \partial \varphi},
\end{eqnarray}
then
\begin{eqnarray}
g_* \bfxi &=& \xi^r(r,\theta,-\varphi) {\partial \over \partial r} + 
\xi^\theta(r,\theta,-\varphi) {\partial \over \partial \theta}
- \xi^\varphi(r,\theta,-\varphi) {\partial \over \partial
\varphi}. 
\end{eqnarray}
Since the background star is invariant under the transformation
$\varphi \to -\varphi$, $t \to - t$, it follows that $g_* \bfxi$ lies
in ${\cal V}_R(-\omega)$ \cite{1980MNRAS.190....7S}.

\item By combining the previous properties, it follows that $(g_*
\bfxi)^*$ belongs to both ${\cal V}_R(\omega^*)$ and ${\cal V}_L(\omega)$.
The mapping $\xivec \to (g_* \xivec)^*$ was first written down by
Schutz in Ref.\ \cite{1979ApJ...232..874S} where it was denoted $S$.

\end{itemize}

We next recall the notation used in Sec.\ \ref{sec:degeneracy} above.
We write the distinct eigenfrequencies as $\omega_a$, and the
right eigenvectors as $\xivec_A = \xivec_{a,k}$ where $1 \le k \le
n_a$ and $n_a$ is the degeneracy associated with the eigenfrequency
$\omega_a$.   

Consider first the case of complex $\omega_a$.  In this case it
follows from the above that the mapping $\xivec \to (g_* \xivec)^*$
maps ${\cal V}_R(\omega_a)$ onto ${\cal V}_L(\omega_a)$
\cite{1980MNRAS.190....7S}.  Hence we can write
\beq
\bftau_{a,k} = \sum_{k^\prime=1}^{n_a} \, T_{kk^\prime} \, (g_*
\xivec_{a,k^\prime} )^*,
\label{lc1}
\end{equation}
for some $n_a \times n_a$ matrix $T_{kk^\prime}$.  By substituting
Eq.\ (\ref{lc1}) into the orthogonality relation (\ref{orthog6}) we
can solve for the matrix $T_{kk^\prime}$ and thus obtain the left
eigenvectors $\bftau_{a,k}$.  We note that Schutz uses a different
basis of left eigenmodes for which $\bftau_{a,k} = (g_*
\xivec_{a,k})^*$ [Eq.\ (3.14) of \cite{1980MNRAS.190....7S}], but for
which the orthonormality condition (\ref{orthog6}) does not hold.
This method of obtaining the left eigenmodes for complex
frequencies is discussed in more detail in Sec.\ \ref{sec:generalcase}
below.    

For real frequencies, the situation is much simpler.  Every $\xivec$
in ${\cal V}_R(\omega)$ also lies in ${\cal V}_L(\omega^*) = {\cal
V}_L(\omega)$, and hence the spaces ${\cal V}_R(\omega)$ and ${\cal
V}_L(\omega)$ coincide \footnote{For a mode $(\xivec_A,\omega_A)$ with
no degeneracy, it follows from this that we can choose the phase of
$\xivec_A$ so that $(g_* \xivec_A)^* = \xivec_A$.  Hence, if $m$ is
the azimuthal quantum number and we write $\xivec_A = \exp[i m
\varphi] {\hat \xivec}_A$, the $r$
and $\theta$ components of ${\hat \xivec}_A$ are purely real and the
$\varphi$ component is purely imaginary.}.  Hence we can write
\beq
\bftau_{a,k} = \sum_{k^\prime=1}^{n_a} \, - i {\cal B}^{(a)\,*}_{kk^\prime} \, 
\xivec_{a,k^\prime},
\label{lc2}
\end{equation}
for some $n_a \times n_a$ matrix ${\cal B}^{(a)}_{kk^\prime}$.  (The factor
of $-i$ and the complex conjugation are included for convenience.)  By
substituting 
Eq.\ (\ref{lc2}) with $\bftau_A = \bftau_{a,k}$ and $\xivec_B =
\xivec_{b,l}$ into the orthogonality relation (\ref{orthog6}) we
obtain
\beq
\delta_{ab} \, \delta_{kl} = {\cal B}^{(a)}_{kk^\prime} {\cal
M}_{ak^\prime,bl}, 
\label{Bformula}
\end{equation}
where the matrix ${\cal M}_{ak^\prime,bl}$ is defined in
Eq.~(\ref{calMdef}).  It follows from Eq.\ (\ref{Bformula}) that, first,
the matrix ${\cal M}_{ak,bl}$ is block diagonal in the sense claimed
in Sec.\ \ref{sec:degeneracy}, and second, for a given value of
$a$ the matrix ${\cal B}_{kk^\prime}^{(a)}$ is just the inverse of the
the diagonal block ${\cal M}_{ak,al}$.

Now as explained in Sec.\ \ref{sec:degeneracy}, we can
always choose the basis $\xivec_{a,k}$ of ${\cal V}_R(\omega_a)$ to
diagonalize the matrix ${\cal M}_{ak^\prime,al}$.  In particular, this
will be automatically true when $n_a=1$ and there is no degeneracy.
For such bases, the matrix $B_{kk^\prime}$ will also be diagonal.  It
then follows from Eq.\ (\ref{lc2}) that each 
left eigenvector can be written as
\beq
\bftau_A = - {i \over b_A} \xivec_A
\label{taufinalans}
\end{equation}
for some constant $b_A$.  Substituting this into the orthonormality
relation (\ref{orthog6}) yields the formula (\ref{normalizationdef})
for $b_A$.  Finally, the formula (\ref{taufinalans}) for left
eigenvectors can be combined with the orthogonality relation
(\ref{orthog6}), the inverse mode expansion (\ref{inverseexpansion7})
and equation of motion (\ref{ee6}) to yield the versions 
(\ref{orthonormal1}), (\ref{inverseexpansion}) and (\ref{fa}) of these
relations quoted in the body of the paper
\footnote{The validity of the orthogonality relation
(\ref{orthonormal1}), for those 
values of $A$ and $B$ for which $\omega_A \ne \omega_B$, can also be
derived directly by contracting Eq.\ (\ref{sdf}) on the left with
$\xivec_B$, subtracting from this the same equation complex
conjugated with $A$ and $B$ interchanged, and dividing by $\omega_A -
\omega_B$.}.

Finally, we note that Eq.\ (\ref{Bformula}) shows that the matrix
${\cal M}_{ak,al}$ is non-degenerate.  Thus, if that matrix is
degenerate, then the assumption underlying the derivation of
Eq.~(\ref{Bformula}) --- that there is no Jordan chain associated with the
eigenfrequency $\omega_a$ --- must fail.  In particular, for
non-degenerate modes, if the constant $b_A$ defined by
Eq.~(\ref{normalizationdef}) vanishes, then the mode must be a Jordan chain
mode.  

\subsection{Computation of the left eigenmodes for Jordan chain modes}
\label{sec:generalcase}

In this subsection we show how to generalize the above analysis to obtain
the left eigenvectors $\bfchi_{A,\sigma}$ for Jordan chain modes.  
The construction is useful, for example, in computing the evolution of
stellar differential rotation, which is described by an infinite set
of Jordan chains of length 1 (see Appendix \ref{sec:proofs3}).

The essential idea is to use the phase space variables rather than the
more complicated configuration space variables, and to use an operator
which maps right eigenvectors onto left eigenvectors.  We define the
operator
\begin{eqnarray}
\bfM = i \left[ \begin{array}{cc}  
	0 & {\bf 1} \\
	- {\bf 1} & 0
	\end{array} \right],
\label{Mdef}
\end{eqnarray}
which is essentially the symplectic structure defined by Friedman and
Schutz \cite{1978ApJ...221..937F}.  It satisfies
$\bfM^\dagger = \bfM = {\bf M}^{-1}$ and 
\be
{\bf M}^\dagger \cdot {\bf T} \cdot {\bf M} = - {\bf T}^\dagger.
\label{Ttransform}
\ee
It follows from Eqs.\ (\ref{Ttransform}), (\ref{rtevec}) and
(\ref{leftevec}) that if $\bfzeta$ is a right 
eigenvector of ${\bf T}$ with eigenfrequency $\omega$, then ${\bf M}
\cdot \bfzeta$ is a left eigenvector with eigenfrequency $\omega^*$.  

Now the set of vectors ${\bf M} \cdot \bfzeta_{A,\sigma}$ forms a
basis, since ${\bf M}$ is invertible and the set of vectors
$\bfzeta_{A,\sigma}$ is a basis.
Therefore we can write the left eigenvectors and left Jordan chain vectors
as
\begin{equation}
\bfchi_{A,\sigma} = \sum_{B,\lambda} {\cal B}_{A\sigma,B\lambda}^* \, \bfM
\cdot \bfzeta_{B,\lambda},
\label{le_ansatz}
\end{equation}
for some matrix ${\cal B}_{A\sigma,B\tau}$, where the complex
conjugation is included for later convenience.  Inserting this into
the orthogonality relation (\ref{orthog1}) gives
\begin{equation}
{\cal B}_{A\sigma,C\tau} \, {\cal M}_{C\tau,B\lambda} = \delta_{AB}
\, \delta_{\sigma+\lambda,p_A},
\label{calBgendef}
\end{equation}
where 
\be
{\cal M}_{C\tau,B\lambda} = \left< \bfzeta_{C,\tau} \, , \, \bfM
\cdot \bfzeta_{B,\lambda} \right>
\label{calMdefgeneral0}
\ee
are the matrix elements of the operator $\bfM$.  Equation
(\ref{calBgendef}) says that the matrices ${\cal B}$ and ${\cal M}$
are inverses of each other (up to the index permutation $\sigma \to
p_A - \sigma$).

The reason the ansatz (\ref{le_ansatz}) is useful is that the matrix
${\cal M}$ is almost diagonal, in the sense that it satisfies the
identity
\be
( \omega_A^* - \omega_B ) {\cal M}_{A\sigma,B\lambda} = 0.
\label{calMidentity}
\ee
It follows from Eq.\ (\ref{calMidentity}) that ${\cal M}$ is block
diagonal, with one block for 
each real frequency and one block for each pair $(\omega, \omega^*)$ of
complex frequencies.  Therefore, we can obtain the left eigenvectors
by inverting each diagonal block of ${\cal M}_{A\sigma,B\lambda}$ to
obtain ${\cal B}_{A\sigma,B\lambda}$, and by using Eq.\ (\ref{le_ansatz}).

To derive the identity (\ref{calMidentity}) we 
use Eqs.\ (\ref{Ttransform}) and (\ref{rtjordan}) to show that the
vectors $\bfM \cdot \bfzeta_{A,\sigma}$ form a left Jordan chain with
frequency $\omega_A^*$:
\be
\left[ {\bf T}^\dagger - i \omega_A \right] \cdot ( \bfM \cdot
\bfzeta_{A,\sigma} ) = - \bfM \cdot \bfzeta_{A,\sigma-1}.
\label{ljcid}
\ee
Next, we have
\begin{eqnarray}
\omega_A^* \, {\cal M}_{A\sigma,B\lambda} &=& \omega_A^* \, \left<
\bfzeta_{A,\sigma} \ , \ \bfM \cdot \bfzeta_{B,\lambda} \right>
\nonumber \\
\mbox{} &=& \left< \omega_A \bfM \cdot \bfzeta_{A,\sigma} \ , \ 
\bfzeta_{B,\lambda} \right> \nonumber \\
\mbox{} &=& \left< -i \bfM \cdot \bfzeta_{A,\sigma-1} 
-i {\bf T}^\dagger \cdot \bfM \cdot \bfzeta_{A,\sigma} \ , \ 
\bfzeta_{B,\lambda} \right> \nonumber \\
\mbox{} &=& i \left< \bfzeta_{A,\sigma-1}  \ , \ 
\bfM \cdot \bfzeta_{B,\lambda} \right> 
+ i \left< \bfM \cdot \bfzeta_{A,\sigma}  \ , \ 
{\bf T} \cdot \bfzeta_{B,\lambda} \right> 
\nonumber \\
\mbox{} &=& i \left< \bfzeta_{A,\sigma-1}  \ , \ 
\bfM \cdot \bfzeta_{B,\lambda} \right> 
+ i \left< \bfM \cdot \bfzeta_{A,\sigma}  \ , \ 
- i \omega_B \bfzeta_{B,\lambda} + \bfzeta_{B,\lambda-1} \right>,
\label{calMidentity1}
\end{eqnarray}
where we have used Eqs.\ (\ref{rtjordan}) and (\ref{ljcid}).
The result (\ref{calMidentity1}) can be written as
\be
( \omega_A^* - \omega_B) {\cal M}_{A\sigma,B\lambda} = i \left( {\cal
M}_{A(\sigma-1),B\lambda} + {\cal M}_{A\sigma,B(\lambda-1)}\right),
\label{calMidentity2}
\ee
and the identity (\ref{calMidentity}) follows from iterating the
identity (\ref{calMidentity2}).

Schutz \cite{1980MNRAS.190....7S} suggested a different general method
of constructing left Jordan chains.  That method is based on the
anti-linear operator 
${\bf S}_2$ defined by
\begin{eqnarray}
{\bf S}_2 \cdot \left[ \begin{array}{c}  
	\xivec\\
	\bfpi
	\end{array} \right] =
\left[ \begin{array}{c}  
	(g_* \xivec)^*  \\
	- (g_* \bfpi)^*
	\end{array} \right],
\label{S2def}
\end{eqnarray}   
which maps right Jordan chains $\bfzeta_{A,\sigma}$ of ${\bf T}$ with
frequency $\omega_A$ onto right Jordan chains ${\bf S}_2 \cdot
\bfzeta_{A,\sigma}$ 
with frequency $\omega_A^*$.  Therefore composing this map with the
operator $\bfM$ yields a mapping $\bfzeta \to \bfM \cdot {\bf S}_2
\cdot \bfzeta$ which takes right Jordan chains of frequency $\omega$ to
left Jordan chains of frequency $\omega$.  The approach here omits the
mapping ${\bf S}_2$ and is simpler to use when the eigenfrequencies
are real.

For Jordan chains of length zero (ordinary eigenvectors), the
method (\ref{le_ansatz}) reduces to
the method (\ref{lc2}) of Sec.\ \ref{sec:Specialization-to-no} above.  
Dropping the Jordan chain indices $\sigma$ and $\lambda$, the ansatz
(\ref{le_ansatz}) reduces to $\bfchi_A = \sum_B {\cal B}_{AB}^* \bfM
\cdot \bfzeta_B$ with ${\cal B}_{AC} \, {\cal M}_{CB} = \delta_{AB}$,
which using Eqs.\ (\ref{jk1}), (\ref{jk2}) and 
(\ref{Mdef}) implies 
\be
\bftau_A = - i \sum_B {\cal B}_{AB}^* \xivec_B,
\ee 
cf.\ Eq.\ (\ref{lc2}) above.  Similarly, the definition
(\ref{calMdefgeneral0}) can 
be simplified using Eqs.\ (\ref{jk1}) and (\ref{Mdef}) to give
\begin{eqnarray}
{\cal M}_{AB} &=& \left< \bfzeta_A, \bfM \cdot \bfzeta_B \right>
\nonumber \\
\mbox{} &=& \left< \xivec_A \, , \, i {\bf B} \cdot \xivec_B
\right> + ( \omega_A^* + \omega_B ) \left< \xivec_A \, , \xivec_B \right>,
\label{calMdefgeneral}
\end{eqnarray}
which is a generalization of the definition (\ref{calMdef}) valid for
complex frequencies.  

As an example, consider now the case of a mode $(\xivec_1,\omega_1)$
with complex 
eigenfrequency which is non-degenerate.  Then there is an associated
mode $(\xivec_2,\omega_2)$ with $\omega_2 = \omega_1^*$, and by
choosing the normalization of $\xivec_2$ we can without loss of
generality take
\be
\xivec_2 = (g_* \xivec_1)^*,
\ee
cf.\ Sec.\ \ref{sec:explicit} above.  The corresponding $2 \times 2$
block of the matrix ${\cal M}$ is from Eqs.\ (\ref{calMidentity}) and
(\ref{calMdefgeneral}) of
the form 
\begin{eqnarray}
{\cal M}_{AB} = \left[ \begin{array}{cc}  
	0 & {\cal D} \\
	{\cal D}^* & 0
	\end{array} \right],
\label{M2def}
\end{eqnarray}
where ${\cal D} = {\cal M}_{12} = N[\xivec_1,\omega_1]$, where
the functional $N$ is 
\be
N[\xivec,\omega] \equiv \left< \xivec \, , \, i {\bf B} \cdot (g_*
\xivec)^* \right> + 2 \omega^* \left< \xivec \, , \,  (g_*
\xivec)^* \right>.
\ee
The corresponding left eigenmodes are therefore
\be
\bftau_1 = - {i \over {\cal D}} \, \xivec_2 = - {i \over {\cal D}} \, (g_*
\xivec_1)^*
\ee
and
\be
\bftau_2 = - {i \over {\cal D}^*} \, \xivec_1.
\ee

\subsubsection{Zero frequency Jordan chains of length one}
\label{sec:jordanone}

We now carry through the explicit computation of the left Jordan
chains for the case that arises in practice in stable rotating stars,
that of zero-frequency Jordan chains of length one.
Consider first the case of no degeneracy.  Dropping the index $A$, we
have from Eq.\ (\ref{rtjordan}) two Jordan chain vectors $\bfzeta_0$
and $\bfzeta_1$ which 
satisfy
\be
{\bf T} \cdot \bfzeta_0 = 0, \ \ \ \ {\bf T} \cdot \bfzeta_1 =
\bfzeta_0.
\label{dr}
\ee
The $2 \times 2$ matrix ${\cal M}_{\sigma\tau} = \left<
\bfzeta_\sigma, {\bf M} \cdot \bfzeta_\tau \right>$ satisfies the
identity 
\be
{\cal M}_{(\sigma-1)\tau} + {\cal M}_{\sigma(\tau-1)} =0
\ee
from Eq.\ (\ref{calMidentity2}).  Using this identity together with
the fact that ${\cal M}$ is Hermitian gives
\begin{eqnarray}
{\cal M}_{\sigma\tau} = \left[ \begin{array}{cc}  
	0 & i \beta \\
	-i \beta & \gamma
	\end{array} \right],
\label{Mvalue}
\end{eqnarray}
where $\beta$ and $\gamma$ are real.  Under the transformation
$\bfzeta_1 \to \bfzeta_1 + \chi \bfzeta_0$, which preserves the
defining relations (\ref{dr}), we have 
\be
\beta \to \beta, \ \ \ \ \gamma \to \gamma + 2 \beta {\rm Im} (\chi).
\ee
It follows that we can choose $\bfzeta_1$ to make $\gamma=0$.  Now
combining Eqs.\ (\ref{le_ansatz}), (\ref{calBgendef}) and
(\ref{Mvalue}) gives for the left Jordan chain vectors
\begin{eqnarray}
\label{ans1}
\bfchi_0 &=& {i \over \beta} \, \bfM \cdot \bfzeta_0 \\
\bfchi_1 &=& -{i \over \beta} \, \bfM \cdot \bfzeta_1.
\label{ans2}
\end{eqnarray}

We now write these results in terms of configuration space variables,
using the notation
\begin{eqnarray}
\bfzeta_{\sigma} = \left[ \begin{array}{c}  
	\bfxi_{\sigma} \\
	\bfpi_{\sigma}
	\end{array} \right],
\label{sjk1a}
\end{eqnarray}   
\begin{eqnarray}
\bfchi_{\sigma} = \left[ \begin{array}{c}  
	\bfsigma_{\sigma} \\
	\bftau_{\sigma}
	\end{array} \right],
\label{sjk2a}
\end{eqnarray}   
for $\sigma = 0,1$.  First, the relations (\ref{dr}) can be written as
\be
{\bf C} \cdot \xivec_0 =0, \ \ \ {\bf C} \cdot \xivec_1 = - {\bf B}
\cdot \xivec_0, \ \ \ \bfpi_0 = {\bf B} \cdot \xivec_0/2, \ \ \
\bfpi_1 = \xivec_0 + {\bf B} \cdot \xivec_1/2.
\label{dr1}
\ee
Second, the formula for $\beta$ is, from Eqs.\ (\ref{Mdef}),
(\ref{sjk1a}) and (\ref{dr1})
\begin{eqnarray}
\beta &=& - i \left< \bfzeta_0 \, , \, \bfM \cdot \bfzeta_1 \right>
\nonumber \\
&=& \left< \xivec_0 \, , \, \xivec_0 \right> + \left< \xivec_0 \, , \,
{\bf B} \cdot \xivec_1 \right>.
\end{eqnarray}
Third, the relations (\ref{ans1}) and (\ref{ans2}) together with Eqs.\
(\ref{Mdef}), (\ref{sjk1a}) and (\ref{sjk2a}) imply that
\be
\bftau_0 = {1 \over \beta} \, \xivec_0, \ \ \ \bftau_1 = - {1 \over
\beta} \, \xivec_1.
\ee
The corresponding equations of motion are, from Eqs.\ (\ref{ee3}) and
(\ref{ee4}), 
\begin{eqnarray}
{\dot c}_1 &=& {1 \over \beta} \left< \xivec_0 \, , \, {\bf a}_{\rm ext}
\right> \\
{\dot c}_0 &=& - {1 \over \beta} \left< \xivec_1 \, , \, {\bf a}_{\rm
ext} \right> + c_1.
\end{eqnarray}

We now generalize these results to allow for degeneracy, since the
space of zero-frequency Jordan chain modes in rotating stars is highly
degenerate (see Appendix \ref{sec:proofs3}).  We use the notation 
\be
A = (a,k)
\ee
of Sec.\ \ref{sec:degeneracy} above, where $a$ labels the distinct
eigenfrequencies and $k$ the eigenvectors associated with each
eigenfrequency.  Therefore the Jordan chain vectors can be written as
\be
\bfzeta_{A\sigma} = \bfzeta_{ak\sigma}
\ee
where $\sigma$ labels the Jordan chain vectors associated with each
eigenvector.  For simplicity we drop the label $a$ in what follows.
The defining relations for Jordan chains of length one are
\be
{\bf T} \cdot \bfzeta_{k0} = 0, \ \ \ \ {\bf T} \cdot \bfzeta_{k1} =
\bfzeta_{k0}.
\label{qdr}
\ee
The matrix ${\cal M}_{k\sigma,l\tau} = \left<
\bfzeta_{k\sigma}, {\bf M} \cdot \bfzeta_{l\tau} \right>$ satisfies the
identity 
\be
{\cal M}_{k(\sigma-1),l\tau} + {\cal M}_{k\sigma,l(\tau-1)} =0
\ee
from Eq.\ (\ref{calMidentity2}).  Using this identity together with
the fact that ${\cal M}$ is Hermitian gives
\begin{eqnarray}
\label{late1}
{\cal M}_{k0,l0} &=& 0 \\
{\cal M}_{k0,l1} &=& - {\cal M}_{k1,l0} = i \beta_{kl} \\
{\cal M}_{k1,l1} &=& \gamma_{kl},
\label{late2}
\end{eqnarray}
where the matrices $\beta_{kl}$ and $\gamma_{kl}$ are Hermitian.

Next, the transformation
\begin{eqnarray}
\bfzeta_{k0} &\to& \bfzeta_{k0}^\prime = F_{kl}^* \bfzeta_{l0}
\nonumber \\
\bfzeta_{k1} &\to& \bfzeta_{k1}^\prime = F_{kl}^* \bfzeta_{l1}
\end{eqnarray}
preserves the relations (\ref{qdr}).  Under this transformation the
matrix $\beta_{kl}$ transforms as ${\bfbeta} \to {\bfbeta}^\prime
= {\bf F} \cdot {\bfbeta} \cdot {\bf F}^\dagger$, and it follows
that we can choose the basis $\bfzeta_{k0}$ to diagonalize
$\beta_{kl}$, so that 
\be
\beta_{kl} = \beta_k \delta_{kl}.
\ee
Similarly the transformation
\begin{eqnarray}
\bfzeta_{k0} &\to& \bfzeta_{k0}^\prime = \bfzeta_{k0}
\nonumber \\
\bfzeta_{k1} &\to& \bfzeta_{k1}^\prime = \bfzeta_{k1} + F_{kl}^* \bfzeta_{l0}
\end{eqnarray}
preserves the relations (\ref{qdr}).  Under this transformation the
matrices transform as 
\begin{eqnarray}
\bfbeta &\to& \bfbeta \nonumber \\
\bfgamma &\to& \bfgamma^\prime = \bfgamma + i {\bf F} \cdot \bfbeta - i
{\bfbeta} \cdot {\bf F}^\dagger.
\label{late3}
\end{eqnarray}
Now it follows from Eqs.\ (\ref{late1})--(\ref{late2}) and the fact that the
matrix ${\cal M}_{k\sigma,l\tau}$ is invertible that $\beta_{kl}$ is
invertible.  Therefore we can find a transformation matrix $F_{kl}$
which achieves $\bfgamma^\prime=0$ in Eq.\ (\ref{late3}), namely 
\be
{\bf F} = i \bfgamma \cdot \bfbeta^{-1} /2.
\ee
Therefore we can choose the basis $\bfzeta_{k1}$ to make
$\gamma_{kl}=0$.  The rest of the analysis now proceeds exactly as for
the non-degenerate case above, and we obtain the equations of motion
\begin{eqnarray}
\label{ira1}
{\dot c}_{k1} &=& {1 \over \beta_k} \left< \xivec_{k0} \, , \, {\bf
a}_{\rm ext} 
\right> \\
{\dot c}_{k0} &=& - {1 \over \beta_k} \left< \xivec_{k1} \, , \, {\bf
a}_{\rm ext} \right> + c_{k1}. 
\label{ira2}
\end{eqnarray}

\section{Linear Independence of a Subset of the Right Eigenvectors}
\label{sec:proofs1}

In this appendix we show that the of non-Jordan-chain right eigenmodes
$\xivec_A$ for which 
the constant $b_A$ is positive 
are linearly
independent.  We start by recalling the notation used in Sec.\
\ref{sec:reality} above: 
the index $\alpha$ labels the distinct right eigenvectors $\xivec_A$
for which $b_A >0$, 
and we write $A = (\alpha,\epsilon)$ where
$\epsilon$ takes on the values $\epsilon=+$ and $\epsilon=-$.  Then
$\xivec_\alpha \equiv \xivec_{\alpha,+}$ and $\xivec_{\alpha,-} =
\xivec_\alpha^*$, while $\omega_{\alpha,\pm} = \pm \omega_\alpha$ and
$b_{\alpha,\pm} = \pm b_\alpha$.  

By combining the orthogonality relation (\ref{orthog6}) with the formula
(\ref{taufinalans}) for left eigenvectors $\bftau_A$, and choosing $A
= (\alpha,+)$ and $B = (\beta,+)$ we obtain the following form of the
orthogonality relation: 
\beq
b_\alpha \, \delta_{\alpha\beta} = \left< \xivec_\alpha \, , \, i {\bf
B} \cdot \xivec_\beta \right> + ( \omega_\alpha + \omega_\beta )
\left< \xivec_\alpha \, , \xivec_\beta \right>. 
\label{orthog10}
\endeq
Suppose now that the vectors $\xivec_\alpha$ are not linearly
independent.  Then, one can find coefficients $c_\alpha$, not all
zero, so the vector 
\beq
{\bf q} \equiv \sum_\alpha c_\alpha \, \xivec_\alpha
\end{equation}
is zero.  Now contract Eq.\ (\ref{orthog10}) with $c_\alpha^* c_\beta$
and sum over $\alpha$ and $\beta$.  The result is
\begin{eqnarray}
\sum_\alpha \, | c_\alpha |^2 \, b_\alpha &=& \left< {\bf q} \, , \, i
{\bf B} \cdot {\bf q} \right > + 2 \left< {\bf q} \, , \, \sum_\alpha
c_\alpha \omega_\alpha \xivec_\alpha \right> \nonumber \\
&=&0,
\end{eqnarray}
where the second line follows from ${\bf q}=0$.  Since on the left
hand side $b_\alpha > 0$ for all $\alpha$, this forces $c_\alpha=0$
for all $\alpha$, which contradicts our assumption above.

\section{Alternative Form of Equations of Motion}
\label{sec:proofs2}

In this appendix we present a version of the general
equations of motion which is second order in time and which is more
similar in form to the
standard equations of motion for non-rotating stars.  We start with
the equations derived in Sec.\ \ref{sec:reality} for the coefficients
$c_{\alpha,+}(t)$ and $c_{\alpha,-}(t)$:
\begin{eqnarray}
\label{firstorder0}
{\dot c}_{\alpha,+} + i \omega_\alpha c_{\alpha,+} &=&
{i \over b_\alpha} f_{\alpha,+}(t),  \\
{\dot c}_{\alpha,-} - i \omega_\alpha c_{\alpha,+} &=&
- {i \over b_\alpha} f_{\alpha,-}(t),
\label{firstorder}
\end{eqnarray}
where $f_{\alpha,\pm}(t) \equiv \left< \xivec_{\alpha,\pm} \, , \,
{\bf a}_{\rm ext}(t) \right>$.  As noted in Sec.\ \ref{sec:reality}, the
Lagrangian displacement $\xivec({\bf x},t)$ will be real if and only
if $c_{\alpha,+} = c_{\alpha,-}^*$ and $f_{\alpha,+} =
f_{\alpha,-}^*$.  Here, however, we allow arbitrary complex
$\xivec({\bf x},t)$.  We define 
\begin{eqnarray}
\label{edef1}
e_\alpha(t) &\equiv& c_{\alpha,+}(t) + c_{\alpha,-}(t) \\
&=& {2 \omega_\alpha \over b_\alpha} \left< {\rm Re}(\xivec_\alpha)\, ,
\, \xivec(t) \right> 
+ {2 \over b_\alpha} \left< {\rm Im}(\xivec_\alpha) \, , \, {\dot
\xivec}(t) + {\bf B} \cdot \xivec(t) \right>,
\end{eqnarray}
where we have used Eqs.\ (\ref{inverseexpansion1}) and
(\ref{inverseexpansion2}). 
It is now straightforward to show using the definition (\ref{edef1})
that the equations of motion (\ref{firstorder0})--(\ref{firstorder}) are 
equivalent to
\beq
{\ddot e}_\alpha(t) + \omega_\alpha^2 e_\alpha(t) = {\cal
F}_\alpha(t),
\label{ee20}
\end{equation}
which is the standard forced harmonic oscillator equation of motion.  The
forcing term here is 
\begin{eqnarray}
\label{calFdef1}
{\cal F}_\alpha & = & - {i \over b_\alpha} ( {\dot f}_{\alpha,-} + i
\omega_\alpha f_{\alpha,-} ) + {i \over b_\alpha} ( {\dot f}_{\alpha,+} - i
\omega_\alpha f_{\alpha,+} ) \\
&=& {2 \omega_\alpha \over b_\alpha} \left< {\rm Re} (\xivec_\alpha)
\, , \, {\bf a}_{\rm ext} \right> +  {2  \over b_\alpha} \left< {\rm Im} (\xivec_\alpha)
\, , \, {\dot 
{\bf a}_{\rm ext}} \right>.
\label{calFdef2}
\end{eqnarray}
It is clear that $e_\alpha(t)$ and ${\cal F}_\alpha(t)$ will be real
if and only if ${\bf a}_{\rm ext}(t)$ and $\xivec({\bf x},t)$ are real.
We can express the Lagrangian displacement $\xivec({\bf x},t)$ in
terms of $e_\alpha(t)$ and ${\dot e}_\alpha(t)$ at any time $t$ by
using the first component of the mode expansion
(\ref{goodexpansion1}), and by using the relations
\begin{eqnarray}
c_{\alpha,+} &=& {1 \over 2} \left( e_\alpha + {i \over \omega_\alpha}
{\dot e}_\alpha \right) + {1 \over 2 \omega_\alpha b_\alpha} \left(
f_{\alpha,+} - f_{\alpha,-} \right), \\
c_{\alpha,-} &=& {1 \over 2} \left( e_\alpha - {i \over \omega_\alpha}
{\dot e}_\alpha \right) - {1 \over 2 \omega_\alpha b_\alpha} \left(
f_{\alpha,+} - f_{\alpha,-} \right).
\end{eqnarray}

In this formalism, the number of ``modes'' $e_\alpha(t)$ is the same
for rotating stars as for non-rotating stars.  In particular, 
for non-rotating stars the expansion coefficients $e_\alpha$ coincide
with the standard expansion coefficients $q_\alpha$ given by
Eq.~(\ref{qalphadef}), when the mode functions 
$\xivec_\alpha({\bf x})$ are chosen to be real.

Note that the equation of motion (\ref{ee20}) for a given mode has a
characteristic feature which is peculiar to rotating stars, namely 
the forcing term depends on the time derivative ${\dot {\bf a}_{\rm ext}}(\xvec,t)$ of
the externally applied force per unit mass, as well as on ${\bf a}_{\rm ext}(\xvec,t)$ 
itself.  For non-rotating stars one can choose a real mode basis
$\xivec_\alpha$ which removes the dependence on ${\dot {\bf a}_{\rm ext}}$ [cf.\
the second term in Eq.\ (\ref{calFdef2})], but for rotating stars
it is not possible to find real mode bases.  

\section{A Set of Zero-Frequency, Jordan-Chain Modes}
\label{sec:proofs3}

In this appendix we justify the claim made in Sec.\
\ref{sec:linear_rot} above that there are always Jordan-chain modes
present in rotating zero-buoyancy stars, even for stable stars.  We show
that the well known class of purely axial, zero frequency modes that
move the star to  
nearby equilibrium states with different angular velocity are
Jordan-chain modes.  Note that in this appendix we do not need to
assume that the angular velocity $\Omega$ of the star is small.
For non-rotating stars, the existence of these Jordan chains was
previously noted in a footnote by Ref.~\cite{1980MNRAS.190....21S}.

It is most convenient to work directly with the linearized Eulerian
equations of motion (\ref{continuity0l}) and (\ref{euler0l}).
Substitute into these equations the ansatz $\delta \rho({\bf x},t) =
\delta \rho({\bf x})$ and
\beq
\delta {\bf u}({\bf x},t) = \delta {\bf u}({\bf x}) 
= r_\perp \delta \Omega(r_\perp) {\bf e}_\varphi,
\label{sss}
\end{equation}
where $(r_\perp,\varphi,z)$ are cylindrical coordinates with 
$r_\perp$ the distance from the rotation axis, 
and ${\bf e}_\varphi$ is the unit vector in the $\varphi$ direction.
One finds that this yields a time-independent solution, and that one
can specify the perturbation $\delta \Omega(r_\perp)$ arbitrarily and
solve for the density perturbation $\delta \rho({\bf x})$
\footnote{The density perturbation always vanishes if the background
star is non-rotating but not in general.}.  

Now switch back to the language of Lagrangian perturbation
theory.  The equations for a Jordan chain of length one with
$\omega_A=0$ are as follows.  [We drop the index $A$ and keep only the
index $\sigma$ for convenience.]  The phase space right eigenvector is,
from Eq.\ (\ref{jk1}), 
\begin{eqnarray}
\bfzeta_0 = \left[ \begin{array}{c}  
	\bfxi_0 \\
	{1 \over 2} {\bf B} \cdot \xivec_0
	\end{array} \right],
\label{jk11}
\end{eqnarray}   
and the associated right Jordan chain vector is
\begin{eqnarray}
\bfzeta_1 = \left[ \begin{array}{c}  
	\bfxi_1 \\
	{1 \over 2} {\bf B} \cdot \xivec_1 + \xivec_0
	\end{array} \right].
\label{jk11a}
\end{eqnarray}   
Here $\xivec_0$ and $\xivec_1$ satisfy, from Eq.\ (\ref{rtjordan}),
\beq
{\bf L}(0) \cdot \xivec_0 = {\bf C} \cdot \xivec_0 =0,
\label{ek0}
\end{equation}
and 
\beq
{\bf L}(0) \cdot \xivec_1 = {\bf C} \cdot \xivec_1 = - {\bf B} \cdot
\xivec_0.
\label{ek1}
\end{equation}

Now let $\xi_0({\bf x})$ be a mode function of the form
(\ref{sss}).  Then one finds that $\nabla \cdot \xivec_0 =0$ and 
$\nabla \cdot (\rho \xivec_0) =0$, and hence from the definition  
(\ref{Cdef-simple})--(\ref{Cdef-b}) of the operator ${\bf C}$,
Eq.~(\ref{ek0}) is satisfied.  One can then always solve Eq.\ (\ref{ek1})
to determine $\xivec_1$, since the left hand side is a pure gradient
for zero-buoyancy stars and the curl of the right hand side vanishes by
Eq.\ (\ref{sss}).
Thus the solutions (\ref{sss}) correspond to a Jordan chains of length one.
The vector $\xivec_1$ 
encodes the density perturbation.
The general solution $\xivec({\bf x},t)$ associated with $\xivec_0$ is
then, from Eq.\ (\ref{gensoln}), 
\beq
\xivec({\bf x},t) = \gamma_0 \xivec_0({\bf x}) + \gamma_1 \left[
\xivec_1({\bf x}) + \xivec_0({\bf x}) t \right],
\end{equation}
where $\gamma_0$ and $\gamma_1$ are constants of integration.

The piece of the solution proportional to $\gamma_0$ has vanishing
Eulerian density and velocity perturbations, and is pure gauge in the
sense of Ref.~\cite{1978ApJ...221..937F}.  By contrast, the piece
proportional to $\gamma_1$ is a physical, non-gauge mode.

For non-rotating stars, the situation is a little different: {\it all}
zero frequency modes are associated with Jordan chains of length one.
This can be seen from the fact that Eq.\ (\ref{ek1}) always has solutions
for non-rotating stars, since ${\bf B}=0$, whereas for rotating stars
a solution only exists if the curl of ${\bf B} \cdot \xivec_0$ vanishes.
The associated solutions of Eq.\ (\ref{basic0}) are of the form
$\xivec(t) = \xivec_1 + t \xivec_0$ where $\xivec_0$ and $\xivec_1$
satisfy ${\bf C} \cdot \xivec_0=0$ and ${\bf C} \cdot \xivec_1=0$. 

Finally, Ref.~\cite{1980MNRAS.190....21S} shows that in
situations where there are Jordan-chain modes, if one adds any small
perturbation to the system then unstable modes are generically
created.  Hence,  
Jordan-chain modes should generically not occur in stable stars.
However, Schutz notes that a necessary condition for this argument to
apply is that a certain matrix element of the perturbation not
vanish.  Specifically, if $\Delta {\bf T}$ is the perturbation to the
operator (\ref{Tdef}) and $\bfzeta_{A,\sigma}$ is a right Jordan
chain with associated left Jordan chain $\bfchi_{A,\sigma}$, then the
argument requires
\beq
\left< \bfchi_{A,0} \, , \, \Delta {\bf T} \cdot \bfzeta_{A,0} \right>
\ne 0.
\label{necc0}
\end{equation}
In the current context, the argument is evaded because of the fact
that the matrix element (\ref{necc0}) vanishes for arbitrary linear
perturbations $\Delta {\bf B}$ and $\Delta {\bf C}$ to the operators
${\bf B}$ and ${\bf C}$, from Eqs.\ (\ref{Tdef}) and (\ref{jk11}).
Generic physical perturbations to the system do not correspond to
generic mathematical perturbations to the operator ${\bf T}$, because of
the structure of Eq.~(\ref{Tdef}).

\section{The Zero Frequency Subspace for a Non-Rotating Star}
\label{sec:zero-freq}
This appendix characterizes the
space ${\cal H}_0$ of zero frequency modes for non-rotating
stars.  The proofs in this appendix are Newtonian versions of the
fully relativistic proofs given in Sec. III of Ref.~\cite{gr-qc0008019}.

The Lagrangian displacements $\bfxi({\bf x})$ for zero frequency modes
satisfy  
\begin{eqnarray}
{\bf C} \cdot \bfxi=0,
\label{zfeqn}
\end{eqnarray}
where the operator ${\bf C}$ is defined by Eq.\ (\ref{Cdef}) specialized
to a non-rotating background star.
From Eqs.\ (\ref{Cdef})--(\ref{Cdef-b}),
Eq.~(\ref{zfeqn}) can be written as
\beq
{\bf \nabla} \left[ g_1(r) \delta \rho({\bf x}) + \delta \phi({\bf
x}) + g_2(r) \xi^r({\bf x}) \right] -  \beta({\bf x}) g_2(r) {\bf
e}_{{\hat r}} =0,  
\label{E1}
\end{equation}
where
\beq
\beta({\bf x}) \equiv {\bf \nabla} \cdot {\xivec},
\end{equation}
the radial component of $\xivec$ is $\xi^r {\bf e}_r$, and the
functions $g_1(r)= \Gamma_1 p / \rho^2 
$ and $g_2(r) = \Gamma_1 p A^r / \rho$
are fixed functions determined by the 
background stellar model and by the adiabatic index $\Gamma_1$ of the
perturbations.  The last two terms involving
the function $g_2$ vanish for zero-buoyancy stars.

The argument of Lockitch and Friedman \cite{gr-qc0008019}, simplified
for Newtonian 
stars, is as follows.  The perturbed pressure ${\bar p} \equiv p +
\delta p$ and perturbed density ${\bar \rho} = \rho + \delta \rho$
obey the equation of hydrostatic balance and Poisson's equation, to
linear order in the perturbation.  Hence, the pair of functions
$({\bar \rho}, {\bar p})$ describe a static self-gravitating
perfect fluid configuration, if one assumes that to each solution of
the linearized fluid equations there corresponds a solution of the
exact equations.  However, it is known that all static,
self-gravitating perfect fluid configurations are spherically
symmetric\footnote{Note that the spherical-symmetry
theorem does not require the barotropic assumption ${\bf \nabla} {\bar
\rho} {\bf \times } {\bf \nabla} {\bar p}=0$.  It requires only that the
perturbed pressure ${\bar p}$ and perturbed density ${\bar \rho}$ satisfy 
the equation of hydrostatic balance and Poisson's equation.}.  
Hence, all perturbations $(\delta \rho, \delta p)$ must 
correspond to displacements to nearby static 
spherical configurations.  Solutions of this type include (i)
perturbations taking the star 
to a nearby static, spherically symmetric configuration with a
different total mass $M$.  If one assumes that the change $\delta M$
to the total mass of the star vanishes, such solutions are
disallowed.  (ii) Solutions with $l=1$ of the form $\bfxi = {\rm
const}$ corresponding to the displacement of the center of mass of the
star.  If one demands that the perturbation leaves fixed the center of
mass of the star, these solutions are disallowed.  (iii) Solutions with
$l=0$
where the background star is marginally stable to radial collapse.
Henceforth we will assume the
background star is such that there are no $l=0$ zero-frequency
solutions of this type. 

If one makes sufficient assumptions to outlaw the cases (i),
(ii) and (iii) above, it follows that all solutions to Eq.\ (\ref{E1})
must satisfy
\beq
\delta \rho = \delta p =0.
\label{vanishing}
\end{equation}
We now consider the cases of zero-buoyancy and nonzero-buoyancy stars.

\subsection{Zero-buoyancy stars}

We showed above that all solutions in ${\cal H}_0$ have $\delta
\rho=0$.  However, 
for $\Avec = 0$, Eq.\ (\ref{E1}) is always
satisfied when $\delta \rho=0$, since the last two terms vanish.  
Hence, the space ${\cal H}_0$ consists precisely
of the perturbations with 
\begin{eqnarray}
\delta \rho = - {\bf \nabla} \cdot (\rho \bfxi) =0.
\label{H0condt}
\end{eqnarray}
Note that this characterization does not hold for rotating stars,
where there are zero frequency modes with $\delta \rho \ne 0$ (see
Appendix \ref{sec:proofs3}).

We next derive a property of ${\cal H}_0$ that is used in Sec.\
\ref{sec:sec3} above.  Let ${\bf P}_0$ to be the operator that
projects orthogonally onto the subspace ${\cal H}_0$.  We now show
that 
\begin{eqnarray}
{\bf P}_0 \cdot \bfxi = 0 \ \ \mbox{if\ and\ only\ if} \ \ \ {\bf
\nabla} \times \bfxi =0.
\label{ans4}
\end{eqnarray}
Let $T^{ij}({\bf x})$ be any antisymmetric tensor that vanishes on
the boundary of the star.  Then an integration by parts shows that
\beq
\int d^3 x \, \nabla_{[i} \, \xi^*_{j]} \, T^{ij} = \left< \xivec \,
, \, {\bf v} \right>
\label{foundation}
\end{equation}
where the vector ${\bf v}$ is given by
\beq
v^i = {1 \over \rho} \nabla_j T^{ij}.
\label{vecdef}
\end{equation}
Now if ${\bf P}_0 \cdot \xivec=0$, then $\xivec$ is orthogonal to all
vectors ${\bf v}$ in ${\cal H}_0$.  But vectors of the form
(\ref{vecdef}) are automatically in ${\cal H}_0$, by
Eq.~(\ref{H0condt}).   Hence both sides of Eq.\ (\ref{foundation})
vanish.  Since this is true for all $T^{ij}({\bf x})$ that vanish on
the boundary, and since $\nabla_i \xi_j$ is continuous, it follows
that $\nabla \times \xivec=0$.  It can be checked that the argument
also carries through in the other direction.

\subsection{Non-zero buoyancy stars}

Turn now to stars with buoyancy forces.  For these stars,
the Lockitch-Friedman argument shows that the space of zero frequency
modes is precisely the set of axial vectors (except possibly in the
$l=0$ sector). 

We use the expansion (\ref{lagpert}) of Lagrangian perturbation on a
basis of vector spherical harmonics.
Here the first two terms are the polar part of $\xivec$ and the last
term is the axial part.  
It is easy to see that the axial part gives a vanishing contribution to both
$\delta \rho$ and $\beta = {\bf \nabla} \cdot \xivec$, and hence all
axial vectors lie in ${\cal H}_0$, from Eq.\ (\ref{E1}). 
Thus the $C_{lm}$ terms do not contribute at all to Eq.~(\ref{E1}) and so we
can drop them and focus on the 
$A_{lm}$ and $B_{lm}$ terms.  
The vanishing of the density perturbation $\delta \rho = - {\bf \nabla}
\cdot( \rho \xivec)$ from Eq.\ (\ref{vanishing}) can be written as
\beq
{1 \over \rho} (\rho A_{lm})^\prime +
2 A_{lm} / r - l (l+1) B_{lm} /r^2 =0.
\label{zd}
\end{equation}
Also, the relation between the Eulerian density and pressure perturbations
for $\Avec \neq 0$ is, from Eq.\ (\ref{Gamma1def}),
\beq
\delta p = {\Gamma_1 p \over \rho} 
\left( \delta \rho + \rho \xivec \cdot \Avec \right).
\end{equation}
Substituting from Eq.\ (\ref{vanishing}) now shows that the radial component
of $\xivec$ vanishes.  Hence
$A_{lm}=0$ for all $l,m$, and it follows from Eq.\ (\ref{zd}) that
$B_{lm}=0$ for all $l,m$ with $l > 0$ too.
Since we are restricting attention to stars for which there are no
$l=0$ zero-frequency modes (see above), it follows that the
space ${\cal H}_0$ of zero frequency modes is therefore precisely the
set of axial vectors.    

Lastly, let ${\bf P}_0$ be the orthogonal projection operator that
projects onto ${\cal H}_0$.  In this case it follows from the
definition of axial [Eq.\ (\ref{lagpert}) with $A_{lm} = B_{lm}=0$] that
\begin{eqnarray}
{\bf P}_0 \cdot \bfxi = 0 \ \ \mbox{if\ and\ only\ if} \ \ \ {\bf r}
\cdot \left({\bf \nabla} \times \bfxi\right) =0.
\label{ans44}
\end{eqnarray}

\section{Non-Existence of Modes Whose Frequencies Scale as the
Square Root of the Rotational Frequency}
\label{sec:noexist}

In this appendix, we show that there are no modes
whose frequencies and mode functions can be expressed as power series
in the square root $\sqrt{\Omega}$ of the star's angular velocity
and whose frequency in the non-rotating limit is zero.
As discussed in Sec.\ \ref{sec:slow_rot1} above, one might
suspect the existence of such modes from the general perturbation
theory analysis of \cite{1980MNRAS.190....21S} and the fact that 
non-rotating stars have Jordan chains of length one.
The result is valid for both zero-buoyancy and nonzero-buoyancy
stars.  The non-existence of such modes is related to the vanishing of
the matrix element (\ref{necc0}) discussed in Appendix \ref{sec:proofs3}
above.  

We assume a one parameter family of modes $\xivec(\Omega),
\omega(\Omega)$ with expansions of the form
\begin{eqnarray}
\omega(\Omega) = \sqrt{\Omega} \,
\omega^{(1/2)} + \Omega \, \omega^{(1)} + O(\Omega^{3/2}),
\label{lambdaexpansionf}
\end{eqnarray}
and 
\begin{eqnarray}
\bfxi(\Omega) = \bfxi^{(0)} + \sqrt{\Omega} \,
\bfxi^{(1/2)} + \Omega \bfxi^{(1)} + O(\Omega^{3/2}).
\label{xiexpansionf}
\end{eqnarray}
If we substitute the expansions
(\ref{lambdaexpansionf})--(\ref{xiexpansionf}) and also the expansions  
(\ref{Bexpansion})--(\ref{Cexpansion}) for the operators ${\bf B}$ and
${\bf C}$ into the quadratic eigenvalue equation (\ref{basic1}) we get
at orders $O(\Omega^0)$, $O(\sqrt{\Omega})$, and $O(\Omega)$ the equations
${\bf C}^{(0)} \cdot \bfxi^{(0)} =0$, 
${\bf C}^{(0)} \cdot \bfxi^{(1)} =0$, and
\beq
{\bf C}^{(0)} \cdot \xivec^{(1)} - \omega^{(1/2)\,2} \xivec^{(0)} =0.
\label{key1}
\end{equation}
We now multiply Eq.\ (\ref{key1}) by the projection operator ${\bf
P}_0$ into the space ${\cal H}_0$ of zero-frequency modes, and use the
fact that ${\bf P}_0 \cdot \xivec^{(0)} =\xivec^{(0)}$ and ${\bf P}_0
\cdot {\bf C}^{(0)} = 0$.  This gives
\beq
w^{(1/2)\,2} \xivec^{(0)} =0,
\end{equation}
which shows that $\omega^{(1/2)}=0$.

\section{Derivation of Selection Rules}
\label{sec:vanishing}

In this appendix we derive the two axial selection rules
(\ref{axialsel1}) and (\ref{axialsel2}) and the selection rule
(\ref{incompsel}) for an incompressible star.  Throughout this
appendix we will drop all terms that are not of zeroth  
order in $\Omega$.  Therefore we will use the pressure, density and
potential profiles of a spherical background star.  

\subsection{Axial selection rules}

We assume that the modes $\xivec_B$
and $\xivec_C$ are axial, and therefore 
\beq
\grad \cdot \xivec_B = \grad \cdot \xivec_C = \delta\rho_B = \delta
\rho_C = \xi_B^r = \xi_C^r =0.
\label{axialidentities}
\endeq
In addition 
we assume that the Eulerian density perturbation $\delta \rho_A$
of mode $\xivec_A$ vanishes, and so from Eq.\
(\ref{deltarhodef}) we have
\beq
{\bf \nabla} \cdot \xivec_A = - {1 \over \rho} {d \rho \over d r}
\xivec_A^r.
\label{divxiidentity}
\endeq
Inserting these results into the expression (\ref{eq:23}) for the
coupling coefficient we obtain
\baray
2 \kappa_{ABC}&=&
\int{\volx \, p \left[(\Gamma_1-1)(\grad\cdot\xivec_A)\Xi_{BC}
+ \chi_{ABC}+\chi_{ACB} \right]}
- \int d^3x \rho \xi^i_A \xi^j_B \xi^k_C \phi_{;ijk} \,.
\label{kappa1}
\earay
Now the background potential $\phi$ is a function of $r$ only, so we
have
\beq
\phi_{;jki}={1\over r}{d\over dr}\biggl({1\over r}{d\phi\over dr}\biggr)
(x_i\delta_{jk}+x_k\delta_{ij}+x_j\delta_{ik})
+{1\over r}{d\over dr}\biggl[{1\over r}{d\over dr}\biggl({1\over r}
{d\phi\over dr}\biggr)\biggr]x_i x_j x_k.
\label{phixxx}
\endeq
Substituting Eq.\  (\ref{phixxx}) into Eq.\
(\ref{kappa1}) and using Eq.\ (\ref{axialidentities}) gives
\baray
2 \kappa_{ABC}&=&
\int{\volx \, p \left( \chi_{ABC}+\chi_{ACB} \right)}
+ \int{ \volx \, p (\Gamma_1-1)(\grad\cdot\xivec_A)\Xi_{BC}} \nonumber \\
&& 
- \int{ \volx \, \rho {d \over d r} \left( {1 \over r} {d \phi
  \over d r} \right) \, \xi_A^r \, \left( \xivec_B \cdot \xivec_C
  \right) }.
\label{kappa2}
\earay

We now evaluate the first term in Eq.\ (\ref{kappa2}) using the
definition (\ref{eq:chiabcdef}) of $\chi_{ABC}$.  This gives
\be
\int{\volx\,{p}\,\chi_{ABC}}=
\int{\volx\,{p}\, \xi^i_{A\,;j}\xi^j_{B\,;k}\xi^k_{C\,;i}}
=\int{\volx\,{p}(\xi^i_A\xi^j_{B\,;k})_{;j}\xi^k_{C\,;i}},
\ee
where the second form of the integral was obtained using 
$\xi^j_{B\,;kj}= \xi^j_{B\,;jk}=0$. 
Upon integrating by parts\footnote{
Throughout this appendix we discard surface terms that arise
from integrations by parts which involve
the radial component $\xi^r_A$ of the mode function $\xivec_A$
evaluated on the star's surface, because the quantity
$$
{\Delta p \over \rho}= {\Gamma_1 p \over \rho^2} {d \rho \over d r}
\xi^r_A = - {\Gamma_1 \over \Gamma} \phi^\prime \, \xi_A^r
$$
must vanish on the surface \cite{1999ApJ...521..764L}.}
we find
\be
\int{\volx\,{p}\, \chi_{ABC}}=
-\int{\volx (p_{,j}\xi^i_A\xi^j_{B\,;k}\xi^k_{C\,;i}
+p\xi^i_A\xi^j_{B\,;k}\xi^k_{C\,;ij})}.
\ee
We can rewrite this using the relationship that follows from $p = p(r)$
\be
(p_{,j}\xi^j_B)_{;k}=0=p_{,j}\xi^j_{B\,;k}+p_{;jk}\xi^j_B;
\ee
the result is
\baray
\int{\volx\,{p}\chi_{ABC}}&=&
\int{\volx \left[p_{;jk}\xi^i_A\xi^j_B\xi^k_{C\,;i}
-{p}\xi^i_A\xi^j_{B\,;k}\xi^k_{C\,;ij} \right] }\nonumber\\
&=&-\int{\volx \left[ p_{;jki}\xi^i_A\xi^j_B\xi^k_C
+ p_{;jk} \xi^i_{A\,;i} \xi_B^j \xi_C^k
+p_{;jk}\xi^i_A\xi^j_{B\,;i}\xi^k_C
+{p}\xi^i_A\xi^j_{B\,;k}\xi^k_{C;ij} \right] },
\earay
where we have again integrated by parts.
We can evaluate the term involving $p_{;jki}$ using an equation
analogous to Eq.\ (\ref{phixxx}), which gives
\baray
\int{\volx\,{p}\chi_{ABC}}
&=&-\int{\volx \left[ 
{d \over d r} \left( {1 \over r} {d p \over d r} \right) \xi_A^r
\left( \xivec_A \cdot \xivec_B \right)
+ p_{;jk} \xi^i_{A\,;i} \xi_B^j \xi_C^k
+p_{;jk}\xi^i_A\xi^j_{B\,;i}\xi^k_C
+{p}\xi^i_A\xi^j_{B\,;k}\xi^k_{C;ij} \right] },
\earay

Next, we symmetrize over the indices $B$ and $C$.  This gives
\baray
\int{\volx\,{p} \left( \chi_{ABC} + \chi_{ACB} \right)}
&=&-2 \int{\volx \left[ 
{d \over d r} \left( {1 \over r} {d p \over d r} \right) \xi_A^r
\left( \xivec_A \cdot \xivec_B \right)
+ p_{;jk} \xi^i_{A\,;i} \xi_B^j \xi_C^k \right]} \nonumber \\
&& -
\int{\volx \left[ 
p_{;jk}\xi^i_A\xi^j_{B\,;i}\xi^k_C
+{p}\xi^i_A\xi^j_{B\,;k}\xi^k_{C;ij} 
+p_{;jk}\xi^i_A\xi^k_{C\,;i}\xi^j_B
+{p}\xi^i_A\xi^k_{C\,;j}\xi^j_{B;ik} 
\right] },
\label{eq:G11}
\earay
where on the right hand side we have relabeled some indices.
We can rewrite the second line of Eq.\ (\ref{eq:G11}) as
\begin{eqnarray}
&& -\int{\volx \left[ 
p_{;jk}\xi^i_A \left( \xi^j_{B}\xi^k_C \right)_{;i}
+{p}\xi^i_A \left( \xi^j_{B\,;k}\xi^k_{C;j} \right)_{;i} 
\right] }, \nonumber \\
&=& \int{\volx \left[ 
p_{;jki}\xi^i_A \xi^j_{B}\xi^k_C 
+ p_{;jk}\xi^i_{A\,;i} \xi^j_{B}\xi^k_C 
+{p},i\xi^i_A \xi^j_{B\,;k}\xi^k_{C;j}
+{p} \xi^i_{A\,;} \xi^j_{B\,;k}\xi^k_{C;j}\right] },
\label{eq:G11a}
\end{eqnarray}
where we have again integrated by parts.
Rewriting the terms involving $p_{;jki}$ and $p_{;jk}$ the same way as
before, and using the definition (\ref{Xiabdef}) of $\Xi_{BC}$, we can
write the expression (\ref{eq:G11}) as
\baray
\int{\volx\,{p} \left( \chi_{ABC} + \chi_{ACB} \right)}
&=&- \int{\volx \left[ 
{d \over d r} \left( {1 \over r} {d p \over d r} \right) \xi_A^r
\left( \xivec_B \cdot \xivec_C \right)
+ {1 \over r} {d p \over d r} \left( {\bf \nabla} \cdot \xivec_A
\right) \left( \xivec_B \cdot \xivec_C \right)
\right] }
\nonumber \\
&& +
\int{\volx \left[ 
p_{,i} \xi_A^i \Xi_{BC} + p \xi^i_{A\,;i} \Xi_{BC} \right] }.
\label{eq:G12}
\earay

Now since the Eulerian density perturbation 
$-(\rho \xi^i_A)_{;i}$ vanishes, we have
\be
(p\xi^i_A)_{;i}={dp\over d\rho}\rho_{;i}\xi^i_A
+p\xi^i_{A\,;i}=p\xi^i_{A\,;i}\biggl(-{d\ln p\over
d\ln\rho}+1\biggl)=-p(\Gamma-1)\xi^i_{A\,;i},
\ee
where $\Gamma$ is the adiabatic index of the background spherical star.
Combining Eqs.\ (\ref{kappa2}), (\ref{eq:G11a}), and (\ref{eq:G12}) now yields
\begin{eqnarray}
2 \kappa_{ABC} &=& \int d^3 x \, \left( \xivec_B \cdot \xivec_C \right)
\left[ - \rho \xi_A^r {d \over d r} \left( {1 \over r} {d \phi \over d
r} \right) - \xi_A^r {d \over d r} \left( {1 \over r} {d p \over d r}
\right) - {1 \over r} {d p \over d r} \left( {\bf \nabla} \cdot
\xivec_A \right) \right] \nonumber \\
&& + \int d^3 x p (\Gamma_1 - \Gamma) \left( {\bf \nabla} \cdot
\xivec_A \right) \Xi_{BC}.
\label{kappa5}
\end{eqnarray}
Using Eq.\ (\ref{divxiidentity}) and the equation of hydrostatic
equilibrium in the background star one can show that the first line of
the right hand side of Eq.\ (\ref{kappa5}) vanishes.  Therefore we
find
\beq
\kappa_{ABC}  = {1 \over 2} \int d^3 x \, p (\Gamma_1 - \Gamma) \, \left(
{\bf \nabla} \cdot \xivec_A \right) \, \Xi_{BC}.
\label{kappa6}
\ee

We can derive two different selection rules from Eq.\ (\ref{kappa6}).  First,
if $\xivec_A$ is axial, then ${\bf \nabla} \cdot \xivec_A =0$ and therefore
$\kappa_{ABC} =0$; this proves the selection rule (\ref{axialsel1})
for three axial modes.  Second, if the perturbations obey the same
equation of state as the background star, then the adiabatic index
$\Gamma_1$ of the perturbations coincides with the adiabatic index
$\Gamma$ of the background star, and $\kappa_{ABC}$ again vanishes.
This proves the second axial selection rule (\ref{axialsel2}).

\subsection{Incompressible selection rule}

We can consider an incompressible star to be the limit $\Gamma_1 \to
\infty$ of a zero-buoyancy polytrope with $\Gamma= \Gamma_1$.
The expressions (\ref{eq:23}) and (\ref{eq:23aa}) for
the coupling coefficient $\kappa_{ABC}$ are valid for incompressible
stars, since they are valid for finite $\Gamma_1$ and they have a
finite limit as $\Gamma_1 \to \infty$.  Note however that one must be
careful to include all the distributional contributions in the
integrands at the stellar surface that arise in the $\Gamma_1 \to
\infty$ limit.

Consider now a linear mode $\xivec({\bf x})$ of the large-$\Gamma_1$ star,
whose $\Gamma_1 \to \infty$ limit is a mode of the incompressible
star.   Clearly the Lagrangian perturbation of the density must scale
as
\be
{ \Delta \rho \over \rho} \propto {1 \over \Gamma_1},
\ee
which implies from Eq. (\ref{delrho}) that 
\be
\grad \cdot \xivec \propto {1 \over \Gamma_1}.
\ee
It follows that to zeroth order in $1/\Gamma_1$, the Eulerian density
perturbation is, from Eq.\ (\ref{deltarhodef}) 
\begin{eqnarray}
	\delta\rho &=& -\nabla\cdot(\rhonot \xivec) = 	
		-\xivec\cdot\nabla\rhonot-\rhonot\delxi\nonumber\\
	&=& \xi^r \tilde\rhonot\delta(r-R)
		\label{cddeltarho}
\end{eqnarray}
where $\tilde\rhonot = {3M/ 4\pi R^3}$ and $M$ and $R$ are the stellar
mass and radius.  

Consider now three modes $\xivec_A$, $\xivec_B$, $\xivec_C$ that
satisfy the hypothesis of the selection rule (\ref{incompsel}).  That
is, they all have vanishing Eulerian density and pressure
perturbations to zeroth order in $1/\Gamma_1$.  It follows from Eq.\
(\ref{cddeltarho}) that they all have vanishing radial components at
the stellar surface,
\be
\xi^r(r=R) =0.
\label{radialv}
\ee
Consider now the expression (\ref{eq:23aa}) for the coupling
coefficient.  In this expression, the second, third and fifth terms
vanish since $\grad \cdot \xivec=0$ for each mode, and the fourth and
sixth terms vanish since each mode has vanishing $\delta \rho$,
$\delta \phi$ and $\delta p$.  The remaining terms yield the following
expression for the coupling coefficient:
\be
\kappa_{ABC}
 =  - {1 \over 2} \int d^3x  \,
\xi^i_A \xi^j_B \xi^k_C 
\left[ p_{;ijk} + \rho \phi_{;ijk} \right].
\label{eq:23ab}
\ee
The integrand in Eq. (\ref{eq:23ab}) vanishes in the interior of the
star, since $p$ and $\phi$ are quadratic functions of position for a
constant density star.  However, there are distributional
contributions to the integral at the stellar surface, since $p_{;i}$
and $\phi_{;ij}$ are discontinuous across the surface.  To evaluate
these contributions, substitute in the equation of hydrostatic
equilibrium in the background star, $p_{;i} + \rho \phi_{;i}=0$, which
yields 
\be
\kappa_{ABC}
 =   {1 \over 2} \int d^3x  \,
\xi^i_{(A} \xi^j_B \xi^k_{C)} 
\left[ 2 \rho_{;i} \phi_{;jk} + \phi_{;i} \rho_{;jk}\right].
\label{eq:23ac}
\ee
The first term vanishes since $\phi_{;ji}$ is finite on the stellar
surface, and for any of the modes we have
\be
\xi^k \rho_{;k} = - {\tilde \rho} \, \xi^r(r=R) =0
\ee
from Eq.\ (\ref{radialv}).  Integrating the second term by parts gives
\be
\kappa_{ABC}
 =   - {1 \over 2} \int d^3x  \,
\left[ \xi^i_{(A} \xi^j_B \xi^k_{C)} \phi_{;i} \right]_{;j} \, \rho_{;k}.
\label{eq:23ad}
\ee
When the derivative of the square bracket is expanded out, each term
will contain either a factor $\xi^i \rho_{;i}$ or a factor $\xi^i
\phi_{;i}$, both of which vanish on the stellar surface.  Therefore we
obtain
\be
\kappa_{ABC} =0
\ee
to zeroth order in $1/\Gamma_1$ and in $\Omega$.  Note that the
cancellation of the distributional components in the expression
(\ref{eq:23ab}) would not be obtained if one uses the Cowling
approximation of neglecting the gravitational terms in the coupling
coefficient, as in Ref.\ \cite{Thesis:Wu}.

\section{Calculating Covariant Derivatives Using the Spin Weighted
Spherical Harmonics and the $\eth$ Operator}
\label{app:ccdswsh}

We define the ``spin-weight'' $s$ of any tensor as in \cite{newmanpenrose}
(also see \cite{jmp...12...1763} for a very readable reference),
in terms of how it transforms under rotations around the radial
direction.  For example, the 
vector $\mvec$ of Sec.~\ref{sec:An-Appr-Choice} transforms
under these types of rotations as $\mvec \to \mvec^\prime$, where
\begin{eqnarray}
	\mvec^\prime = e^{i\psi}\mvec\label{eq:36}
\end{eqnarray}
where $\psi$ is the angle of rotation.  We say that this vector has 
spin-weight $s=1$.  In general, a quantity $\vartheta$
is said to have spin-weight $s$ if it transforms as $\vartheta \to
\vartheta^\prime$ where
\begin{eqnarray}
	\vartheta^\prime = e^{is\psi}\vartheta.\label{eq:37}
\end{eqnarray}
In particular, the basis vectors $\lvec$,$\mvec$ and $\mbarvec$
defined in 
Sec.~\ref{sec:An-Appr-Choice} have spin-weights $0,1$ and $-1$ 
respectively.
Thus we can decompose any  given vector function, $\bvec\ofrthph$,  
into three components with
definite spin-weight by dotting it with one of the three basis vectors 
$(\lvec,\mvec,\mbarvec)$. Each scalar component can then be expanded in the
 complete set of basis functions $\Ylms$, e.g. 
\begin{eqnarray}
	\bvec\cdot\lvec &\equiv&\zeta_0\ofrthph = 
		\sumlam\bzlam\ofr\,\,\Ylamz\nonumber\\
	\bvec\cdot\mvec &\equiv&\zeta_1\ofrthph =
		\sumlam\bplam\ofr\,\,\Ylamo\nonumber\\
	\bvec\cdot\mbarvec&\equiv&\zeta_{-1}\ofrthph =
		\sumlam\bmlam\ofr\,\,\Ylammo.\label{eq:32}
\end{eqnarray}
For notational brevity we note that $\mvec$ and $\mbarvec$, and
consequently $\zeta_1$ and $\zeta_{-1}$,  are 
complex conjugates of each other.  Thus, in subsequent calculations we
will write $+ c.c.$ when convenient.\footnote{Note 
that in Section~\ref{sec:An-Appr-Choice} we choose to implement this 
by simply adding terms with  $s$ replaced by  $-s$.}   The connection
coefficients on this basis are
\begin{eqnarray}
	&\Gullmmbar&  = -\Gumllm
		 = -{1\over r}\nonumber\\
	&\Gumlmmbar& = -\Gumlmm 
		 = -\rtti {\cot\theta\over r},\label{connection}
\end{eqnarray}
where all the other components are zero except for complex conjugates
of the above, e.g. 
\begin{eqnarray}
	(\Gumlmm)^* = \Gumbarlmbarmbar = \rtti {\cot\theta\over r}.
		\label{eq:38}
\end{eqnarray}
It follows that the components of the covariant derivative of ${\bf
B}$ are 
\begin{eqnarray}
	\Bullcdl &=& \Bulldl\nonumber\\
	\Bumlcdm &=& \Bumldm +\gamI\Bul +\gamII\Bum\nonumber\\
	\Bumlcdmbar &=& \Bumldmbar  -\gamII\Bum\nonumber\\
	\Bumlcdl &=& \Bumldl \nonumber\\
	\Bullcdm &=& \Bulldm -\gamI\Bumbar,\label{eq:39}
\end{eqnarray}
where all others are zero except for complex conjugates of the above.
Now consider the $\eth$ (``edth'') and $\bar\eth$ operators of
Ref.~\cite{newmanpenrose}, 
defined by their action on a function, $\zeta_s$ of spin-weight $s$,
\begin{eqnarray}
	\eth\zeta_s &=& -(\sin\theta)^s\rttwo(\mvec\cdot\opL)(\sin\theta)^{-s}
	\zeta_s
	= -\left(\dlth +i\csc\theta\dlph 
		-s\cot\theta\right)\zeta_s\nonumber\\
	\bar\eth\zeta_s&=&(\sin\theta)^{-s}
		\rttwo(\mbarvec\cdot\opL)(\sin\theta)^{s}
	\zeta_s= \left(-\dlth +i\csc\theta\dlph 
		-s\cot\theta\right)\zeta_s.
\end{eqnarray}
Using these definitions in Eq.~(\ref{eq:39}) we get
\begin{eqnarray}
	\Bumlcdm &=& -\ovrttr\eth\zeta_{-1} +\ovr\zeta_0\nonumber\\
	\Bullcdm &=& -\ovrttr\eth\zeta_0-\ovr\zeta_1 \nonumber\\
	\Bumlcdmbar &=& -\ovrttr\bar\eth\zeta_{-1} \nonumber\\
	\Bumlcdl &=& \dlr\zeta_{-1} \nonumber\\
	\Bullcdl &=& \dlr\zeta_0\nonumber
\end{eqnarray}
and complex conjugates of the above.  The $\eth$ ($\bar\eth$) operator, 
when acting on the spin-weighted spherical harmonics, raises (lowers) 
the spin-weight by one,
\begin{eqnarray}
		\eth\Ylms &=& \sqrt{({l}-s)({l}+s+1)}\Ylmspo\nonumber\\
	\bar\eth\Ylms &=& -\sqrt{({l}+s)({l}-s+1)}\Ylmsmo.
		\label{edthonylms}
\end{eqnarray}
Thus when calculating a covariant derivative such as $\Bumlcdm$ we can
write
\begin{eqnarray}
	\Bumlcdm = \ovr\sum_\Lambda\left(\bzlam\ofr
		-{\Lambda\over\sqrt{2}}\bmlam\ofr\right)\,\,\Ylamz,
		\label{eq:41}
\end{eqnarray}
and similarly for the other covariant derivatives.  This allows us
to define the functionals as in Eq.~(\ref{gfhdefs}).
For decomposing vectors of the form~(\ref{genformmodes}), note that
\begin{eqnarray}
	\mvec\cdot\nabla \enskip\Ylamz&=& -{\Lambda\over \sqrt{2} r} 
	\enskip\Ylamo
		\nonumber\\  
	\mbarvec\cdot\nabla \enskip\Ylamz&=& -{\Lambda\over \sqrt{2} r} 
		\enskip\Ylammo
		\nonumber\\
	\mvec\cdot\opL\enskip\Ylamz&=& -{\Lambda\over \sqrt{2} }
		\enskip \Ylamo
		\nonumber\\
	\mbarvec\cdot\opL\enskip\Ylamz&=& -{\Lambda\over \sqrt{2} } 
		\enskip\Ylammo.
		\label{decompwmmbar}
\end{eqnarray}
Using these relations together with Eqs.\ (\ref{axialgen}) and
(\ref{polargen}) we find the functions $\fzlam$, $\fplam$ and
$\fmlam$ for $z-$parity odd modes
\begin{eqnarray}
	\fzlm\ofr&=& {\Wjpmpo\over r}\,\delljpmpo\nonumber\\
	\fplm\ofr &=& -{\Vjpmpo\over r}\,\lamjpmpo\,\delljpmpo
			\nonumber\\
		& &+{\Ujpm\over r}\,\lamjpm\,\delljpm
			\nonumber\\
	\fmlm\ofr&=&{\Vjpmpo\over r}\,\lamjpmpo\,\delljpmpo
			\nonumber\\
		& &+{\Ujpm\over r}\,\lamjpm\,\delljpm
			,\label{axialfs}
\end{eqnarray}
and for $z-$parity even modes
\begin{eqnarray}
	\fzlm\ofr&=& {\Wjpm\over r}\,\delljpm\nonumber\\
	\fplm\ofr &=& -{\Vjpm\over r}\,\lamjpm\,\delljpm
			\nonumber\\
		& &+{\Ujpmpo\over r}\,\lamjpmpo\,\delljpmpo
			\nonumber\\
	\fmlm\ofr&=&{\Vjpm\over r}\,\lamjpm\,\delljpm
			\nonumber\\
		& &+{\Ujpmpo\over r}\,\lamjpmpo\,\delljpmpo
			.\label{polarfs}
\end{eqnarray}

\section{Direct Derivation of Second Order 
Equations of Motion}
\label{app:nvpd}
There are at least
three ways to derive the equation of motion for the second order
Lagrangian perturbations: (1) Expand the variational principle to third
order in the Lagrangian displacement $\xibf$, and then vary with respect
to $\xibf$; (2) Vary with respect to $\xibf$ to get the equation of motion,
then expand to second order; (3) Directly perturb the equation of motion
to second order in $\xibf$:
\begin{eqnarray}\label{eqmo}
\Delta\left(\frac{du_i}{dt}+\frac{1}{\rho}\nabla_i p + \nabla_i \phi\right)
=0.
\end{eqnarray}
The second method is carried out in Appendix \ref{variational}.
Here we carry out the third method.


Recall that the Eulerian and Lagrangian perturbations of a quantity
$Q$ are defined as
\begin{eqnarray}
	\delta Q & =&  Q(\xbf,t) - Q_0(\xbf,t), \\
	\Delta Q & = & Q(\xbf+\xibf(\xbf,t),t) - Q_0(\xbf,t),
\end{eqnarray}
where $Q_0$ is the value of $Q$ in the unperturbed solution. 
[In this appendix we shall sometimes use 
$\rho_0$, $p_0$ and $\phi_0$ to denote the background density, pressure and
potential; these quantities were denoted simply $\rho$, $p$ and $\phi$
in the body of the paper.]  Suppressing
the time dependence of the quantities, we can write to second order
\begin{eqnarray}
	\Delta Q & =&  Q(\xbf+\xibf) - Q_0(\xbf) \nonumber \\
        	 & = & Q(\xbf+\xibf) - Q_0(\xbf+\xibf) 
			+ Q_0(\xbf+\xibf) - Q_0(\xbf)
                 	\nonumber \\
         	& = & \delta Q(\xbf+\xibf) + Q_0(\xbf+\xibf) 
			- Q_0(\xbf) \nonumber \\
         	& = & \delta Q + \xi^i\nabla_i\delta Q +\xi^i\nabla_i Q
               + \half \xi^i \xi^j \nabla_i \nabla_j Q.\label{delQ}
\end{eqnarray}
In the last line we have dropped the subscript 0 on the last two terms.
Note that $\delta Q$ in the second term on the right-hand side
need only be evaluated to first order. Substituting the first order relation
\begin{eqnarray}
\label{1storder}
\delta Q = \Delta Q - \xi^i\nabla_i Q
\end{eqnarray}
in this term gives the alternative relation
\begin{eqnarray}\label{delQ2}
\Delta Q = \delta Q + \xi^i\nabla_i\Delta Q +\xi^i\nabla_i Q
           - \xi^i\nabla_i \xi^j\nabla_j Q
               - \half \xi^i \xi^j \nabla_i \nabla_j Q.
\end{eqnarray}

By definition of the Lagrangian displacement vector
the unperturbed and perturbed fluid trajectories are related by
\begin{eqnarray}
x^i_0=c^i(t),\quad\quad x^i=c^i(t)+\xi^i(c^j(t),t).
\end{eqnarray}
Thus
\begin{eqnarray}
u^i_0=\frac{dc^i}{dt},\quad\quad u^i=\frac{dc^i}{dt}+\frac{dc^j}{dt}
\frac{\partial\xi^i}{\partial x^j}+\frac{\partial\xi^i}{\partial t}.
\end{eqnarray}
Hence,
\begin{eqnarray}
\Delta u^i = \frac{d\xi^i}{dt},
\label{deltavi}
\end{eqnarray}
an equation that is exact to all orders in $\xibf$
\footnote{\label{Lie}
Note that we have
defined the Lagrangian perturbation of the vector $v^i$ by subtracting the
components at two locations separated by $\xibf$. This procedure is valid
for Cartesian vector components in flat spacetime. We can derive formulas
valid in arbitrary coordinate systems by working first in Cartesian
coordinates, and then rewriting all partial derivatives as covariant
derivatives at the end.

Friedman and Schutz \cite{1978ApJ...221..937F}
have suggested defining the Lagrangian perturbation
by mapping the perturbed vector back to the original location so that
vectors are subtracted only at the same point (Lie derivative). This gives
different intermediate values for the Lagrangian perturbations of various
quantities than the procedure followed here.
However, when the final equations of motion are written
in terms of $\xibf$, the results are the same since one is using the same
definition of $\xibf$. While the Lie derivative is essential in relativistic
applications, the present approach is simpler here.} (see also Ref.\
\cite{lbo}).


An expression for $\Delta \rho$ follows from conservation of mass.
The variation in the total mass due to the perturbation is
\begin{eqnarray}
	\delta M &=& \int_{V+\Delta V}\rho(\xbf)\,d^3x -
		\int_{V}\rho_0(\xbf)\,d^3x \nonumber \\
		& =& \int_V [\rho(\xbf+\xibf)J-\rho_0(\xbf)]\,d^3x.\label{delM}
\end{eqnarray}
Here we have made the transformation $\xbf'=\xbf-\xibf(\xbf)$ in the first
integral to make the volumes the same, and then dropped the primes.
The Jacobian of the transformation is
\begin{eqnarray}
	J=\frac{\partial(\xbf+\xibf)}{\partial(\xbf)}
		=\det(\delta^i_j+L^i{}_j),\label{eq:1}
\label{Jdef}
\end{eqnarray}
where
\begin{eqnarray}
	L^i{}_j=\frac{\partial\xi^i}{\partial x^j}.\label{eq:2}
\label{Ldef}
\end{eqnarray}
Setting $\delta M= 0$ and substituting
$\rho(\xbf+\xibf)=\rho_0(\xbf)+\Delta \rho$ in Eq.~(\ref{delM}), we get
\begin{eqnarray}
	\frac{\Delta \rho}{\rho}=\frac{1-J}{J},\label{eq:3}
\end{eqnarray}
exact to all orders in $\xibf$. Keeping terms through second
order from Eq.~(\ref{eq:15}), we find
\begin{eqnarray}
\label{delrho}
	\frac{\Delta \rho}{\rho}=-\nabla_i\xi^i+\half[(\nabla_i\xi^i)^2+
		\nabla_i\xi^k\nabla_k \xi^i].
\end{eqnarray}


Consider an equation of state 
$p=p(\rho,\mu)$ where $\mu$ denotes quantities such as entropy
or composition.
For adiabatic perturbations in which the composition does not have time
to change $\Delta \mu = 0$, so to second
order
\begin{eqnarray}
	\Delta p = \left.\frac{\partial p}{\partial \rho}\right|_\mu
		\Delta \rho +
		\frac{1}{2}\left.\frac{\partial^2 p}
		{\partial \rho^2}\right|_\mu (\Delta \rho)^2.\label{eq:5}
\end{eqnarray}
In terms of the adiabatic index governing the perturbations,
\begin{eqnarray}
	\Gamma_1=\left.\frac{\partial\ln p}{\partial\ln\rho}\right|_\mu,
			\label{eq:6}
\end{eqnarray}
we get
\begin{eqnarray}
\label{delp}
	\frac{\Delta p}{p} = \Gamma_1 \frac{\Delta \rho}{\rho}
			 + \frac{1}{2}\left[
		\Gamma_1(\Gamma_1-1)+\rho\frac{\partial \Gamma_1}
			{\partial\rho}\right]\left(
			\frac{\Delta \rho}{\rho}\right)^2.
\end{eqnarray}
Here $\Delta\rho$ need be evaluated only to first order in the last term.


To derive the equation of motion for the perturbations, we need to perturb
various derivatives of quantities. Consider for example $\partial Q /
\partial t$. By Eq.~(\ref{delQ}),
\begin{eqnarray}\label{Deldel2}
	\Delta\frac{\partial Q}{\partial t} =  
		\delta \frac{\partial Q}{\partial t}
 		+ \xi^i\nabla_i\delta \frac{\partial Q}{\partial t} 
		+\xi^i\nabla_i
		\frac{\partial Q}{\partial t} 
               	+ \half \xi^i \xi^j \nabla_i \nabla_j 
		\frac{\partial Q}{\partial t}.
\end{eqnarray}
The Eulerian perturbation operator commutes with partial differentiation:
\begin{eqnarray}
	\delta \frac{\partial}{\partial t}=
			\frac{\partial}{\partial t}\delta ,\label{eq:7}
\end{eqnarray}
so the first term on the right-hand side of Eq.~(\ref{Deldel2}) is
\begin{eqnarray}
	\frac{\partial}{\partial t}\delta Q 
		= \frac{\partial}{\partial t}\left(
         	\Delta Q - \xi^i\nabla_i\delta Q -\xi^i\nabla_i Q
               - \half \xi^i \xi^j \nabla_i \nabla_j Q\right),\label{eq:8}
\end{eqnarray}
where we have used Eq.~(\ref{delQ}) again. Thus
\begin{eqnarray}\label{Deldel}
	\Delta\frac{\partial Q}{\partial t} 
		=  \frac{\partial}{\partial t} \Delta Q
  		-\frac{\partial\xi^i}{\partial t}\left(\nabla_i\delta Q 
		+\nabla_i Q
               + \xi^j \nabla_i \nabla_j Q\right).
\end{eqnarray}
Here we have used the symmetry of $\nabla_i\nabla_j$ to rewrite the last term
without the factor of $1/2$.
An alternative form of Eq.~(\ref{Deldel}) follows from substituting
Eq.~(\ref{1storder}):
\begin{eqnarray}\label{Deldel3}
	\Delta\frac{\partial Q}{\partial t} 
		=  \frac{\partial}{\partial t} \Delta Q
  	-\frac{\partial\xi^i}{\partial t}\left(\nabla_i\Delta Q +\nabla_i Q
               - \nabla_i \xi^j \nabla_j Q\right).
\end{eqnarray}
A similar equation holds if $\partial/\partial t$ is replaced by $\partial
/\partial x^i$, and then we can let $\partial /\partial x^i \to \nabla_i$
to get a tensor equation valid in any coordinate system.


For the perturbation of the acceleration, we have
\begin{eqnarray}
	\Delta \frac{du_i}{dt}&=&\Delta\left(\frac{\partial}{\partial t} 
			+ u^k\nabla_k
		\right)u_i \nonumber \\
		& = &\Delta \frac{\partial}{\partial t}u_i 
		+ u^k\Delta \nabla_k u_i + \Delta u^k
		\nabla_k u_i + \Delta u^k\Delta \nabla_k u_i.
\label{acc1}
\end{eqnarray}
Note that our formulas hold in non-Cartesian coordinates
whether we perturb the covariant components
$u_i$ or the contravariant components $u^i$, for the reason explained in
footnote \ref{Lie}.
Commute the operators $\Delta \partial/\partial t$ and $\Delta \nabla_k$ in
Eq.~(\ref{acc1}) using Eq.~(\ref{Deldel}) and the corresponding equation for
$\nabla_k$. Many terms cancel, and we are left with
\begin{eqnarray}
\label{acc2}
	\Delta \frac{du_i}{dt}=\frac{d^2 \xi_i}{dt^2}.
\end{eqnarray}
For a time-independent unperturbed star $\partial u^i/\partial t = 0$, and so
\begin{eqnarray}
	\Delta \frac{du_i}{dt}& =& \left(\frac{\partial}{\partial t} 
		+ u^j\nabla_j
	\right) \left(\frac{\partial}{\partial t} + u^k\nabla_k \right)\xi_i
	\nonumber \\
	& = &\frac{\partial^2 \xi_i}{\partial t^2}+2u^j\nabla_j
		\frac{\partial \xi_i}{
	\partial t} + u^j\nabla_j(u^k\nabla_k\xi_i).
	\label{acc3}
\end{eqnarray}


Using Eq.~(\ref{delQ}) and commuting $\delta$ and $\nabla_i$, we get
\begin{eqnarray}
\label{phi1}
	\Delta\nabla_i\phi=  \nabla_i\delta \phi 
		+ \xi^j\nabla_j\nabla_i\delta \phi
	+\xi^j\nabla_j \nabla_i\phi
               + \half \xi^j \xi^k \nabla_j \nabla_k \nabla_i\phi.
\end{eqnarray}
The Eulerian perturbation is found from Poisson's equation:
\begin{eqnarray}
	\nabla^2\phi = 4\pi G\rho \qquad  \Longrightarrow \qquad 
		\nabla^2\delta\phi  =
		4\pi G \delta \rho,\label{eq:9}
\end{eqnarray}
which implies
\begin{eqnarray}
\label{phi2}
	\delta\phi  = -G\int\frac{\delta\rho'}{|\xbf-\xbf'|} \, d^3 x'.
\end{eqnarray}
We get an expression for $\delta \rho$ by substituting Eq.~(\ref{delrho})
in Eq.~(\ref{delQ2}):
\begin{eqnarray}
	\delta \rho &=&
		-\rho\nabla_i\xi^i+\half\rho[(\nabla_i\xi^i)^2+
	\nabla_i\xi^k\nabla_k \xi^i]\nonumber\\
 	& &+ \xi^i\nabla_i(\rho\nabla_j\xi^j) -\xi^i\nabla_i \rho
           + \xi^i\nabla_i \xi^j\nabla_j \rho
               + \half \xi^i \xi^j \nabla_i \nabla_j \rho.
\end{eqnarray}
After some algebra, we can simplify this to \footnote{Note that the
second term in Eq.\ (\ref{phi3}) typically includes a
$\delta$-function at the stellar surface coming from the second
derivative of the density.  This $\delta$-function is physical and
should be included; it corrects for the fact that the expression
(\ref{phi3}) vanishes outside the surface of the unperturbed star
whereas in reality $\delta \rho$ can be nonvanishing there.}
\begin{eqnarray}
\label{phi3}
	\delta \rho = -\nabla_i(\rho\xi^i) +
		\half\nabla_i\nabla_j(\rho\xi^i\xi^j).
\end{eqnarray}


We have
\begin{eqnarray}
	\Delta\left(\frac{1}{\rho}\nabla_i p\right) & =& 
	\Delta\left(\frac{1}{\rho}\right)\nabla_i p +
	\frac{1}{\rho}\Delta\nabla_i p +
	\Delta\left(\frac{1}{\rho}\right)\Delta\nabla_i p \nonumber \\
	& =& \left[-\frac{1}{\rho^2}\Delta \rho +
		 \frac{1}{\rho^3}(\Delta \rho)^2
	\right]\nabla_i p +
	\frac{1}{\rho}\Delta\nabla_i p -
	\frac{1}{\rho^2}\Delta \rho\Delta\nabla_i p.
\end{eqnarray}
Substitute Eq.~(\ref{Deldel3}) with $\nabla_i p$ replacing $\partial Q/\partial
t$:
\begin{eqnarray}
\label{p1}
	\rho\Delta\left(\frac{1}{\rho}\nabla_i p\right)
	& =& \left[-\frac{\Delta \rho}{\rho} + \left(\frac{\Delta \rho}{\rho}
	\right)^2
	\right]\nabla_i p +
	\nabla_i \Delta p\nonumber\\
	& -&
	\nabla_i \xi^j(\nabla_j\Delta p 
		-\nabla_j\xi^k\nabla_k p +\nabla_j p)\nonumber \\
	&\qquad &-\frac{\Delta \rho}{\rho}(\nabla_i\Delta p 
		-\nabla_i\xi^j\nabla_j p)
\end{eqnarray}
Recall the definitions of the tensors $\Theta^j{}_i$ and $\Xi^j{}_i$
and their traces from Eq.~(\ref{thetaxipsidef}).
Their divergences satisfy the identity
\begin{eqnarray}
\label{identity}
	\nabla_j\Xi^j{}_i - \nabla_j\Theta^j{}_i = \half\nabla_i(\Xi - \Theta).
\end{eqnarray}
Substitute  Eqs.~(\ref{delrho}) and (\ref{delp}) in Eq.~(\ref{p1}) to get
\begin{eqnarray}
\label{p2}
	\rho\Delta\left(\frac{1}{\rho}\nabla_i p\right)
	& = &\nabla_j \xi^j\nabla_i p -\nabla_i\xi^j\nabla_j p 
		-\nabla_i(p\Gamma_1
		\nabla_j\xi^j) \nonumber\\
	& \quad& +\half(\Theta - \Xi)\nabla_i p
	\nonumber\\
		&+&\nabla_i\left[p\Gamma_1\half(\Theta+\Xi)+\half p\Theta
		\left(\Gamma_1^2
	-\Gamma_1 + \rho\frac{\partial\Gamma_1}{\partial\rho}\right)\right]
 		\nonumber\\
	& &+ (\Xi^k{}_i-\Theta^k{}_i) \nabla_k p 
	+\nabla_i\xi^j\nabla_j(p\Gamma_1\nabla_k\xi^k)
	-\nabla_j\xi^j\nabla_i(p\Gamma_1\nabla_k\xi^k)\nonumber
\end{eqnarray}

It will be convenient to rewrite the second order terms on the
right-hand side of this equation as a divergence. (The first order terms
can also be written as a divergence.)
The last two terms are
\begin{eqnarray}
	\nabla_i\xi^j\nabla_j(p\Gamma_1\nabla_k\xi^k)
	-\nabla_j\xi^j\nabla_i(p\Gamma_1\nabla_k\xi^k)
	& =& \nabla_j(\nabla_i\xi^j p\Gamma_1\nabla_k\xi^k)
	\nonumber\\
		&&-\nabla_i(\nabla_j\xi^j p\Gamma_1\nabla_k\xi^k) \nonumber \\
	& =& \nabla_j(p\Gamma_1\Theta^j{}_i)-\nabla_i(p\Gamma_1\Theta).
\end{eqnarray}
Also, by Eq.~(\ref{identity}) we have
\begin{eqnarray}
	(\Xi^k{}_i-\Theta^k{}_i) \nabla_k p & =& 
		\nabla_k[p(\Xi^k{}_i-\Theta^k{}_i)]
	-p\nabla_k(\Xi^k{}_i-\Theta^k{}_i) \nonumber \\
	&=& \nabla_k[p(\Xi^k{}_i-\Theta^k{}_i)] - \half p\nabla_i(\Xi-\Theta).
\end{eqnarray}
Thus Eq.~(\ref{p2}) becomes
\begin{eqnarray}
\label{p3}
	\rho\Delta\left(\frac{1}{\rho}\nabla_i p\right)
		& =& \nabla_j \xi^j\nabla_i p -\nabla_i\xi^j\nabla_j p 
		-\nabla_i(p\Gamma_1
	\nabla_j\xi^j)\nonumber \\
	& \quad& + \nabla_j\Bigl\lbrace
	p(\Gamma_1-1)\Theta^j{}_i + \half\delta^j{}_i
		p\Bigl[(\Gamma_1-1)^2 
	+ \rho
	\frac{\partial\Gamma_1}{\partial\rho}\Bigr]\Theta\nonumber\\
		& &
		+ p\Xi^j{}_i + \half\delta^j{}_i p(\Gamma_1-1)\Xi\Bigr\rbrace.
\end{eqnarray}


Multiplying Eq.~(\ref{eqmo}) by $\rho$, and using eqs.~(\ref{acc2}), (\ref{p3}),
and (\ref{phi3}), we obtain the equation of motion for the perturbation
to second order:
\begin{eqnarray}
\label{final}
	\rho\frac{d^2 \xi_i}{dt^2}& =& -\nabla_j \xi^j\nabla_i p +
	\nabla_i\xi^j\nabla_j p +\nabla_i(p\Gamma_1 \nabla_j\xi^j)
	-\rho\nabla_i\delta^{(1)} \phi -\rho\xi^j\nabla_j \nabla_i\phi
		\nonumber\\ 
	& & - \nabla_j\Bigl\lbrace
	p(\Gamma_1-1)\Theta^j{}_i + \half\delta^j{}_i
	p\Bigl[(\Gamma_1-1)^2 + \rho
	\frac{\partial\Gamma_1}{\partial\rho}\Bigr]\Theta\nonumber\\
 	& &+ p\Xi^j{}_i + \half\delta^j{}_i p(\Gamma_1-1)\Xi\Bigr\rbrace 
	-\rho \nabla_i\delta^{(2)} \phi\nonumber\\ 
		& &-\rho \xi^j\nabla_j\nabla_i\delta^{(1)} \phi
         - \half \rho \xi^j \xi^k \nabla_j \nabla_k \nabla_i\phi.
\label{sauleq}
\end{eqnarray}
Here we have split the quantity $\delta\phi$ defined by eqs.~(\ref{phi2})
and (\ref{phi3}) into first and second order pieces $\delta^{(1)} \phi$
and $\delta^{(2)} \phi$. Accordingly the first line on the right-hand
side of Eq.~(\ref{final}) gives the first order terms in the equation
of motion, while the remaining lines give the second order terms.

Finally, we note that the above derivation was carried out in the
inertial frame.  To obtain the equation of motion in the rotating
frame, we replace the inertial-frame Euler equation (\ref{eqmo})
at the start of the calculation by the rotating-frame Euler equation
(\ref{euler0}).  This involves adding to the left hand side the
term
\beq
2 {\bf \Omega} \times {\bf u},
\label{extraterm}
\endeq
and replacing the background potential $\phi$ by $\phi + \phi_{\rm
rot}$, where the centrifugal potential $\phi_{\rm rot}$ is given by
Eq.\ (\ref{phirotdef}).  Taking the variation of the term
(\ref{extraterm}) using Eq.\ (\ref{deltavi}) yields
\beq
2 {\bf \Omega} \times {d \xivec \over d t},
\endeq
which should be added to the left hand side of Eq.\ (\ref{sauleq}),
and substituting $\phi \to \phi + \phi_{\rm rot}$ on the right hand
side results in the additional term $-\rho \xi^j \nabla_j \nabla_i
\phi_{\rm rot}$ on the right hand side.
In the application of this paper, the background velocity in the
rotating frame vanishes, and so we can replace $d \xivec / d t$ with
$\partial \xivec / \partial t$ and $d^2 \xivec / d t^2$ with
$\partial^2 \xivec / \partial t^2$.  With these modifications, Eq.\
(\ref{sauleq}) reduces to the result quoted in the body of the paper,
given by Eqs.~(\ref{Bdef}), (\ref{Cdefa}), (\ref{basicnonlin0}),
(\ref{basicnonlin1}), (\ref{eq:17}) and (\ref{eq:16}).

\section{Variational Derivation of Second Order Equations of Motion}
\label{variational}

Derivations of the hydrodynamical equations directly from a variational
principle have a long history
\cite{eckart,herivel,yourgrau,schutzsorkin}, and have been applied to
perturbation theory for nonrotating stars (e.g. 
Ref. \cite{1989ApJ...342..558K}).
Hydrodynamical equations can be derived from a Lagrangian viewpoint
by extremizing the action
\be
S=\int{dt\,\lag (t)}=\int{dt\,[T(t)-V(t)]},
\label{action0}
\ee
where the Lagrangian is the difference between kinetic and potential
energies for the fluid. Included in the potential energy are both
the gravitational potential (both the self potential and external
potentials) and the thermal energy. 
The specific variational principle we use here is the following
\cite{yourgrau}.  Label fluid elements by their positions ${\bf y}_0$ at
some initial time $t = t_0$, and let the position of fluid element
${\bf y}_0$ at time $t$ be
$
{\bf y}({\bf y}_0,t).
$
Let $\rho_0({\bf y}_0)$ be the mass density at time $t_0$, then the
element of mass in the fluid is
$
d M = \rho_0 d^3 y_0 = \rho d^3 y
$
where $\rho = \rho_0 /J$ and $J$ is the Jacobian 
\beq
J = {\rm det} \left( {\partial y_i \over \partial y_{0\,j}} \right)
\ee
Then the kinetic energy is
\beq
T = {1 \over 2} \int dM \, \left( {\partial {\bf y} \over \partial t}
\right)_{{\bf y}_0}^2,
\label{ke0}
\endeq
and the potential is
\beq
V = \int dM \, {\cal E}(\rho) - {G \over 2} \int d^3 y_0 \int d^3
y_0^\prime \rho_0({\bf y}_0) \rho_0({\bf y}_0^\prime) { 1 \over \left|
{\bf y}({\bf y}_0,t) - {\bf y}({\bf y}_0^\prime,t) \right| }.
\label{pot0}
\endeq
Here $\rho$ is understood to be the functional of ${\bf y}({\bf y}_0,t)$
given by $\rho = \rho_0 / J$, and ${\cal E}$ is the internal energy
per unit mass of the fluid, related to the equation of state by $d
{\cal E} / d \rho = p(\rho) / \rho^2$.  With these definitions, varying
the action (\ref{action0}) with respect to ${\bf y}({\bf y}_0,t)$ gives
the Euler equation \cite{yourgrau}.

We can generalize the above variational principle slightly by allowing
the energy per unit mass to depend on additional hydrodynamic
variables such as specific entropy or composition.  We denote the set
of such variables by the vector ${\vec \mu}$, so that
\beq
{\cal E} = {\cal E}(\rho,{\vec \mu}).
\endeq
In this case, the action principle (\ref{action0}) does not determine
the evolution of the variables ${\vec \mu}$.  However, if we restrict
attention to situations in which the variables ${\vec \mu}$ are
determined by the function ${\bf y}({\bf y}_0,t)$,
then the action (\ref{action0}) does determine the motion of the
fluid.  An example of such a situation is where each fluid element
conserves its values of ${\vec \mu}$; see below.

We now apply this principle to perturbations of a uniformly rotating
star.  We denote the background quantities by $\rho_0$, $p_0$ and
$\phi_0$; these were denote by $\rho$, $p$ and $\phi$ in the body of
the paper.  We can reformulate the variational principle (\ref{action0})
by transforming to the co-rotating frame.  Define ${\bf
y}_0 = {\bf U}(t) \cdot {\bf x}_0$ and ${\bf y} = 
{\bf U}(t) \cdot {\bf x}$, where ${\bf U}(t)$ is the appropriate orthogonal
rotation matrix.  The form of the piece (\ref{pot0}) of the action
does not change under this transformation, so we merely have to
transform the kinetic energy (\ref{ke0}).  
In the rotating frame, the position at time $t$ of a mass
element with comoving coordinate $\xzero$ is
\be
\xvec=\xvec(\xzero,t)\equiv\xzero+\xivec(\xzero,t).
\label{xdef} 
\ee
[Note that the $\xzero$ of this appendix was denoted ${\bf x}$ in the
body of the paper.]  The position as seen in the inertial frame in
which the star is 
at rest is
\be
y_i=U_{ij}(t)[x_{0\,j}+\xi_j(\xvec_0,t)].
\label{eq:J7}
\ee
The inertial-frame fluid velocity is found by differentiating this
equation with respect to time:
\be
u_i={dU_{ij}\over dt}[x_{0\,j}+\xi_j(\xvec_0,t)]
+U_{ij}(t) {\partial \xi_j(\xvec_0,t) \over\partial t}.
\ee
Using the identity
$
{dU_{ij}/dt}=\epsilon_{ipq}\Omega_pU_{qj}(t)
$,
we find
\be
{\bf U}^{-1}(t) \cdot {\bf u} = {\bf \Omega} \times \left[ {\bf x}_0 +
\xivec(\xvec_0,t) \right] + {\partial \xivec(\xvec_0,t) \over \partial t},
\label{eq:J9}
\ee
where we have used $\Omega_m=U_{pm}(t)\Omega_p$.
This yields
\be
\vert\uvec\vert^2=
\vert\Omvec\times[\xzero+\xivecofxt]\vert^2
-2[\xzero+\xivecofxt]\cdot
\biggl(\Omvec\times{\partial\xivecofxt\over\partial t}
\biggr)+\biggl\vert{\partial\xivec\over\partial t}
\biggr\vert^2.
\ee
To find the equations of motion for the perturbations of
the star, we can dispense with all terms in the Lagrangian
that are either zeroth or first order in $\xivec$ and its
time derivatives\footnote{Note that there is a term proportional
to $\dot\xivec$ in $\vert\uvec\vert^2$, but varying the action
with respect to this term does not contribute to the equations
of motion. This is because the result of varying this term is
a total time derivative.} We will denote with subscripts 2 quantities
from which the zeroth and first order pieces in $\xivec$ have been
subtracted.  Thus, the relevant piece of the kinetic energy is
\be
\ttwo=\int dM{ \left[{1\over 2}\biggl\vert{\partial\xivec\over\partial t}
\biggr\vert^2+{1\over 2}[\Omega^2\vert\xivec\vert^2
-(\Omvec\cdot\xivec)^2]-
\xivec\cdot\biggl(\Omvec\times{\partial\xivecofxt\over\partial t}
\biggr) \right] }.
\label{kin3}
\ee
We also note that we can define the canonical momentum
\be
\pivec={\partial\kin_2\over\partial(\partial\xivec/\partial t)}=
{\partial\xivec\over\partial t}+
\Omvec\times\xivec,
\ee
where $\kin_2$ is the integrand in Eq.\ (\ref{kin3}),
and that 
\be
\ttwo-\pivec\cdot{\partial\xivec\over\partial t}
=-{\vert\pivec\vert^2\over 2}+\pivec\cdot(\Omvec\times\xivec)
=-{\vert\pivec-\Omvec\times\xivec\vert^2\over 2}
+{\vert\Omvec\times\xivec\vert^2\over 2}.
\ee
A Hamiltonian for the system is thus given by
\be
H_2 = {1 \over 2} \int{dM\,\biggl[{\vert\pivec
-\Omvec\times\xivec\vert^2}
-{\vert\Omvec\times\xivec\vert^2}\biggr]} + V_2(t),
\ee
where $V_2$ is the potential (\ref{pot0}) with zeroth order and first
order terms dropped, which only
depends on $\xivec$, $\grad_0\xivec$ and the background star,
but not on $\pivec$.  Hamilton's equations then yield the equations of
motion 
\baray
\dot\xivec&=&\pivec-\Omvec\times\xivec\nonumber\\
\dot\pivec&=&{\fvec\over\rho_0}-\Omvec\times\pivec,
\label{hameqs}
\earay
where the accelerations $\fvec$ is determined from the variation of
the potential $V_2$. In the linear approximation,
these results are equivalent to the first six equations in
Appendix A, with no external force. We can combine Eqs.~(\ref{hameqs})
into the single equation
\be
{\partial^2\xivec\over\partial t^2}+2\Omvec\times{\partial\xivec\over\partial t}
={\fvec\over\rho_0}-\Omvec\times(\Omvec\times\xivec).
\label{eqa22}
\ee

It remains to compute the variations of the potential, $V_2$, which
contains two parts, internal energy and gravitational energy.
The internal energy is
\be
\Ugh=\int{dM\,\eperm(v,\muvec)}
\ee
where $v$ is the volume per unit mass.  (We express $\eperm$ here as a
function of $v$ instead of $\rho = 1/v$).  As explained above, 
$\muvec$ is a set of variables such as the specific entropy and
composition of the fluid. The volume per unit mass in the fluid is
related to its 
value in the unperturbed state, $v_0$, by
\be
v=\Jxxzero v_0,
\ee
where $\Jxxzero=\det(1+L)$, with $L^i_j=\delta_{ij}+\partial\xi^i/\partial
x_{0,j}$ [cf.~Eqs.~(\ref{Jdef}) and (\ref{Ldef})].
The Jacobian $J$ is given by the exact formula
\be
	\Jxxzero = 1+\covdel\cdot\xivec +\half\lpar \Theta
			-\Xi\rpar
		+\sixth(\covdel\cdot\xivec)\lpar\Theta
			 -3
			\Xi\rpar +\third\chi\label{eq:15}
\ee
where $\Theta$, $\Xi$ and $\chi$ are the traces of the tensors
defined in Eq.\ (\ref{thetaxipsidef}).
This formula is essentially just the characteristic equation for the
$3\times 3$ matrix $\xiuilcdj$.

To go further, we need to specify how $\muvec$ changes in the presence
of fluid displacements.  We assume that for each fluid element
$\muvec$ remains frozen at its 
background value $\muvec_0$ \footnote{More generally, we could assume
that $\muvec$ changes only in response to gradients in $\xivec$, and,
since the various components of $\muvec$ are scalars, presumably
would depend only on scalar 
quantities that can be constructed from $\grad_0\xivec$.}. This means
that we focus on perturbations that preserve for each fluid element
the composition, specific entropy, etc. of the unperturbed star.
We are essentially assuming 
that the timescales for the fluid to relax to local composition are
relatively long.\footnote{To do better, we could introduce equations
of heat transport, composition and so on.}

Although it is easy to subtract the contributions to $\Ugh$ that are
zeroth and first order in $\xivec$, for computing its variation it is
easier to work with $\Ugh$ rather than $\Ugh_2$, and subtract off the
displacement field through $\Jxxzero$ only. The result is
\baray
\delta\Ugh&=&
-\int{dM\,v_0p[v_0\Jxxzero,\muvec_0]\delta\Jxxzero}\nonumber\\
&=&-\int{\volzero\,p[v_0\Jxxzero,\muvec_0]\delta\Jxxzero},
\earay
where $p(v)=-\partial\eperm(v,\muvec)/\partial v$ is the pressure.
To simplify further, we substitute
$
\delta\Jxxzero=
\Jxxzero(1+L)^{-1}_{ji}
\partial\delta\xi_i/\partial{x}_{0\,j}$,
and then integrate by parts, and impose the surface boundary
condition $p=0$; after subtracting the pressure gradient force
in the background star, the result is
\begin{eqnarray}
\delta\Ugh_2&=&
\int{\volzero\,\delta\xi_i{\partial\over\partial{x}_{0\,j}}
\biggl[p[v_0\Jxxzero,\muvec_0]\Jxxzero
(1+L)^{-1}_{ji}-\delta_{ij}p(v_0,\muvec_0)\biggr]}
\label{delU}
\end{eqnarray}
Note the exact relation
\be
J(1+L)^{-1}_{ji}=[1+\covdel\cdot\xivec+\textstyle{1\over 2}(\Theta - \Xi)]
\delta_{ji} - \xi_{j;i} + \Xi_{ji} - \Theta_{ji}.
\ee
(For a derivation, see e.g. Ref.\ \cite{silva}.)
Expanding the quantity in square brackets in Eq.~(\ref{delU})
to second order, we find
\baray
& &p[v_0\Jxxzero,\muvec_0]\Jxxzero
(1+L)^{-1}_{ji}-\delta_{ij}p(v_0,\muvec_0)=
p_0(v_0,\muvec_0)\biggl\{
\delta_{ij}(1-\Gamma_1)\grad_0\cdot\xivec-{\partial\xi_j\over\partial
x_{0\,i}}
\nonumber\\& &
+{\delta_{ij}\over 2}\biggl[\Theta\biggl((\Gamma_1-1)^2+\rho_0{\partial
\Gamma_2\over\partial\rho_0}\biggr)+(\Gamma_1-1)\Xi\biggr]
+(\Gamma_1-1)\Theta^j_i+\Xi^j_i+\cdots \biggr\},
\label{ugheq}
\earay
which gives the same result, to second order, as
Eq.~(\ref{sauleq}). Here, we have defined a generalized version of the
adiabatic index of the perturbation, 
\be
\Gamma_1=-\biggl({\partial\ln p_0(v_0,\muvec_0)\over\partial\ln v_0}\biggr)_{\muvec_0}
=\biggl({\partial\ln p_0(\rho_{0},\muvec_0)\over\partial\ln \rho_0}\biggr)_{\muvec_0},
\label{gammaonedef}
\ee
which also applies in a zero entropy star, for perturbations whose composition,
and possibly other characteristics, remain fixed at their unperturbed values.
It is straightforward
to extend the expansion to any desired order, and also to expand
the internal energy per unit mass, $\eperm(v_0J,\muvec_0)$, to any
desired order.

Finally, we need to consider the gravitational potential energy,
\be
\Vgh=-{G\over 2}\int{\dM\,\dMp\over\vert\xzero-\xzerop
+\xivec(\xzero,t)-\xivec(\xzerop,t)\vert}.
\ee
Varying with respect to the displacement field gives
\be
\delta\Vgh={G\over 2}\int{\dM\,\dMp\,
[\delta\xivec(\xzero,t)-\delta\xivec(\xzerop,t)]_i
{[\xzero-\xzerop+\xivec(\xzero,t)-\xivec(\xzerop,t)]_i\over
\vert(\xzero-\xzerop+\xivec(\xzero,t)-\xivec(\xzerop,t)
\vert^3}};
\ee
by switching primed and unprimed variables, and subtracting off the
zeroth order gravitational force, we find
\be
\delta{\Vgh}_2=G\int{\dM\,\delta\xivec(\xzero,t)}\cdot
\int{\dMp\,\biggl[
{\xzero-\xzerop+\xivec(\xzero,t)-\xivec(\xzerop,t)\over
\vert(\xzero-\xzerop+\xivec(\xzero,t)-\xivec(\xzerop,t)
\vert^3}-{\xzero-\xzerop\over\vert\xzero-\xzerop\vert^3}\biggr]}.
\ee
Gathering results, we see that the acceleration in
Eq.(\ref{hameqs}), computed to {\it any} desired order in the
amplitude of the Lagrangian perturbation field, $\xivecofxt$,
is $\fvec/\rho_0=\fvec_P/\rho_0+\fvec_G/\rho_0$, where
\baray
\label{forces0}
{\fvec^i_P}&=&
-{\partial\over\partial x_{0,j}}\biggl[
p[v_0\Jxxzero,\muvec_0]\Jxxzero
(1+L)^{-1}_{ji}-\delta_{ij}p(v_0,\muvec_0)\biggr] 
\\ 
{\fvec_G\over\rho_0}&=&-G\int{\dMp\,\biggl[
{\xzero-\xzerop+\xivec(\xzero,t)-\xivec(\xzerop,t)\over
\vert(\xzero-\xzerop+\xivec(\xzero,t)-\xivec(\xzerop,t)
\vert^3}-{\xzero-\xzerop\over\vert\xzero-\xzerop\vert^3}\biggr]}.
\label{forces}
\earay

Equations (\ref{ugheq}) and (\ref{forces0}) allow us to calculate
$\fvec_P$ to second order in the perturbation; the result is Eq.~(\ref{eq:16}).
Similarly, let us expand $\fvec_G=\fvec_G^{(1)}+\fvec_G^{(2)}
+\cdots$;
the contributions from various orders can be found by expanding
Eq.~(\ref{forces}).  To first order, the result of the calculation is that
\footnote{An integration by parts is required to get this form of the gravitational
force. The outer boundary has been taken to be outside the star, so if the mass
density goes to zero discontinuously there, a surface term must be included.}
\be
\Fi^{(1)}=-\rho_0\xij\gradi\gradj\phi_0(\xzero)
-\rho_0\gradi\delta^{(1)}\phi\ofxt,
\ee
where
\be
\delta^{(1)}\phi\ofxt=G\int{\volzerop\,{\gradjp[\rho_0(\xzerop)\xijp]
\over\vert\xzero-\xzerop\vert}}.
\ee
The second order contribution to the gravitational force is
(compare Eq.[\ref{sauleq}])
\be
\Fi^{(2)}=-{\rho_0\xivec_j\xivec_k\over 2}\gradj\gradk\gradi
\phi_0-\rho_0\xivec_j\gradj\gradi\delta\phi-\rho_0\gradi
\delta^{(2)}\phi,
\label{iraeq}
\ee
where 
\be
\delta^{(2)}\phi\equiv -{G\over 2}\int{\volzerop
\,\rho_0(\xzero)\xijp\xikp\gradjp\gradkp\vert
\xzero-\xzerop\vert^{-1}}.
\ee
It is possible to integrate by parts to change the form of
$\delta^{(2)} \phi$ to show it coincides with the expression
(\ref{eq:18aa}).

\section{Energy and Angular Momentum of stellar perturbations}
\label{sec:eam}

In Sec.\ \ref{bfzetaexprs} of this appendix we give general formulae
for the physical energy 
$E_{\rm phys}$ and physical ($z$-component of) angular momentum
$J_{\rm phys}$ of a perturbation, and also the rotating-frame canonical energy
$E_{\rm can,rot}$
\cite{1978ApJ...221..937F}.      
General expressions for these quantities have been derived by Friedman
and Schutz \cite{1978ApJ...221..937F}, in a context more general than that
considered here (uniform rotation of the background star).  However,
the results of Friedman and Schutz are expressed in terms of an
inertial-frame Lagrangian displacement $\xivec_{\rm in}$.  For a
uniformly rotating background star, it is more natural and simpler to
use instead the rotating-frame Lagrangian displacement $\xivec$ as we
do in this paper.  The expressions in Sec.\ \ref{bfzetaexprs} below
can be obtained by specializing and translating the results of
Friedman and Schutz, or, as we outline below, can also be easily
obtained directly.

We also derive, in Sec.\ \ref{modeexprs}, expressions for these quantities
in terms of the mode coefficients 
$c_{A\sigma}$, and in Sec.\ \ref{sec:energydeposited} compute the
physical energy 
deposited in a star by 
an externally applied acceleration  ${\bf a}_{\rm ext}$ to second
order in ${\bf a}_{\rm ext}$.

Friedman and Schutz \cite{1978ApJ...221..937F} derived several general
properties of the quantities $E_{\rm phys}$, $J_{\rm phys}$
and $E_{\rm can,rot}$.  First, the
physical energy $E_{\rm phys}[\bfzeta]$ of a perturbation $\bfzeta$
\footnote{This is defined to be the difference between the energy of
the perturbed configuration and the energy of the background, unperturbed
configuration.} has a piece that is linear in $\bfzeta$ as well as
pieces that are quadratic and higher order in $\bfzeta$:
\be
E_{\rm phys}[\bfzeta] = E_{\rm phys}^{(1)}[\bfzeta] + E_{\rm
phys}^{(2)}[\bfzeta,\bfzeta] + O(\bfzeta^3),
\ee
and similarly for the physical angular momentum:
\be
J_{\rm phys}[\bfzeta] = J_{\rm phys}^{(1)}[\bfzeta] + J_{\rm
phys}^{(2)}[\bfzeta,\bfzeta].
\ee
Second, one has the identity
\be
E_{\rm phys}^{(1)}[\bfzeta] = \Omega J_{\rm phys}^{(1)}[\bfzeta],
\label{FSidentity}
\ee
and so the physical energy in the rotating frame, defined as
\be
E_{\rm phys,rot} \equiv E_{\rm phys} - \Omega J_{\rm phys}
\label{Erotdef}
\ee
has no linear term in $\bfzeta$:
\be
E_{\rm phys,rot}[\bfzeta] = E_{\rm phys,rot}^{(2)}[\bfzeta,\bfzeta] 
+ O(\bfzeta^3)
\label{Erotdef0}
\ee
where
\be
E_{\rm phys,rot}^{(2)}[\bfzeta,\bfzeta] = 
E_{\rm phys}^{(2)}[\bfzeta,\bfzeta] - \Omega J_{\rm
phys}^{(2)}[\bfzeta,\bfzeta].
\label{Erotdef1}
\ee
Third, this quantity coincides with the so-called rotating frame
canonical energy $E_{\rm can,rot}[\bfzeta]$ defined in
Ref.\ \cite{1978ApJ...221..937F}, that is, the conserved quantity related to
the time translation symmetry in the rotating frame by Noether's theorem.

\subsection{Explicit expressions in terms of $\bfzeta$}
\label{bfzetaexprs}

The physical energy of a perturbation can be computed using the
procedure of Appendix \ref{variational} above, and adding the kinetic
and potential energies $T(t)$ and $V(t)$.  To linear order, the result is 
\begin{eqnarray}
E_{\rm phys}^{(1)}[\bfzeta] &=& \int d^3 x \, \rho_0({\bf x}) ( {\bf
\Omega} \times {\bf x} ) \cdot \left[ {\dot \xivec} + 2 {\bf \Omega}
\times \xivec \right] \\
&=& \left< \xivec_{r} \, , \, {\dot \xivec} + {\bf B} \cdot \xivec
\right>
 = \left< \xivec_{r} \, , \, \bfpi +  {\bf B} \cdot \xivec /2
\right>,
\label{Ephys1}
\end{eqnarray}
where
\be
\xivec_r({\bf x}) = {\bf \Omega} \times {\bf x}
\label{xivecrdef}
\ee
is the mode function corresponding to uniform rotation and we have
used Eq.\ (\ref{pidef}).  The second order piece is
\begin{eqnarray}
E_{\rm phys}^{(2)}[\bfzeta,\bfzeta] 
&=& 
{1 \over 2} \left< {\dot \xivec} \, , \, {\dot \xivec} \right>
+ {1 \over 2} \left< \xivec \, , \, {\bf C} \cdot \xivec \right>
+ {1 \over 2} \left< {\dot \xivec} \, , \, {\bf B} \cdot \xivec \right>
+ {1 \over 4} \left< {\bf B} \cdot {\xivec} \, , \, {\bf B} \cdot
\xivec \right> \nonumber \\
&=& 
{1 \over 2} \left< \bfpi \, , \, \bfpi \right>
+ {1 \over 2} \left< \xivec \, , \, {\bf C} \cdot \xivec \right>
+ {1 \over 8} \left< {\bf B} \cdot \xivec \, , \, {\bf B} \cdot
\xivec \right>.
\label{Ephys2}
\end{eqnarray}
Here we have used the fact that $\xivec$ and ${\dot \xivec}$ are real
to eliminate terms such as $\left< {\dot \xivec} \, , \, {\bf B} \cdot
{\dot \xivec} \right>$.
The physical angular momentum is given by
\be
{\bf J}_{\rm phys} = \int dM \, {\bf y} \times {\bf u}({\bf y}).
\label{thisg}
\ee
Here we are using the notation of Appendix \ref{variational},
where ${\bf y}$ and ${\bf u}$ are the inertial-frame location and
velocity of a fluid element.  Using Eqs.\ (\ref{eq:J7}) and
(\ref{eq:J9}) in Eq.\ (\ref{thisg}) gives 
\be
{\bf U}(t)^{-1} \cdot {\bf J}_{\rm phys} = \int d^3 x \, \rho_0({\bf
x}) ( {\bf x} + \xivec ) \times \left[ {\bf \Omega} \times {\bf x} +
{\bf \Omega \times \xivec} + {\dot \xivec} \right],
\ee
and the $z$-component is given by
\begin{eqnarray}
\Omega J_{\rm phys}[\xivec] &=& {\bf \Omega} \cdot
{\bf J}_{\rm phys} = {\bf \Omega} \cdot {\bf U}(t)^{-1} \cdot
{\bf J}_{\rm phys} \nonumber \\
&=& \int d^3 x \, \rho_0({\bf
x}) \left[ {\bf \Omega} \times {\bf x} + {\bf \Omega} \times \xivec
\right]  \cdot \left[ {\bf \Omega} \times {\bf x} +
{\bf \Omega \times \xivec} + {\dot \xivec} \right].
\label{Jphystot}
\end{eqnarray}
Expanding this in powers of $\xivec$ one obtains at linear order using
Eq.\ (\ref{Ephys1}) the identity (\ref{FSidentity}).  At second order
we obtain
\begin{eqnarray}
\Omega J_{\rm phys}^{(2)}[\bfzeta,\bfzeta] &= &
 {1 \over 2} \left< {\dot \xivec} \, , \, {\bf B} \cdot \xivec \right>
+ {1 \over 4} \left< {\bf B} \cdot {\xivec} \, , \, {\bf B} \cdot
\xivec \right> \nonumber \\
&=& {1 \over 2} \left< \bfpi \, , \, {\bf B} \cdot \xivec \right>.
\label{Jphys2}
\end{eqnarray}
The physical energy in the rotating frame is therefore, from Eqs.\
(\ref{Erotdef1}), (\ref{Ephys2}), and (\ref{Jphys2})
\begin{eqnarray}
\label{Ephysrot2config}
E_{\rm phys,rot}[\bfzeta] 
&=& 
{1 \over 2} \left< {\dot \xivec} \, , \, {\dot \xivec} \right>
+ {1 \over 2} \left< \xivec \, , \, {\bf C} \cdot \xivec \right> +
(\bfzeta^3)  \\
&=& 
{1 \over 2} \left< \bfpi \, , \, \bfpi \right>
+ {1 \over 2} \left< \xivec \, , \, {\bf C} \cdot \xivec \right>
+ {1 \over 8} \left< {\bf B} \cdot \xivec \, , \, {\bf B} \cdot
\xivec \right>
- {1 \over 2} \left< \bfpi \, , \, {\bf B} \cdot \xivec \right> + O(\bfzeta^3).
\label{Ephysrot2}
\end{eqnarray}
The expression (\ref{Ephysrot2config}) agrees with the expression Eq.\
(43) of Ref.\ \cite{1978ApJ...221..937F} for rotating-frame canonical
energy, as expected \footnote{Equation (43) of Ref.\
\cite{1978ApJ...221..937F} is actually an expression for
inertial-frame canonical energy.  The operators which appear in that
expression, which we will denote by ${\bf A}_{\rm in}$, ${\bf
B}_{\rm in}$ and ${\bf C}_{\rm in} $, are defined such that the
linearized equation of motion is ${\bf A}_{\rm in} \cdot {\ddot \xivec}_{\rm
in} + {\bf B}_{\rm in} \cdot {\dot \xivec} + {\bf C}_{\rm in} \cdot
\xivec=0$, where $\xivec_{\rm in}$ is the inertial-frame Lagrangian
displacement.  However, one can obtain an expression for the
rotating-frame canonical energy, as defined by Eq.\ (64) of Ref.\
\cite{1978ApJ...221..937F}, by replacing in Eq. (43) the inertial
frame quantities ${\bf A}_{\rm in}$, ${\bf B}_{\rm in}$, ${\bf C}_{\rm
in}$ and $\xivec_{\rm in}$ by their rotating-frame counterparts used
in this paper, ${\bf A}$, ${\bf B}$, ${\bf C}$, and $\xivec$.  Note
that in our notational convention the operator ${\bf A}$ is the
identity operator.}. 

\subsection{Expressions in terms of mode coefficients}
\label{modeexprs}

From the above discussion we can write the total energy of the
perturbation, to quadratic order, as the sum of three terms, a
quadratic rotating-frame energy (physical or canonical), a linear
angular momentum term, and a quadratic angular momentum term:
\be
E_{\rm phys}[\bfzeta] = E_{\rm phys,rot}^{(2)}[\bfzeta,\bfzeta] +
\Omega J_{\rm phys}^{(1)}[\bfzeta] + \Omega J_{\rm
phys}^{(2)}[\bfzeta,\bfzeta] + O(\bfzeta^3).
\label{Ephystotal}
\ee
Of these three terms, the first is diagonalized by the mode basis
$\bfzeta_{A\sigma}$, the second can be expressed as the coefficient
$c_{A\sigma}$ of a single Jordan chain mode (corresponding to uniform
rotation), and the third is not diagonalized by the mode basis
$\bfzeta_{A\sigma}$ and contains cross terms between different modes.

\subsubsection{Rotating-frame energy}

Consider first the rotating-frame energy.  From Eqs.\ (\ref{zetadef}),
(\ref{Tdef}), (\ref{Mdef}) and (\ref{Ephysrot2}) we can write this
as
\be 
E_{\rm phys,rot}^{(2)}[\bfzeta,\bfzeta] = {i \over 2} \left< \bfzeta
\, , \, {\bf M} \cdot {\bf T} \cdot \bfzeta \right>.
\ee
Inserting the mode expansion (\ref{zetaexpand}) and using the definition
(\ref{rtjordan}) of a right Jordan chain and the definition
(\ref{calMdefgeneral0}) of the matrix ${\cal M}_{A\sigma,B\tau}$ gives 
\be
E_{\rm phys,rot}^{(2)}[\bfzeta,\bfzeta] = \sum_{A\sigma,B\tau} \,
c_{A\sigma}^* \, c_{B\tau} \left[ { \omega_B \over 2} {\cal
M}_{A\sigma,B\tau} + {i \over 2} {\cal M}_{A\sigma,B(\tau-1)} \right].
\ee
Since the energy is real and ${\cal M}$ is Hermitian we can rewrite this as
\be
E_{\rm phys,rot}^{(2)}[\bfzeta,\bfzeta] = {1 \over 4}
\sum_{A\sigma,B\tau} \, c_{A\sigma}^* \, c_{B\tau} \left[ (
\omega_A^* + \omega_B )  {\cal 
M}_{A\sigma,B\tau} + i {\cal M}_{A\sigma,B(\tau-1)} -
i {\cal M}_{A(\sigma-1),B\tau} \right].
\label{Erotfinalans}
\ee
Since the matrix ${\cal M}$ is block diagonal (see Sec.\
\ref{sec:generalcase} above), this rotating-frame energy can be written
as a sum of three contributions, from (i) real-frequency modes,
(ii) complex-frequency (unstable) modes, and (ii) non-trivial Jordan
chain modes.  The term corresponding to real frequencies is 
\be
{1 \over 2} \sum_A \varepsilon_A | c_A|^2,
\label{ErotrealA}
\ee
where we have assumed that one is using a basis which diagonalizes the
real-frequencies portion of the matrix ${\cal M}_{AB}$ (cf.\ Sec.\
\ref{sec:degeneracy} above), and $\varepsilon_A$ is the mode
rotating-frame-energy at unit amplitude, given 
from Eqs.\ (\ref{normalizationdef}) and (\ref{calMdef}) by
\be
\varepsilon_A =  \omega_A \, {\cal M}_{AA} = { \omega_A b_A }.
\label{varepsilondef}
\ee
Using the notations (\ref{gnotations}) we can rewrite the energy
(\ref{ErotrealA}) as
\be
\sum_\alpha \, | c_\alpha |^2 \omega_\alpha b_\alpha = 
\sum_\alpha \, | c_\alpha |^2 \varepsilon_\alpha.
\label{Erotreal}
\ee

To write down the portion of the expression (\ref{Erotfinalans})
corresponding to complex frequencies, we need to introduce some more
notation.  The modes come in pairs $(\xivec,\omega)$ and $(g_*
\xivec^*, \omega^*)$; see Secs.\ \ref{sec:Specialization-to-no} and
\ref{sec:generalcase} above. 
We write $A = ({\hat a}\delta k)$, where ${\hat a}$ and
$\delta$ label the frequency $\omega_{{\hat a}\delta k} = \omega_{{\hat
a}\delta}$, and $k$ labels the different
(degenerate) eigenvectors associated with that frequency.  The index
$\delta$ can take on the values $1$ and $-1$ such that
\begin{eqnarray}
\omega_{{\hat a}1} & = & \omega_{\hat a} \\
\omega_{{\hat a}(-1)} & = & \omega^*_{\hat a} \\
\xivec_{{\hat a}1k} &=& \xivec_{{\hat a}k} \\
\xivec_{{\hat a}(-1)k} &=& ( g_* \xivec_{{\hat a}k})^*.
\end{eqnarray}
It is possible to choose the mode basis such that the matrix ${\cal
M}$ takes the form
\begin{eqnarray}
{\cal M}_{{\hat a}\delta k,{\hat b}\epsilon l} &=&  \delta_{{\hat
a}{\hat b}} \, \delta_{\delta,-\epsilon} \, \delta_{kl} \left[
\delta_{\delta,1} {\cal D}_{{\hat a}k} + \delta_{\delta,-1} {\cal
D}_{{\hat a}k}^* \right];
\end{eqnarray}
cf.\ Eq.\ (\ref{M2def}) above.  Inserting this into the expression
(\ref{Erotfinalans}) gives for the contribution to $E_{\rm
phys,rot}^{(2)}$ from complex-frequency modes
\be
\sum_{{\hat a}} \, {\rm Re}
\, \left[ \sum_k \, \omega_{{\hat a}1} \, c_{{\hat a}1k}^* c_{{\hat
a}(-1)k} \, {\cal D}_{{\hat a}k} \right].
\label{Erotcomplex}
\ee
We see that if only a single mode is excited, the corresponding
rotating-frame energy vanishes, in agreement with the result of
Friedman and Schutz \cite{1978ApJ...221..937F}.  However, if a mode
$(\xivec,\omega)$ and also its conjugate pair $(g_* \xivec^*,
\omega^*)$ are both excited, then there is a nonzero contribution to
the rotating-frame energy.

Finally, there is a contribution to $E_{\rm phys,rot}^{(2)}$ from
non-trivial Jordan chain modes.  For simplicity, we restrict attention
here to the case of zero-frequency Jordan chains of length one.
For this case the expression (\ref{Erotfinalans}) reduces to 
\be
{i \over 2} \sum_{AB} \, {\cal M}_{A1,B0} \, c_{A1}^* c_{B1},
\label{Ecanjordan}
\ee
where we have used the identities (\ref{late1})--(\ref{late2}).  As
discussed in Sec.\ \ref{sec:jordanone} above, we can always choose the
mode basis to diagonalize the matrix 
\begin{eqnarray}
\beta_{AB} &=& i {\cal M}_{A1,B0} = i \left< \bfzeta_{A1} \, , \, \bfM
\cdot \bfzeta_{B0} \right> \\
&=& \left< \xivec_{A0} \, , \, \xivec_{B0} \right> - 
\left< \xivec_{A1} \, , \, {\bf C} \cdot \xivec_{B1} \right>,
\end{eqnarray}
and the expression (\ref{Ecanjordan}) therefore becomes
\be
{1 \over 2} \sum_A \, \beta_A \, |c_{A1}|^2,
\label{Erotjordan}
\ee
where $\beta_A = \beta_{AA}$ are the eigenvalues of the matrix
$\bfbeta$.  Note that there is no rotating-frame energy associated
with the $c_{A0}$ coefficients.   

To summarize, the total rotating-frame energy is given by the sum of
the expressions (\ref{Erotreal}), (\ref{Erotcomplex}), and
(\ref{Erotjordan}).

\subsubsection{Linear piece of angular momentum}

The linear piece of the $z$-component of the angular momentum
(proportional to the linear piece of the physical, inertial-frame
energy) can be written as the coefficient of a single Jordan-chain mode.
That Jordan-chain mode corresponds to the degree of freedom of uniform
spin-up $\Omega \to \Omega + \Delta \Omega$ of the star, and is a
zero-frequency Jordan chain of length one.  As explained
in Sec.\ \ref{sec:proofs3} above, the space of such zero-frequency,
length one Jordan-chain modes contains all the differential-rotation
modes and is infinitely degenerate.  In Sec.\ \ref{sec:generalcase} above, we
explained how to compute the basis $\bfchi_{A\sigma}$ of left Jordan
chains of this space, starting from a specified basis
$\bfzeta_{A\sigma}$ of right Jordan chains.  Here, however, it turns
out to be more convenient to specify the basis by first choosing the
basis $\bfchi_{A\sigma}$ of left Jordan chains, satisfying Eq.\
(\ref{leftjordan}), and by defining $\bfzeta_{A\sigma}$ to be the
dual basis defined by Eq.\ (\ref{orthog1}).  

The uniform-rotation mode, specified in terms of its left Jordan
chain $\bfchi_\sigma$, $\sigma=1,2$, is as follows.  Let $\xivec_r$ be
the uniform-rotation mode 
function (\ref{xivecrdef}), and define $\bftau_1$ to be any solution
to the equation
\be
{\bf C} \cdot \bftau_1 = {\bf B} \cdot \xivec_r.
\ee
Define
\begin{eqnarray}
\bfchi_0 = \left[ \begin{array}{c}  
	- {\bf B} \cdot \xivec_r /2 \\
	\xivec_r
	\end{array} \right]
\label{ur0}
\end{eqnarray}
and
\begin{eqnarray}
\bfchi_1 = \left[ \begin{array}{c}  
	\xivec_r - {\bf B} \cdot \bftau_1 /2 \\
	\bftau_1
	\end{array} \right].
\label{ur1}
\end{eqnarray}
Then one can verify using the relation ${\bf C} \cdot \xivec_r=0$ that
the left Jordan chain relations ${\bf T}^\dagger \cdot 
\bfchi_0=0$ and ${\bf T}^\dagger \cdot \bfchi_1 = {\bfchi}_0$ are
satisfied.  The basis $\bfchi_{A\sigma}$ of left Jordan chain modes 
can be completed by adding to the above chain any linearly independent
set of left Jordan chains describing the differential rotation modes. 
From Eqs.\ (\ref{cAsigmaeqn}), (\ref{Ephys1}) and (\ref{ur0}), the coefficient
$c_{\sigma}$ with $\sigma=1$ corresponding to the uniform-rotation
mode is 
\begin{eqnarray}
c_1 &=& \left< \bfchi_0 \, , \, \bfzeta \right> 
 = \left< \xivec_{r} \, , \, \bfpi +  {\bf B} \cdot \xivec /2
\right> \\
&=& E_{\rm phys}^{(1)}[\bfzeta] = \Omega J_{\rm phys}^{(1)}[\bfzeta].
\label{Ephys1ans}
\end{eqnarray}

\subsubsection{Quadratic piece of angular momentum}

From Eq.\ (\ref{Jphys2}), the quadratic piece of the angular momentum can be
written as 
\be
\Omega J_{\rm phys}^{(2)}[\bfzeta,\bfzeta] = 
\left< \bfzeta \, , \, {\bf K} \cdot \bfzeta \right>
= \sum_{A\sigma,B\tau} \, c_{A\sigma}^* c_{B\tau} {\cal K}_{A\sigma,B\tau},
\label{Jphys2a}
\ee
where
\begin{eqnarray}
{\bf K} =  {1 \over 4} \left[ \begin{array}{cc}  
	0 & - {\bf B} \\
	{\bf B} & 0
	\end{array} \right]
\label{Kdef}
\end{eqnarray}
and ${\cal K}_{A\sigma,B\tau} = \left< \bfzeta_{A\sigma} \, , \, {\bf
K} \cdot \bfzeta_{B\tau} \right>$.  For non-Jordan chain,
real-frequency modes we can write the coefficients ${\cal K}_{AB}$,
using the orthogonality relation (\ref{orthonormal1}) and Eq.\
(\ref{lid1}), as 
\begin{eqnarray}
\label{Kab}
{\cal K}_{AB} &=& 
{1 \over 4} (\omega_A + \omega_B) \left< \xivec_A \, , \, i {\bf B}
 \cdot \xivec_B \right> + {1 \over 4} 
\left< {\bf B} \cdot \xivec_A \, , \, {\bf B} \cdot \xivec_B \right>
\\
&=& {1 \over 2} \omega_A b_A \delta_{AB} - {1 \over 4}
(\omega_A + \omega_B)^2 \left< 
\xivec_A \, , \, \xivec_B \right> + {1 \over 4} \left< {\bf B} \cdot
\xivec_A \, , \, {\bf B} \cdot \xivec_B \right>.
\label{Kab1}
\end{eqnarray}
For regular modes in a rotating star, in the limit $\Omega \to 0$, the
second term in Eq.\ (\ref{Kab}) scales as $O(\Omega^2)$ and is
therefore small compared to the first term which is linear in $\Omega$.
Therefore in 
general the off-diagonal elements of ${\cal K}_{AB}$ are non
vanishing; see Eq.\ (\ref{nonorthogeg}) above.

\subsection{Energy deposited by an external force}
\label{sec:energydeposited}

Turn now to the question of how to compute the energy deposited in a
rotating star by an externally applied acceleration ${\bf a}_{\rm
ext}({\bf x},t)$, to quadratic order in ${\bf a}_{\rm ext}$.  
The rotating-frame energy can be easily computed, just as one computes
the energy deposited in a non-rotating star.  Assuming that there are
no unstable modes, one evolves the mode amplitudes using Eq.\
(\ref{fa}) or Eqs.\ (\ref{ira1})--(\ref{ira2}), and one computes the
energy from Eqs.\ (\ref{Erotreal}) and (\ref{Erotjordan}).

The angular momentum contributions to the total inertial-frame energy
(\ref{Ephystotal}) are more difficult to compute.  As shown above, the
quadratic piece of the angular momentum contains cross terms between
different modes.  Also, in evaluating the linear term $E_{\rm
phys}^{(1)}[\bfzeta]$ given by Eq.\ (\ref{Ephys1ans}), one
needs to solve the equation of motion for the coefficient $c_1(t)$ of
the uniform-rotation mode to quadratic order in ${\bf a}_{\rm ext}$,
i.e., one needs to solve for the back reaction of all the other
stellar modes on the uniform-rotation mode.  The linear equation of
motion for the coefficient $c_1(t)$ is given by combining Eqs.\
(\ref{Fdef}), (\ref{ee4}), (\ref{ur0}), which gives the equation for
angular momentum conservation to linear order:
\be
{\dot c}_1 = \left< \xivec_r \, , \, {\bf a}_{\rm ext} \right>.
\label{c1linear}
\ee
To quadratic order in ${\bf a}_{\rm ext}$ (or equivalently in
$\xivec$), one can obtain the equation of motion for $c_1(t)$ by 
using the following form of the second-order equation of motion
\begin{eqnarray}
{\ddot {\bfxi}} + {\bf B} \cdot {\dot {\bfxi}} + {\bf C} \cdot
{\bfxi}= {\bf a}^{(2)}[\xivec,\xivec] + {\bf a}_{\rm ext}({\bf x} +
\xivec,t) + O(\xivec^3),
\label{basicnonlin0a}
\end{eqnarray}
which generalizes Eqs.\ (\ref{basic0}) and (\ref{basicnonlin0}) above.
Replacing the external acceleration on the right hand side of Eq.\
(\ref{c1linear}) with the right hand side of Eq.\
(\ref{basicnonlin0a}), and using the identity 
\be
\left< \xivec_r \, , \, {\bf a}^{(2)}[\xivec,\xivec] \right> = {1
\over 2} \left< {\bf B} \cdot \xivec \, , \, {\bf C} \cdot \xivec \right>
\label{torqueidentity}
\ee
which is derived below, one obtains 
the second-order equation of motion for $c_1$ in the form
\be
{\dot c}_1 = \left< \xivec_r \, , \, {\bf a}_{\rm ext} \right> 
+ \left< \xivec_r \, , \, ( \xivec \cdot \bfnabla ) {\bf a}_{\rm ext}
\right>  + {1 \over 2} \left< {\bf B} \cdot \xivec \, , \, {\bf C}
\cdot \xivec \right> + O(\bfzeta^3).
\ee
The third term on the right hand side here describes the excitation of the
uniform-rotation mode by the other stellar modes.

It is possible to get around these difficulties in the computation of
the angular momentum deposited in the star, by computing the total
torque acting on the star, and by not attempting to compute how the
angular momentum deposited is distributed between the various stellar
modes.  The total torque is given by\footnote{If one takes a time
derivative of the expression (\ref{Jphystot}) for the total angular
momentum and uses the equation of motion (\ref{basicnonlin0a}), one
obtains the result (\ref{totaltorque}) with the two additional terms
$
\left< \xivec_r \, , \, {\bf a}^{(2)}[\xivec,\xivec] \right> - 
 \left< {\bf B} \cdot \xivec \, , \, {\bf C} \cdot \xivec \right>/2.
$
This allows one to derive the identity (\ref{torqueidentity}), which
can also be verified directly using the expressions (\ref{Bdef}),
(\ref{Cdefa}), (\ref{eq:17}) and (\ref{eq:16}).}
\begin{eqnarray}
{\bf \Omega} \cdot {\dot {\bf J}} &=& {\bf \Omega} \cdot \int d^3 x \,
\rho_0({\bf x}) \, \left[ {\bf x} + \xivec({\bf x}) \right] \times
{\bf a}_{\rm ext}({\bf x} + \xivec) \\
&=& \left< \xivec_r \, , \, {\bf a}_{\rm ext} \right> + 
\left< \, {\bf B} \cdot \xivec /2 \, , \, {\bf a}_{\rm
ext} \right> + \left< \xivec_r \, , \, ( \xivec \cdot \bfnabla) {\bf
a}_{\rm ext} \right> + O({\bf a}_{\rm ext}^3,\xivec^3).
\label{totaltorque}
\end{eqnarray}
Suppose now that the external acceleration can be expanded as
\be
{\bf a}_{\rm ext} = \varepsilon {\bf a}_{\rm ext}^{(1)} +
\varepsilon^2 {\bf a}_{\rm ext}^{(2)} + O(\varepsilon^3),
\ee
and that the response of the stellar modes is similarly expanded as
\be
c_{A\sigma}(t) = \varepsilon c_{A\sigma}^{(1)}(t) + 
\varepsilon^2 c_{A\sigma}^{(1)}(t) + O(\varepsilon^3).
\ee
The second-order angular momentum deposited can now be computed as
follows.  First, solve for the linear response $c_{A\sigma}^{(1)}(t)$
of the star.  Second, compute the angular momentum deposited due to
the first term on the right hand side of Eq.\ (\ref{totaltorque}),
which is the standard expression for torque.  Third, add to this the
angular momentum deposited by the remaining two terms in the total
torque (\ref{totaltorque}), which can be written as
\be
\varepsilon^2 \sum_{A\sigma} \, c^{(1)}_{A\sigma}(t) \, {\cal
F}_{A\sigma}(t) + O(\varepsilon^3),
\label{totaltorque1}
\ee
with
\be
{\cal F}_{A\sigma}(t) = \int d^3 x \, \rho_0({\bf x}) \,
\xivec_{A\sigma}({\bf x}) \cdot {\bfnabla} \left[ {\bf \Omega} \cdot {\bf
x} \times {\bf a}_{\rm ext}^{(1)}({\bf x}) \right].
\ee
This procedure yields a expression for the total angular momentum
deposited which is a sum over modes $(A\sigma)$.  Note, however, that
the term $A\sigma$ in the expression (\ref{totaltorque1}) is not the
angular momentum deposited into the mode $\bfzeta_{A\sigma}$; as noted
above the total angular momentum contains cross terms between
different modes.

\newcommand{\apjl}{Astrophys. J. Lett.}
\newcommand{\aap}{Astron. and Astrophys.}
\newcommand{\cmp}{Commun. Math. Phys.}
\newcommand{\grg}{Gen. Rel. Grav.}
\newcommand{\lr}{Living Reviews in Relativity}
\newcommand{\mnras}{ Mon. Not. Roy. Astr. Soc.}
\newcommand{\pr}{Phys. Rev.}
\newcommand{\prsl}{Proc. R. Soc. Lond. A}
\newcommand{\ptrsl}{Phil. Trans. Roy. Soc. London}

\begin{table}
\caption{Coupling coefficient summary for modes of zero-buoyancy
stars, to zeroth order in the star's angular velocity.  Dashes denote
coefficients which are forced to vanish by 
selection rules.  Stars denote nonzero coefficients.}
\label{tab:cdCCsum}
\begin{tabular}{l|cccc} 
   &2 pure $r$ modes& 
2 regular&2 hybrid rotational\\ [6pt]
\hline
pure $r$ mode &---& * & * \\
regular &*& * &* \\
hybrid rotational&---&*&* \\ 
\end{tabular}
\end{table}


\begin{thebibliography}{10}

\bibitem{1998ApJ...502..708A}
N. Andersson, Astrophys.\ J. {\bf 502},  708 (1998).

\bibitem{1998ApJ...502..714F}
J.~L. Friedman and S.~M. Morsink, Astrophys.\ J. {\bf 502}, 714 (1998).

\bibitem{prl...80...4843}
L. Lindblom, B.~J. Owen, and S.~M. Morsink, Phys.\ Rev.\ Lett. {\bf
80},  4843  (1998).  

\bibitem{1998ApJ...501L..89B}
L. {Bildsten}, \apjl \, {\bf 501},  L89  (1998).

\bibitem{physrevD...58...084020}
B.~J. Owen {\it et~al.}, Phys.\ Rev.\ D {\bf 58},  084020  (1998).

\bibitem{1999ApJ...516..307A}
N. Andersson, K.~D. Kokkotas, and N. Stergioulas, Astrophys.\ J. {\bf 516},
  307  (1999).

\bibitem{1999ApJ...510..846A}
N. {Andersson}, K. {Kokkotas}, and B.~F. {Schutz}, Astrophys.\ J. {\bf 510},
  846  (1999).

\bibitem{1999ApJ...517..328L}
Y. Levin, Astrophys.\ J. {\bf 517},  328  (1999).

\bibitem{astro-ph0006028}
Y. Levin and G. Ushomirsky, \mnras {\bf 324}, 917 (2001).

\bibitem{astro-ph9911188}
L. Rezzolla, F.~K. Lamb, and S.~L. Shapiro, Astrophys. J. {\bf 528},
L29 (2000).

\bibitem{gr-qc9909084}
L. Lindblom and G. Mendell, Phys. Rev. D {\bf 61}, 104003 (2000).

\bibitem{1999A&A...125..193}
H.~C. Spruit, \aap {\bf 125}, 193.

\bibitem{diffrot}
Y.\ Levin and G.\ Ushomirsky, \mnras {\bf 322}, 515 (2001).

\bibitem{anderssonreview}
N.\ Andersson and K.D.~Kokkotas, gr-qc/0010102.

\bibitem{astro-ph0006123}
Y. Wu, C.D. Matzner, and P. Arras, \apj {\bf 549}, 1011 (2001).

\bibitem{gr-qc0007086}
N. Stergioulas and J.~A. Font, Phys. Rev. Lett. {\bf 86}, 1148 (2001).

\bibitem{astro-ph0010653}
L. Lindblom, J.~E. Tohline, and M. Vallisneri, Phys. Rev. Lett. {\bf
86}, 1152 (2001).

\bibitem{prd...55..714}
L. Blanchet, Phys. Rev. D {\bf 55},  714  (1997).

\bibitem{ApJ1999...525..939}
L. Rezzolla {\it et~al.}, Astrophys. J. {\bf 525},  935  (1999).

\bibitem{Leenote}
The enhancement factor was given in the paper as $\sim 750$ but in
fact was $\sim 4500$; L. Lindblom, private communication.

\bibitem{1978ApJ...221..937F}
J.~L. {Friedman} and B.~F. {Schutz}, Astrophys.\ J. {\bf 221},  937  (1978).

\bibitem{1978ApJ...222..281F}
J.~L. Friedman and B.~F. Schutz, Astrophys.\ J. {\bf 222},  281  (1978).

\bibitem{1979ApJ...232..874S}
B.~F. {Schutz}, Astrophys.\ J. {\bf 232},  874  (1979).

\bibitem{1979RSLPS.368..389D}
J. {Dyson} and B.~F. {Schutz}, Royal Society of London Proceedings Series {\bf
  368},  389  (1979).

\bibitem{1980MNRAS.190....7S}
B.~F. {Schutz}, \mnras {\bf 190},  7  (1980).

\bibitem{1980MNRAS.190....21S}
B.~F. {Schutz}, \mnras {\bf 190},  21  (1980).

\bibitem{newmanpenrose}
E.~T. {Newman} and R. {Penrose}, J.\ Math.\ Phys. {\bf 7}, 863 (1966).

\bibitem{Thesis:Wu}
Y. Wu, Ph.D. thesis, California Institute of Technology, 1998.

\bibitem{1996ApJ...466..946K}
P. {Kumar} and J. {Goodman}, Astrophys.\ J. {\bf 466},  946  (1996).

\bibitem{unnoetal}
W. {Unno}, Y. {Osaki}, H. {Ando}, and H. {Shibahashi}, {\em Nonradial
  Oscillations of Stars}, 1st  ed. (University of Tokyo Press, Tokyo, 1979).

\bibitem{1990ApJ...355..226I}
J.~R. {Ipser} and L. {Lindblom}, Astrophys.\ J. {\bf 355},  226  (1990).

\bibitem{1996ApJ...460..827B}
L. {Bildsten}, G. {Ushomirsky}, and C. {Cutler}, Astrophys.\ J. {\bf 460},
  827 (1996).

\bibitem{astro-ph9909009}
A. Gautschy, astro-ph/9909009  (1999).

\bibitem{1987A&A...175...81R}
A. {Rocca}, Astron.\ and Astrophys. {\bf 175},  81  (1987).

\bibitem{1981MNRAS.196..371P}
J. {Papaloizou} and J.~E. {Pringle}, \mnras {\bf 196},  371  (1981).

\bibitem{1995ApJ...449..294K}
P. Kumar, C.O. Ao, and E.J. Quataert, Astrophys.\ J. {bf 449},
294 (1995).

\bibitem{1997ApJ...490..847L}
D. {Lai}, Astrophys.\ J. {\bf 490},  847 (1997).

\bibitem{1999MNRAS.308..153H}
W.~C.~G. {Ho} and D. {Lai}, \mnras {\bf 308},  153  (1999).

\bibitem{1995MNRAS.277..471S}
G.~J. {Savonije}, J.~C.~B. {Papaloizou}, and F. {Alberts}, \mnras {\bf 277},
  471 (1995).

\bibitem{1997MNRAS.291..633S}
G.~J. {Savonije} and J.~C.~B. {Papaloizou}, \mnras {\bf 291},  633 (1997).

\bibitem{1999A&A...341..842W}
M.~G. {Witte} and G.~J. {Savonije}, Astron.\ and Astrophys. {\bf 341},  842
  (1999).

\bibitem{lbo}
D. {Lynden-Bell} and J. {Ostriker}, \mnras {\bf 136},  293  (1967).

\bibitem{gmoderefs}
L.S. Finn, \mnras {\bf 227}, 265, (1987); 
A. Reisenegger and P. Goldreich, \apj, {\bf 395}, 240, (1992); {\it
ibid} \apj, {\bf 426}, 688, (1994); D. Lai, \mnras {\bf 307}, 1001
(1999).

\bibitem{Thesis:Lockitch}
K. Lockitch, Ph.D. thesis, University of Wisconsin-Milwaukee, 1999.

\bibitem{1999ApJ...521..764L}
K.~H. {Lockitch} and J.~L. {Friedman}, Astrophys.\ J. {\bf 521},  764  (1999).

\bibitem{1980SSRv...27..653S}
P. {Smeyers}, Space Science Reviews {\bf 27},  653  (1980).

\bibitem{1981A&A....94..126P}
J. {Provost}, G. {Berthomieu}, and A. {Rocca}, \aap {\bf 94}, 126 (1981).

\bibitem{1982ApJ...256..717S}
H. {Saio}, Astrophys.\ J. {\bf 256},  717  (1982).

\bibitem{yoshidaleenonisentropic}
S. Yoshida and U. Lee, Astrophys. J. Supp. {\bf 129}, 353 (2000).

\bibitem{1987AcA....37..313D}
W. {Dziembowski} and A. {Kosovichev}, Acta Astronomica {\bf 37},  313  (1987).

\bibitem{1998A&A...334..911S}
F. {Soufi}, M.~J. {Goupil}, and W.~A. {Dziembowski}, \aap {\bf 334},  911
  (1998).

\bibitem{1978MNRAS.182..423P}
J. {Papaloizou} and J.~E. {Pringle}, \mnras {\bf 182},  423  (1978).

\bibitem{PhysrevD...59...044009}
L. Lindblom and J.~R. Ipser, Phys.\ Rev.\ D {\bf 59},  044009  (1998).

\bibitem{gr-qc9902052}
L. Lindblom, G. Mendell, and B.~J. Owen, Phys. Rev. D {\bf 60}, 064006 (1999).

\bibitem{yoshidaleeisentropic}
S. {Yoshida} and U. {Lee}, Astrophys.\ J. {\bf 529},  997  (2000).

\bibitem{astro-ph9901532}
S. Yoshida, S. Karino, S. Yoshida, and Y. Eriguchi, \mnras {\bf 316},
L1 (2000).

\bibitem{gr-qc0008019}
K.~H. {Lockitch}, N. {Andersson}, and J.~L. {Friedman}, Phys. Rev. D
{\bf 63}, 024019 (2001).

\bibitem{1980A&A....89..314S}
Y. {Sobouti}, Astron.\ and Astrophys. {\bf 89},  314  (1980).

\bibitem{1994A&A...286..879V}
T. {van Hoolst}, Astron.\ and Astrophys. {\bf 286},  879  (1994).

\bibitem{1989ApJ...342..558K}
P. {Kumar} and P. {Goldreich}, Astrophys.\ J. {\bf 342},  558  (1989).

\bibitem{Wuref}
See Appendix G of Ref. \cite{Thesis:Wu}, but note that the coupling
coefficient $\kappa_{\rm there}$ used there is related to the coupling
coefficient $\kappa_{\rm here}$ used here by 
$$
\kappa_{\rm there} = - {1 \over 3} \kappa_{\rm here}.
$$
Note also that the right hand sides of Eqs. (G43) and (G44) of
Ref. \cite{Thesis:Wu} should be multiplied by $-1$.

\bibitem{Graff}
G. Berthomieu, G. Gonczi, Ph. Graff, J. Provost, and A. Rocca, 
\aap, {\bf 70}, 597 (1978).

\bibitem{bishopgoldberg}
R. Bishop and S. Goldberg, {\em Tensor analysis on manifolds.} (McMillan, New
  York, 1968).

\bibitem{sharonprivate}
S. {Morsink}, Private communication  (2000).

\bibitem{edmonds}
A.~R. Edmonds, {\em Angular Momentum in Quantum Mechanics.} (Princeton
  University Press, Princeton, New Jersey, 1968).

\bibitem{jmp...12...1763}
W.~B. Campbell, J.\ Math.\ Phys. {\bf 12},  1763  (1971).

\bibitem{eckart}
C. Eckart, Phys. Rev. {\bf 54}, 920 (1938).

\bibitem{herivel}
J.~W. Herivel, Proc. Camb. Phil. Soc. {\bf 51},  344  (1955).

\bibitem{yourgrau}
W. Yourgrau and S. Mandelstam, {\em Variational Principles in Dynamics and
  Quantum Theory} (Dover, New York, 1968).

\bibitem{schutzsorkin}
B.\ F.\ Schutz and R.\ Sorkin, Ann. Phys. {\bf 107}, 1 (1977).

\bibitem{silva}
R. R. Silva, J. Math. Phys. {\bf 39}, 6206 (1998).

\end{thebibliography}
\end{document}